\documentclass[aps,pra,reprint,amsmath,amssymb,superscriptaddress,longbibliography,preprintnumbers]{revtex4-2}

\usepackage{graphicx}
\usepackage[colorlinks=true,urlcolor=blue,citecolor=blue,linkcolor=blue,bookmarks=false,pdfstartview={FitH}]{hyperref}
\usepackage{mathtools}

\usepackage{enumitem}
\usepackage{ulem} 
\usepackage{xcolor}
\usepackage{hyperref}
\hypersetup{
     colorlinks=true,
     linkcolor=blue,
     filecolor=blue,
     citecolor = blue,      
     urlcolor=blue
     }
\usepackage{url}
\def\TNS{Ta$_2$NiSe$_5$\,}
\def\TNSS{Ta$_2$NiS$_5$\,}
\def\TNSX{Ta$_2$Ni(Se$_{1-x}$S$_x$)$_5$\,}
\def\Tc{$T_c$\,}
\def\Tcx{$T_c(x)$\,}
\def\Tcx{$T_c(x)$\,}
\def\Tcex{$T_c^{ex}$\,}
\def\B2g{B$_{2g}$\,}
\def\Ag{A$_{g}$\,}
\usepackage{color,soul}
\begin{document}

\preprint{This is a preprint of an article published in the Physical Review B with \href{https://journals.aps.org/prb/issues/104/4}{the Editor's Suggestion}.} 
\preprint{The final authenticated version is available online at \doi{10.1103/PhysRevB.104.045102}.}

\title{
	Lattice dynamics of the Ta$_2$Ni(Se$_{1-x}$S$_x$)$_5$ excitonic insulator 
}

\author{Mai Ye}
\email{mye@physics.rutgers.edu}
\affiliation{Department of Physics and Astronomy, Rutgers University, Piscataway, NJ 08854, USA}
\author{Pavel A. Volkov}
\email{pv184@physics.rutgers.edu}
\affiliation{Department of Physics and Astronomy, Rutgers University, Piscataway, NJ 08854, USA}
\author{Himanshu Lohani}
\affiliation{Department of Physics, Technion - Israel Institute of Technology, Haifa 32000, Israel}
\author{Irena Feldman}
\affiliation{Department of Physics, Technion - Israel Institute of Technology, Haifa 32000, Israel}
\author{Minsung Kim}
\affiliation{Department of Physics and Astronomy, Rutgers University, Piscataway, NJ 08854, USA}
\author{Amit Kanigel}
\affiliation{Department of Physics, Technion - Israel Institute of Technology, Haifa 32000, Israel}
\author{Girsh Blumberg}
\email{girsh@physics.rutgers.edu}
\affiliation{Department of Physics and Astronomy, Rutgers University, Piscataway, NJ 08854, USA}
\affiliation{National Institute of Chemical Physics and Biophysics, 12618 Tallinn, Estonia}

\date{\today}

\begin{abstract}
Recently, we employed electronic polarization-resolved Raman spectroscopy to reveal the strongly correlated excitonic insulator (EI) nature of Ta$_2$NiSe$_5$ [Volkov {\it et al.}, \href{https://www.nature.com/articles/s41535-021-00351-4}{npj Quant. Mater. 6, 52 (2021)}], 
and also showed that for Ta$_2$Ni(Se$_{1-x}$S$_x$)$_5$ alloys the critical excitonic fluctuations diminish with sulfur concentration $x$ exposing a cooperating lattice instability that takes over for large $x$ [Volkov {\it et al.}, \href{https://arxiv.org/pdf/2104.07032.pdf}{arXiv:2104.07032}]. 
Here we focus on the lattice dynamics of the EI family Ta$_2$Ni(Se$_{1-x}$S$_x$)$_5$ ($x = 0$, ..., 1). 
We identify all Raman-active optical phonons of $A_g$ (fully symmetric) and $B_{2g}$ ($ac$-quadrupole-like) symmetries ($D_{2h}$ point group) and study their evolution with temperature and sulfur concentration. 
We demonstrate the change of selection rules at temperatures below the orthorhombic-to-monoclinic transition at \Tcx that is related to the EI phase. 
We find that \Tcx decrease monotonically from 328\,K for \TNS to 120\,K for \TNSS and that the magnitude of lattice distortion also decreases with the sulfur concentration $x$.
For $x < 0.7$, the two lowest-frequency \B2g phonon modes show strongly asymmetric lineshapes at high temperatures due to Fano interference with the broad excitonic continuum present in a semimetallic state. 
Within the framework of extended Fano model, we develop a quantitative description of the interacting exciton-phonon excitation lineshape, enabling us to derive the intrinsic phonon parameters and determine the exciton-phonon interaction strength, that affects the transition temperature \Tcx.
While at low temperatures the intrinsic phonon parameters are in good agreement with the {\it ab initio} calculations and the anharmonic decay model, their temperature dependencies show several anomalous behaviors: 
(i) Frequencies of \B2g phonons harden pronouncedly upon cooling in vicinity of \Tcx for $x < 0.7$ semimetals and, in contrast, soften monotonically for Ta$_2$NiS$_5$ semiconductor; 
(ii) The lifetime of certain phonons increases strongly below \Tcx for $x < 0.7$ revealing the gap opening in the broken symmetry phase; 
(iii) For most modes, the intensity shows rather strong temperature dependence hat we relate to the interplay between electronic and phononic degrees of freedom.
For \TNS we also observe signatures of the acoustic mode scattered assisted by the structural domain walls formed below $T_c$. 
Based on our results, we additionally present a consistent interpretation of the origin of oscillations observed in time-resolved pump-probe experiments.
	
\end{abstract}

\maketitle

\section{Introduction\label{sec:Intro}}

In a narrow-gap semiconductor or semimetal, if the exciton binding energy exceeds the band gap, the Coulomb attraction between electrons and holes favors spontaneous formation of a macroscopic number of excitons. These bosonic quasiparticles then form a coherent state, opening an interaction-induced gap and leading to a insulating phase, named excitonic insulator~\cite{Ex1967,Ex1968,Ex1970}. If the electron and hole states belong to bands of different symmetry, the resulting excitonic insulator state is expected to break the crystal lattice symmetry.

Consider the case of a semimetal.
If the band gap is indirect, i.e. the valence band maximum and the conduction band minimum are located at different places in the $k$-space, condensation of excitons leads to a charge-density-wave (CDW) state and accompanying crystal distortion which breaks translational symmetry~\cite{Ex1970}. 
The CDW wavevector is then given by the spanning vector that connects the valence band maximum to the conduction band minimum. 
One such example is semimetal 1T-TiSe$_2$, which has been proposed to hold an excitonic insulating state below 190\,K~\cite{TS2000,TS2007,TS2011,TS2017}. In contrast, if the band gap is direct, excitonic condensation only results in lattice distortion which breaks point-group symmetry. In this situation, the size of the unit cell does not change but its shape changes. 
Ta$_2$NiSe$_5$ is a candidate for excitonic insulator of this kind.

\begin{figure}
\includegraphics[width=0.48\textwidth]{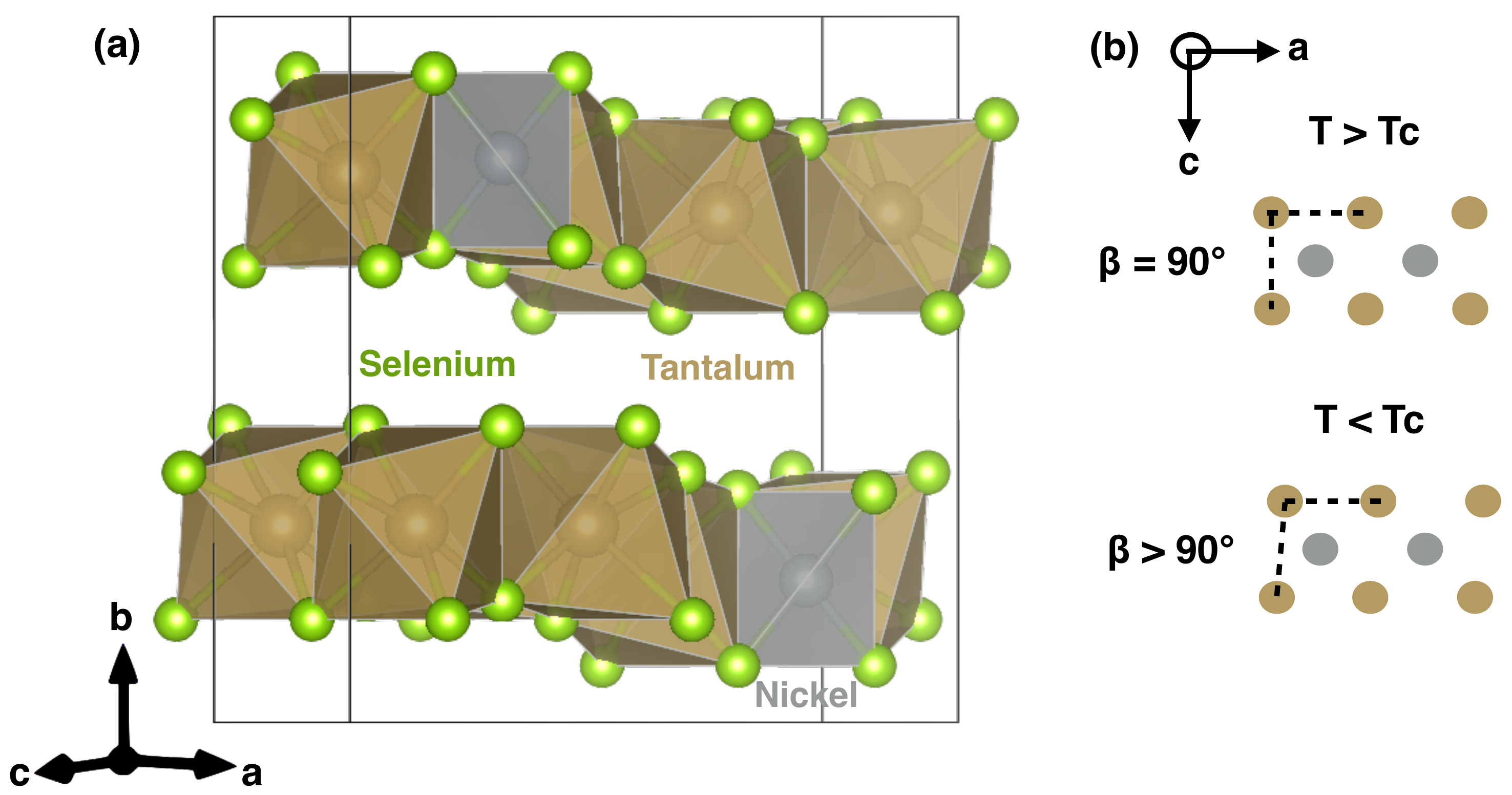}
\caption{\label{fig:Structure} Crystal structure of Ta$_2$NiSe$_5$. The frame in (a) indicates the unit cell.}
\end{figure}

Ta$_2$NiSe$_5$ shows a second-order structural phase transition at $T_c=326$\,K~\cite{Transport2017,Structure1986}. In the high temperature phase, this material has a orthorhombic structure (point group $D_{2h}$); in the low-temperature phase, the structural is monoclinic (point group $C_{2h}$) with the angle $\beta$ between $a$ and $c$ directions deviating from 90$^{\circ}$, Fig.\,\ref{fig:Structure}. Below \Tc, the valence-band top flattens as discovered by angle-resolved photoemission spectroscopy (ARPES) studies~\cite{ARPES2009,ARPES2014,ARPES2020}. Band flatness at low temperature is viewed as a characteristic feature of the excitonic insulator ground state.
%\mpar{Cite also Nanling Wang's Nat Comm}
Moreover, studies of nonequilibrium dynamics have demonstrated photoinduced enhancement of excitonic order~\cite{Fast2017,FastT2017}, photoinduced multistage phase transitions~\cite{Fast2021}, ultrafast reversal of excitonic order~\cite{Ning2020}, and coherent order-parameter oscillations~\cite{Fast2018a}.

The lattice degrees of freedom play an important role in the physics of Ta$_2$NiSe$_5$. 
In particular, the orthorhombic-to-monoclinic structural change can be induced via electron-phonon coupling~\cite{Theory2013} even if the origin of the transition is excitonic~\cite{millis2019}. 
Moreover, a recent study combining ultrafast experiments with calculations proposes that the phase transition in Ta$_2$NiSe$_5$ is structural in its nature~\cite{Baldini2020}. 
Thus, understanding the underlying phsyics of excitonic insulator candidates requires the lattice dynamics to be examined in a detailed way. Interestingly, substitution of Se with S results in a whole family of compounds Ta$_2$Ni(Se$_{1-x}$S$_x$)$_5$, where the orthorhombic-to-monoclinic transition has been suggested to be suppressed with $x$~\cite{Transport2017}. 
While it is known, that S substitution leads to an enhanced direct band gap~\cite{ARPES2019S}, which is detrimental to the EI state, its effects on the lattice modes has not been systematically studied. 

Because the inversion symmetry of Ta$_2$Ni(Se$_{1-x}$S$_x$)$_5$ is preserved across \Tc, the soft mode of this zero-wavevector structural transition, if any, is Raman active. 
Polarization resolved low frequency Raman spectroscopy, therefore, is particularly suitable for studying the physics of the phase transition, and for evaluating the contribution of the lattice vibrations to the transition. 
This experimental method offers both the high energy resolution and the ability to selectively probe bosonic excitations in different symmetry channels~\cite{Hayes2004}.

In this work, we present a systematic study of the lattice dynamics for the family of Ta$_2$Ni(Se$_{1-x}$S$_x$)$_5$ ($x=0$, 0.25, 0.67, and 1) pristine crystals and alloys to quantitatively explore the spectral parameters of the phonons. We find the change of selection rules, indicating an orthorhombic-to-monoclinic structural transition at $T_c(x)$, which is finite for all $x$. The transition temperature $T_c(x)$ and the magnitude of lattice distortion both decrease with sulfur concentration $x$. 

For \TNSX with $x \leq 0.67$, the two lowest-energy symmetry-breaking \B2g phonon modes exhibit strong asymmetric lineshape above the transition temperature, resulting from the coupling between these modes and an excitonic continuum of the same symmetry.
We develop an extended model to analyze the physics of this Fano-type interference effect. 
Using this model, we disentangle the excitonic and phononic contributions to the Raman response, and illustrate the effect of the coupling between them: apart from renormalizing the bare phononic and excitonic responses, this effect also results in an additional interference term in the total Raman response. 
We find that the \B2g phonon modes exhibit no intrinsic softening on cooling towards the transition temperature. 
In contrast, the excitonic response shows clear soft-mode behavior~\cite{Pavel2020,Pavel2021}. 
Additionally, many modes exhibit an anomalous increase in linewidth above $T_c$, suggesting the closing of the EI gap.
For \TNS below \Tc, we also observe signatures of the acoustic excitations with finite momenta, enabled by the formation of a quasi-periodic structure of domain walls.

For \TNSS, on the contrary, the phonon modes in the symmetry-breaking channel maintain symmetric lineshape, but their frequency anomalously decreases upon cooling. 
Because the structural change for \TNSS is very weak, no anomaly of phonon frequency or width is detected around its structural transition temperature. 

For all compositions, we find that most phonons exhibit strong dependence of their intensity on temperature, and present arguments in favor of this being a result of a coupling to electronic excitations.

Additionally, we perform density functional theory (DFT) calculations of the lattice dynamics for \TNS and \TNSS. 
The phonon frequencies agree well with the experiment, and the obtained displacement patterns of the optical modes allow to explain the differences of the couplings of the three non-symmetric phonons to the excitonic continuum.

Finally, we demonstrate that our results are consistent with the modes observed in time-resolved experiments. We show however, that the signatures of the Fano-shaped phonons and the excitonic continuum above $T_c$ may be hard to observe in that type of experiments.

The rest of this paper is organized as follows. In Sec.~\ref{sec:Exp} we describe the sample preparation and experimental setup. In Sec.~\ref{sec:OV} we show an overview of the phonon spectra of Ta$_2$Ni(Se$_{1-x}$S$_x$)$_5$ family, and compare the measured phonon frequencies at low temperatures with the calculated values. In Sec.~\ref{sec:PT} from the temperature dependence of the phonon intensity we obtain phase transition temperatures, which are compared with the resistance measurements. In Sec.~\ref{sec:Fano} we analyze the Fano interference feature of B$_{2g}$-symmetry Raman response. In Sec.~\ref{sec:Para} we present the temperature dependence of phonon-mode parameters: the results for B$_{2g}$-symmetry modes are given in SubSec.~\ref{subsec:ParaB}, and those for A$_{g}$-symmetry modes are given in SubSec.~\ref{subsec:ParaA}. In Sec.~\ref{sec:Time} we convert Raman response from frequency domain to time domain, and discuss relevant time-resolved studies on Ta$_2$NiSe$_5$. In Sec.~\ref{sec:Con} we provide a summary of the observations and their implications. Further discussion is provided in appendices: the methods used for the estimation of laser heating rate in Appendix.~\ref{sec:AHeat}; the mathematical formalism for the generalized Fano model in Appendix~\ref{sec:AFano}, and more illustration of this model in Appendix~\ref{sec:AFanoMore}; the fitting results of the anharmonic decay model for the phonon modes in Appendix~\ref{sec:ATable}.

\section{Experimental\label{sec:Exp}}

Single crystals of Ta$_2$Ni(Se$_{1-x}$S$_x$)$_5$ family were grown by chemical vapor transport method. Elemental powders of tantalum, nickel, selenium and sulfur were mixed with stoichiometric ratio and then sealed in an evacuated quartz ampule with a small amount of iodine as the transport agent. The mixture was placed in the hot end of the ampule ($\sim$950$^\circ$C) under a temperature gradient of about 10$^\circ$C/cm. After about a week mm-sized needle-like single crystals were found at the cold end of the ampule. The crystals are shiny and cleave easily. We used x-ray diffraction and electron dispersive X-ray spectroscopy to verify the exact composition of the crystals and their uniformity. The samples for resistance and Raman measurements are from the same batch.

The resistance is measured along a axis in a four-probe configuration using a Quantum Design PPMS system. 

The TEM images of domain structures are taken on a FEI Titan Themis G2 system using a Gatan double-tilt cryo-stage.

For Raman measurements, the samples were cleaved in ambient conditions to expose the $ac$ plane; the cleaved surfaces were then examined under a Nomarski microscope to find a strain-free area. 
Raman-scattering measurements were performed in a quasi-back-scattering geometry from the samples mounted in a continuous helium-gas-flow cryostat. 

For acquisition of the low frequency Raman response required for this study, we used a custom fast f/4 high resolution 500/500/660\,mm focal lengths triple-grating spectrometer \hl(with 1800\,mm$^{-1}$ master holographic gratings comprised of 
(i) aberration corrected subtractive stage providing a reliable 12 orders-of-magnitude stray light rejection at as low frequency as 4\,cm$^{-1}$ from the elastic line, 
(ii) a third stage monochromator, and (iii) a liquid-nitrogen-cooled charge-coupled device (CCD) detector (Princeton Instruments).  
All the acquired data were corrected for the spectral response of the spectrometer. 
Three slit configurations were used: 
100\,$\mu$m slit width providing 0.19\,meV spectral resolution; 
50\,$\mu$m slit width rendering 0.10\,meV spectral resolution; 
and, for high resolution data, 25\,$\mu$m slit width rendering 0.06\,meV spectral resolution.

For polarization optics, a Glan-Taylor polarizing prism (Melles Griot) with a better than 10$^{-5}$ extinction ratio to clean the laser excitation beam and a  broad-band 50\,mm polarizing cube (Karl Lambrecht Corporation) with an extinction ratio better than 1:500 for the analyzer was used. 
Two polarization configurations were employed to probe excitations in different symmetry channels. The relationship between the scattering geometries and the symmetry channels~\cite{Hayes2004} is given in Table~\ref{tab:Exp1}.

\begin{table}
\caption{\label{tab:Exp1}The Raman selection rules in the high-temperature orthorhombic (point group D$_{2h}$) and low-temperature monoclinic (point group C$_{2h}$) phases. Upon the reduction of symmetry from D$_{2h}$ to C$_{2h}$, the A$_{g}$ and B$_{2g}$ irreducible representations of D$_{2h}$ group merge into the A$_{g}$ irreducible representation of C$_{2h}$ group.}
\begin{ruledtabular}
\begin{tabular}{ccc}
Scattering&Symmetry Channel&Symmetry Channel\\
Geometry&(D$_{2h}$ group)&(C$_{2h}$ group)\\
\hline
aa&A$_{g}$&A$_{g}$\\
ac&B$_{2g}$&A$_{g}$\\
\end{tabular}
\end{ruledtabular}
\end{table}

The 647\,nm line from a Kr$^+$ ion laser was used for excitation. 
Incident light was focused to an elongated along the slit direction 50$\times$100\,$\mu$m$^{2}$ spot. 
For data taken below 310\,K, laser power of 8\,mW was used. To reach temperature above 310\,K, we kept the environmental temperature at 295\,K and increased laser power to reach higher sample temperature in the excitation spot. 
All reported data were corrected for laser heating in two mutually consistent ways [Appendix.~\ref{sec:AHeat}]:
(i) by Stokes/anti-Stokes intensity ratio analysis, based on the principle of detailed balance, 
(ii) by checking laser power that is inducing the phase transition. In addition to that, we performed a thermoconductivity model calculation that suggests a linear scaling of the heating rate with the beam spot size.

The  first-principle  calculations were performed using DFT within Projector Augmented-Wave (PAW) formalism \cite{paw} and Perdew-Burke-Ernzerhof (PBE)  parametrized exchange-correlation energy  functional \cite{gga}. For Ta$_2$NiSe$_5$, we used the implementation of the Quantum ESPRESSO package \cite{espresso}, using the pseudopotentials generated by Dal Corso \cite{espresso_pp} and with grimme-d3 Van-der-Waals correction \cite{grimme} included. For Ta$_2$NiS$_5$, the implementation of the  Vienna  Ab initio  Simulation Package (VASP) code \cite{vasp1,vasp2} was used. The phonon frequencies for both cases were calculated by finite displacement method as implemented in PHONOPY \cite{phonopy}. The initial structures subject to relaxation for \TNS and \TNSS have been taken from experiments on the high-temperature orthorhombic phase, Ref.\cite{Nakano2017} and Ref.\cite{Structure1985}, respectively.

%\section{Results and Discussion\label{sec:Res}}

\section{Overview\label{sec:OV}}

In this section we show an overview of the phonon spectra of Ta$_2$Ni(Se$_{1-x}$S$_x$)$_5$ family, and compare the measured phonon frequencies at low temperatures with the calculated values.

\begin{figure}
\includegraphics[width=0.48\textwidth]{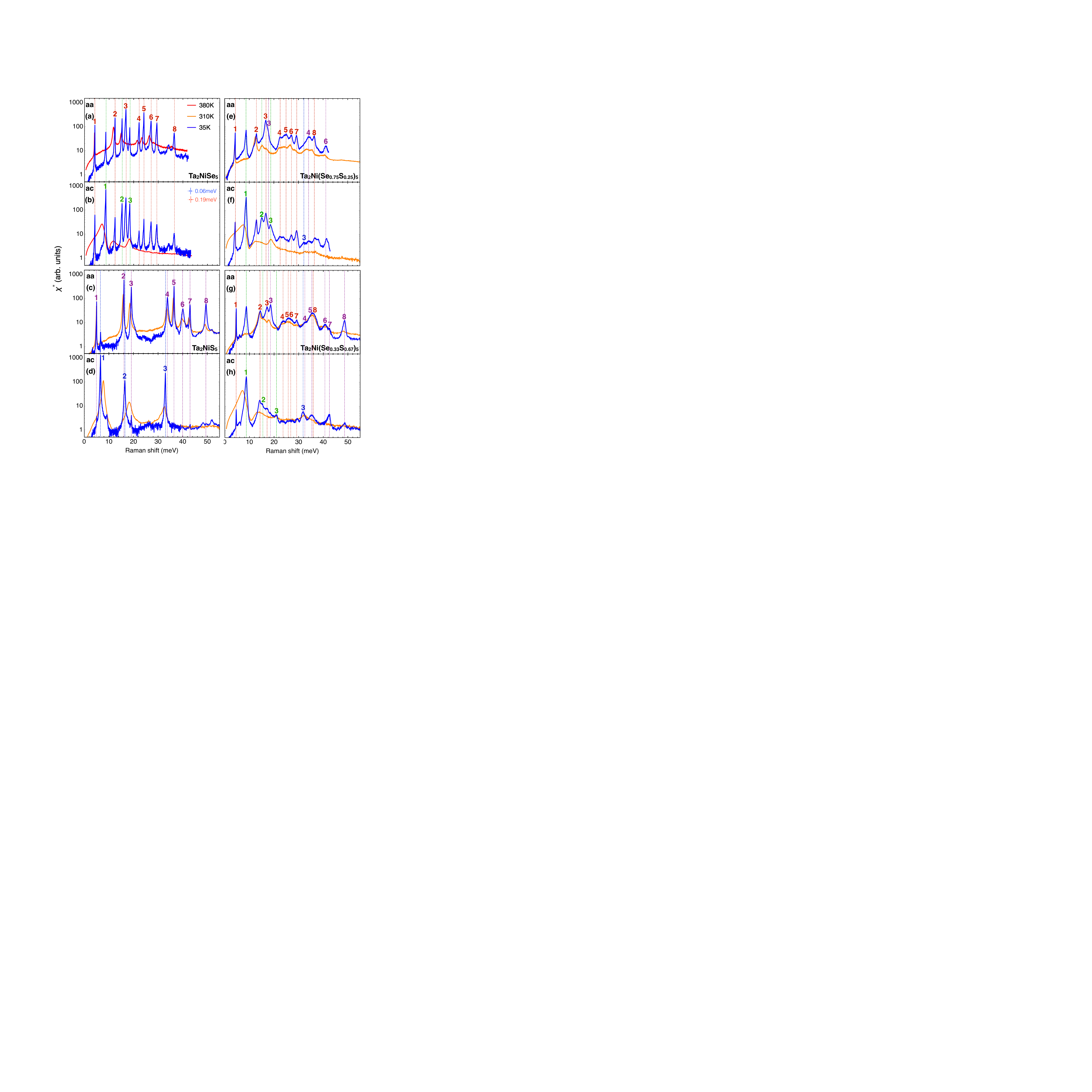}
\caption{\label{fig:OV} 
Raman response $\chi^{\prime\prime}$ in the $aa$ and $ac$ scattering geometries for Ta$_2$Ni(Se$_{1-x}$S$_x$)$_5$ family, plotted in semi-log scale. 
The phonon modes are classified by the A$_{g}$ and B$_{2g}$ irreducible representations of $D_{2h}$ point group. 
The A$_{g}$ and B$_{2g}$ modes of Ta$_2$NiSe$_5$ are labeled by red and green colors, respectively; the A$_{g}$ and B$_{2g}$ modes of Ta$_2$NiS$_5$ are labeled by purple and blue colors, respectively. 
For the samples with $x$ = 0.25 and 0.67, the modes are labeled by the color scheme of Ta$_2$NiSe$_5$ if its energy is close to that of Ta$_2$NiSe$_5$ modes, or the color scheme of Ta$_2$NiS$_5$ if its energy is close to that of Ta$_2$NiS$_5$ modes. The spectral resolution is 0.19\,meV for the $ac$ spectra at high temperature, and 0.06\,meV for the $ac$ spectra at 35\,K and all $aa$ spectra.}
\end{figure}

The phonon spectra of Ta$_2$Ni(Se$_{1-x}$S$_x$)$_5$ family ($x$=0, 0.25, 0.67, and 1) are summarized in Fig.\,\ref{fig:OV}. 
For stoichometric compositions at high temperature, we observe $8$ modes in $aa$ geometry and $3$ modes in $ac$ geometry, in agreement with the 8\,A$_{g}$ and 3\,B$_{2g}$ Raman-active phonon modes expected in the high-temperature orthorhombic phase (space group $Cmcm$): the A$_{g}$ modes appear in $aa$ scattering geometry and the B$_{2g}$ modes in $ac$ scattering geometry [Table. \ref{tab:Exp1}]. 
Except for the \TNSS sample, the two lowest-energy \B2g modes, B$_{2g}^{(1)}$ and B$_{2g}^{(2)}$, exhibit strongly broadened and asymmetric lineshapes at high temperature. This anomalous broadening is almost absent for Ta$_2$NiS$_5$ (note the logarithmic intensity scale in Fig.\,\ref{fig:OV}).

At low temperature, a transition into a monoclinic phase (space group $C2/c$) occurs. 
In the low-temperature phase, because of the mirror symmetry breaking, all Raman-active phonon modes appear in both $aa$ and $ac$ scattering geometries [Table. \ref{tab:Exp1}]. 
All of the expected Raman-active modes were observed: they are identified by dashed lines in Fig.\,\ref{fig:OV}; 
the additional weak spectral features result from the second-order scattering. 
For example, the weak feature at 34.1\,meV in $aa$ scattering geometry for \TNS corresponds to a second-order scattering feature of the 16.8\,meV B$_{2g}^{(2)}$ mode.

The measured phonon frequencies for Ta$_2$NiSe$_5$ and Ta$_2$NiS$_5$ at 35\,K generally match well with the calculated values [Table.~\ref{tab:Energy}]. 
As the calculations were carried out in the orthorhombic phase, this suggests that monoclinicity by itself does not strongly affect the phonon frequencies. 
Larger than 10\% discrepancies between the measured and calculated values appear only for the B$_{2g}^{(1)}$ and B$_{2g}^{(2)}$ modes of Ta$_2$NiSe$_5$ and the B$_{2g}^{(1)}$ mode of Ta$_2$NiS$_5$. 
As discussed in Sec.~\ref{sec:Fano}, these modes exhibit an anomalous behavior that can be attributed to their strong coupling to the excitons.

\begin{table}
	\caption{\label{tab:Energy} Comparison of the experimentally measured phonon energies at 35\,K and the calculated values for Ta$_2$NiSe$_5$ and Ta$_2$NiS$_5$. Unit is meV; modes are identified by their irreducible representations in the orthorhombic phase, where 8\,\Ag modes and 3\,\B2g modes are expected.}
	\begin{ruledtabular}
		\begin{tabular}{ccccc}
			&\multicolumn{2}{c|}{Ta$_2$NiSe$_5$}&\multicolumn{2}{c}{Ta$_2$NiS$_5$}\\
			Mode           & Exp.  & Calc.  & Exp.  & Calc.  \\
			\hline
			A$_{g}^{(1)}$  & 4.22  & 4.22  & 5.01  & 4.82  \\    
			B$_{2g}^{(1)}$ & 8.66  & 7.61  & 6.50  & 7.66  \\
			A$_{g}^{(2)}$  & 12.28 & 11.87 & 16.13 & 15.18 \\
			B$_{2g}^{(2)}$ & 15.25 & 10.14 & 16.41 & 16.20 \\
			A$_{g}^{(3)}$  & 16.76 & 15.53 & 19.13 & 18.27 \\
			B$_{2g}^{(3)}$ & 18.38 & 19.39 & 32.96 & 33.23 \\
			A$_{g}^{(4)}$  & 22.17 & 21.91 & 33.82 & 33.82 \\
			A$_{g}^{(5)}$  & 24.14 & 24.09 & 36.50 & 36.36 \\
			A$_{g}^{(6)}$  & 27.02 & 26.34 & 40.07 & 40.13 \\
			A$_{g}^{(7)}$  & 29.39 & 29.01 & 43.10 & 43.07 \\
			A$_{g}^{(8)}$  & 36.32 & 37.47 & 49.49 & 49.62 \\
		\end{tabular}
	\end{ruledtabular}
\end{table}

\begin{figure}[t]
	\includegraphics[width=0.40\textwidth]{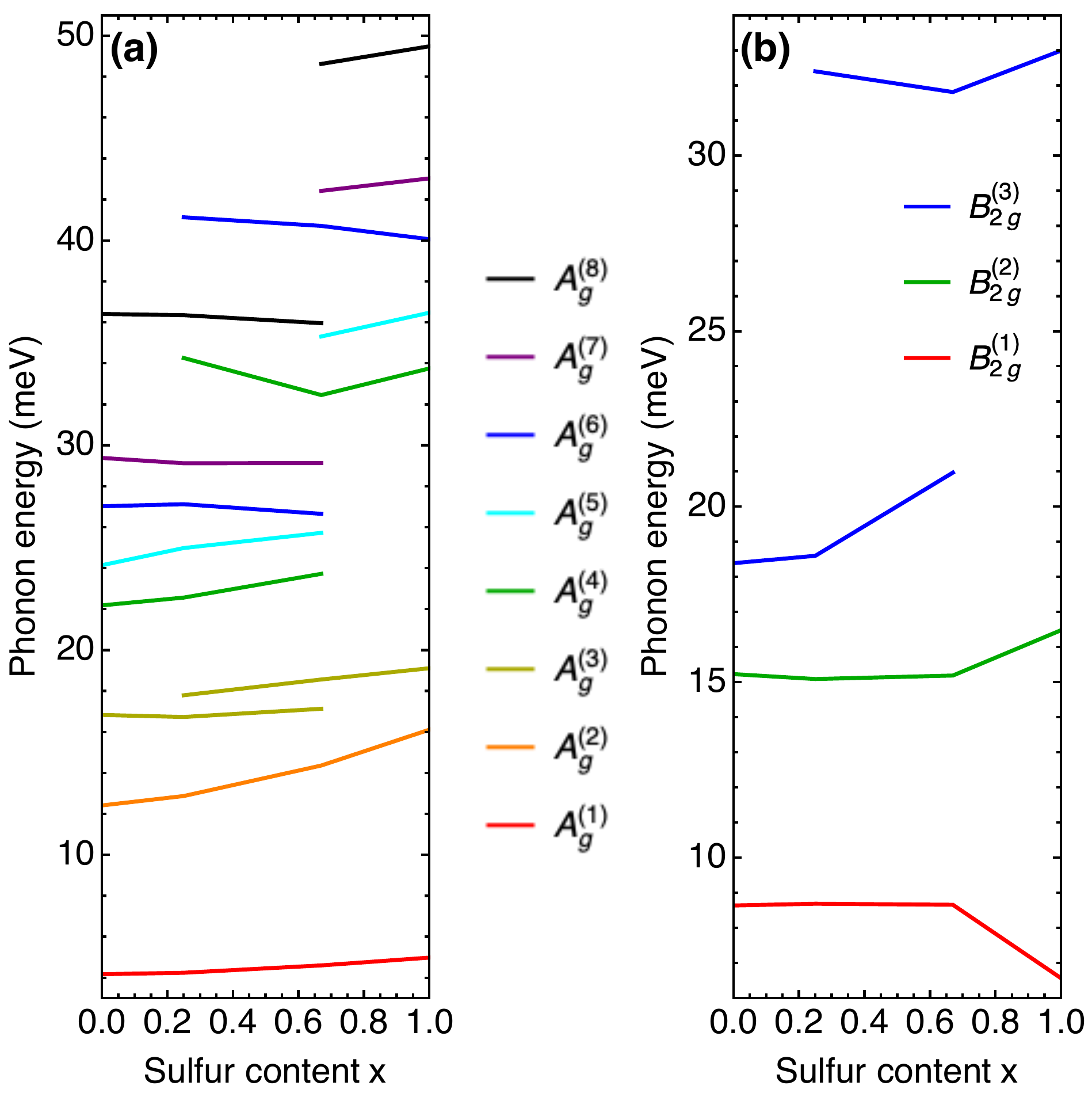}
	\caption{\label{fig:Doping} 
		The doping dependence of the Raman-active phonon frequencies at 35\,K for Ta$_2$Ni(Se$_{1-x}$S$_x$)$_5$ family. 
		For the doped samples, the B$_{2g}^{(3)}$ and A$_{g}^{(3)}$-A$_{g}^{(8)}$ phonons exhibit two-mode behavior (i.e. two peaks in the spectrum), characteristic of alloys \cite{elliot1974,barker1975,Bergman1996}.
	}
\end{figure}

For the alloy compositions ($x=0.25$ and 0.67), the phonon modes show larger linewidth compared with that for the stoichometric compositions. Moreover, the B$_{2g}^{(3)}$ and A$_{g}^{(3)}$-A$_{g}^{(8)}$ modes exhibit two distinct frequencies: one close to the frequency of that mode in Ta$_2$NiSe$_5$, and the other close to the frequency in Ta$_2$NiS$_5$. 
Such behavior is commonly observed for alloys in which the frequencies of the same phonon mode in the two end-point materials differ substantially~\cite{elliot1974,barker1975,Bergman1996}.
The doping dependence of the phonon frequencies measured at 35\,K is presented in Fig.\,\ref{fig:Doping}. 
The frequencies of the phonon modes for Ta$_2$NiSe$_5$ are consistent with the recent experimental studies~\cite{Raman2016,Raman2019,Kaiser2020,Kim2021,Kim2021Correction} and the calculated values~\cite{Subedi2020}. 

\section{Phase transition\label{sec:PT}}
In this section we deduce the phase transition temperatures from the observed change in selection rules in Raman response and compare the results with the resistance measurements.

Below $T_c$, the 11 phonon modes are allowed by symmetry to appear in both scattering geometries [Table.\ref{tab:Exp1}]; thus, 8 additional modes appear in $ac$ geometry and 3 in $aa$. 
We call the appearance of phonon modes below \Tc in the orthogonal to allowed above \Tc scattering geometry as phonon-mode "leakage". 
As shown in Fig.\,\ref{fig:OV}, all four samples exhibit phonon-mode "leakage" at low temperature, indicating presence of a point group symmetry breaking structural phase transition. 

\begin{figure}[b]
\includegraphics[width=0.48\textwidth]{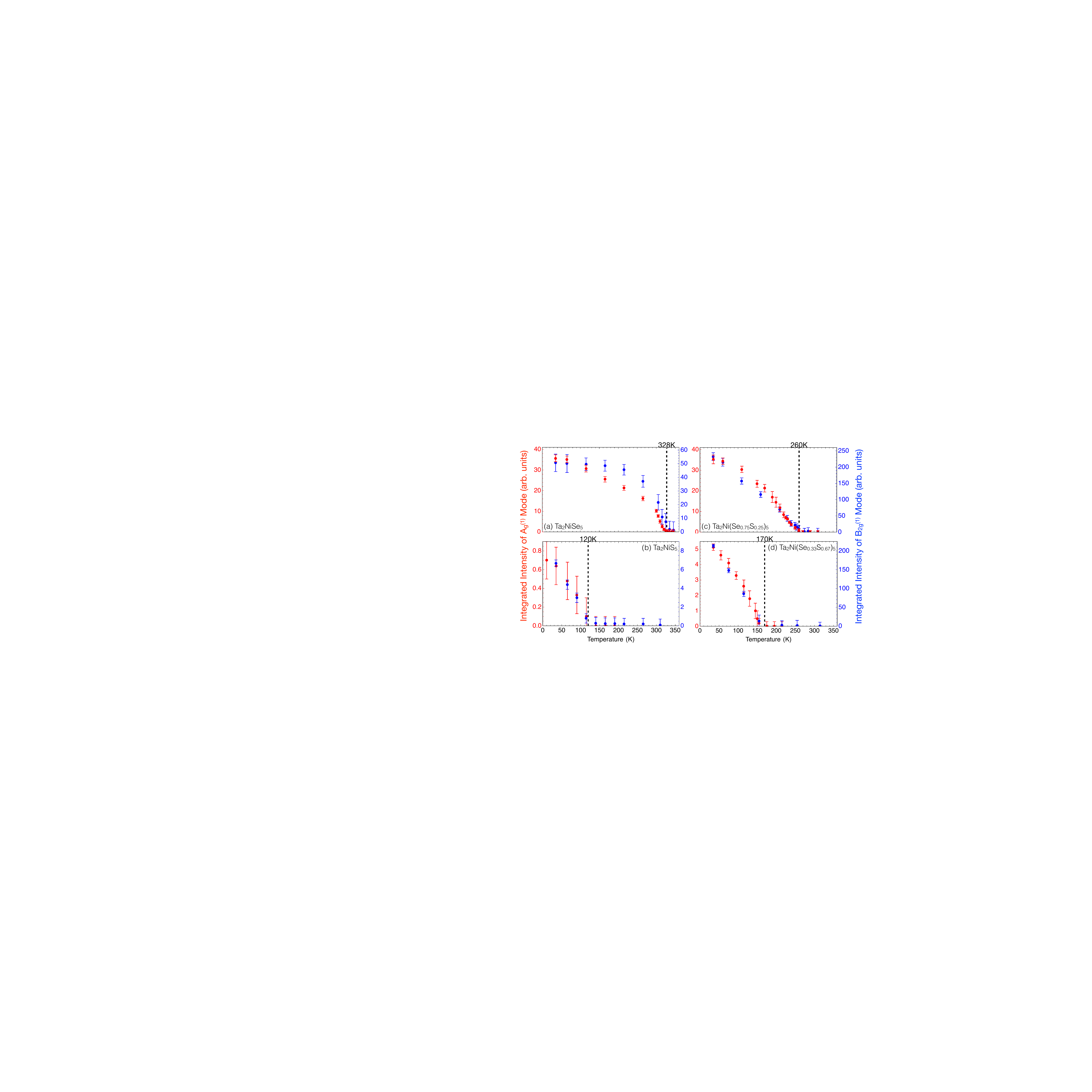}
\caption{\label{fig:Leak} Temperature dependence of the integrated intensity of A$_{g}^{(1)}$ and B$_{2g}^{(1)}$ modes in the 'forbidden' scattering geometry ($ac$ and $aa$, respectively) for (a) Ta$_2$NiSe$_5$, (b) Ta$_2$NiS$_5$, (c) Ta$_2$Ni(Se$_{0.75}$S$_{0.25}$)$_5$, and (d) Ta$_2$Ni(Se$_{0.33}$S$_{0.67}$)$_5$.}
\end{figure}

In Fig.\,\ref{fig:Leak} we show the temperature dependence of the "leakage" of A$_{g}^{(1)}$ and B$_{2g}^{(1)}$ modes for Ta$_2$Ni(Se$_{1-x}$S$_x$)$_5$ family. 
The "leakage" of other modes is given in Sec.~\ref{sec:Para}. 
For each sample, the "leakage" of different modes appear below the same temperature. 
	
We assign the temperature of the "leakage" onset as the structural phase transition temperature \Tc.
The phase transition temperatures for Ta$_2$Ni(Se$_{1-x}$S$_x$)$_5$ family are given in Table\,\ref{tab:Temp}. 
The transition temperatures $T_c(x)$ decrease with sulfur doping $x$.
Moreover, for larger sulfur content $x$ the phonon intensities in the 'forbidden' scattering geometry also decrease. 
This quantity is proportional to the square of the order parameter (OP), which can be represented by, e.g., the deviation of the angle $\beta$ between $a$ and $c$ axes [Fig.\,\ref{fig:Structure} (b)] from 90$^\circ$. 
While the general monotonic growth of the leakage intensity on cooling is consistent with these expectations, the details of temperature dependence can vary for different modes [Fig.\,\ref{fig:Leak}(a)]. 
The decreasing trend of the phonon intensity in the 'forbidden' scattering geometry with sulfur content $x$ therefore indicates that the structural distortion becomes weaker with higher sulfur concentration.

\begin{table}
\caption{\label{tab:Temp}The structural phase transition temperature of Ta$_2$Ni(Se$_{1-x}$S$_x$)$_5$ family determined by observing phonon-mode "leakage" in Raman spectra.}
\begin{ruledtabular}
\begin{tabular}{cc}
Sulfur Content $x$&Transition Temp. (K)\\
\hline
0&328$\pm$5\\
0.25&260$\pm$10\\
0.67&170$\pm$10\\
1&120$\pm$10\\
\end{tabular}
\end{ruledtabular}
\end{table}

\begin{figure}[t]
\includegraphics[width=0.40\textwidth]{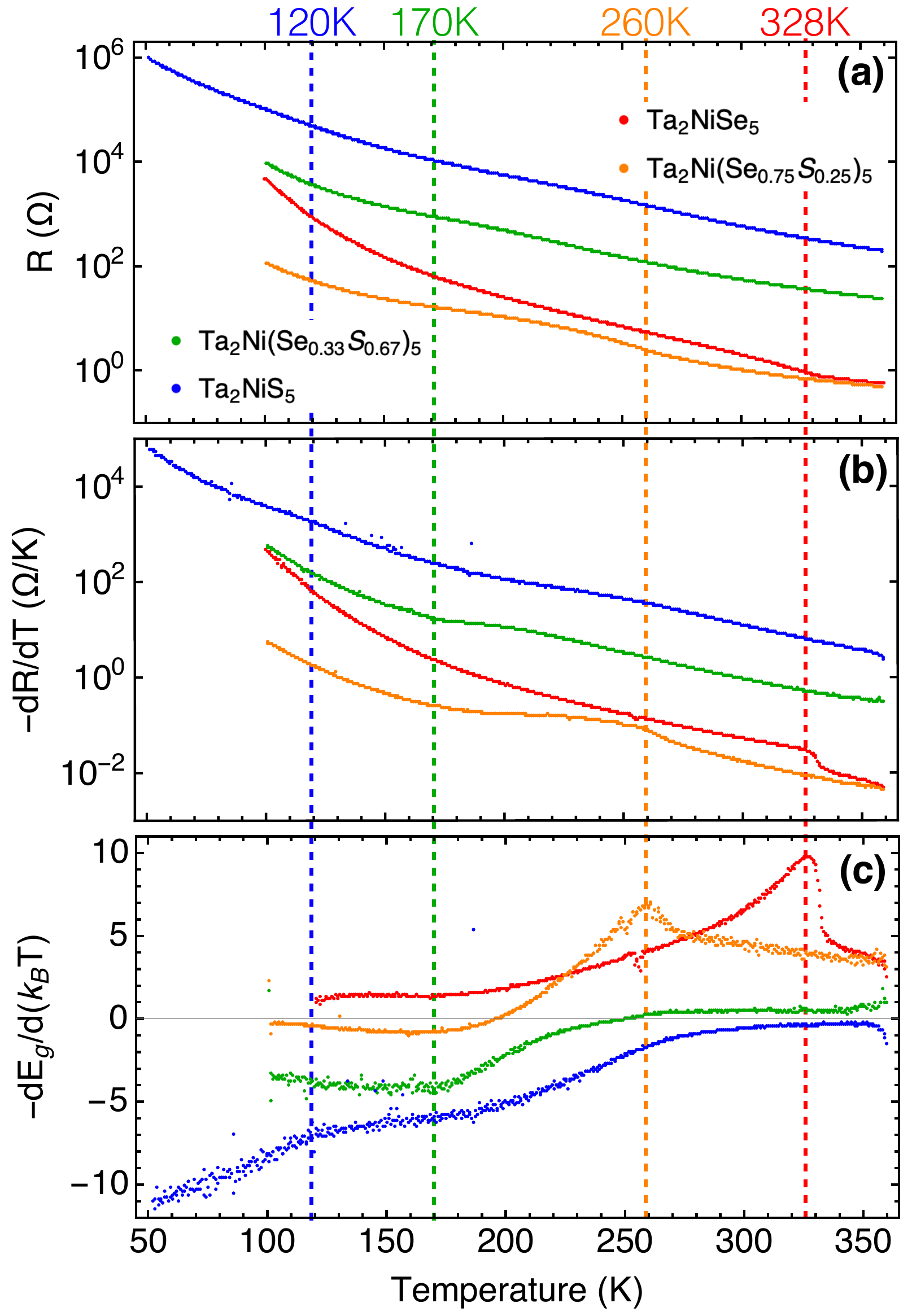}
\caption{\label{fig:Resistance} 
Temperature dependence of the resistance along $a$ axis for Ta$_2$Ni(Se$_{1-x}$S$_x$)$_5$ family. 
(a) The resistance, plotted in semi-log scale.
(b) The temperature derivative of resistance, plotted in semi-log scale. 
(c) The temperature derivative of the transport activation gap $E_{g}=k_BT\log_{10}{[R(T)/R(360K)]}$. 
The transition temperatures determined from Raman measurements are marked by dashed lines for comparison.}
\end{figure}

A complementary way to determine the phase transition temperature was demonstrated in \cite{Transport2017}, where anomalies in the temperature dependence of resistance have been detected at $T_c(x)$.
In Fig.\,\ref{fig:Resistance} we show the resistance data analysis for the samples from the same batch as used for the Raman measurements.
For Ta$_2$NiSe$_5$ and Ta$_2$Ni(Se$_{0.75}$S$_{0.25}$)$_5$, the temperature at which the temperature derivative of resistance displays a kink and that of transport activation gap shows a peak, coincides with the Raman-determined transition temperature. 
For Ta$_2$Ni(Se$_{0.33}$S$_{0.67}$)$_5$ and Ta$_2$NiS$_5$, though, the features in resistance data are less pronounced. The latter can be anticipated from their much more insulating character, potentially masking a small change of resistance due to transition on top of a large background resistance. 

\section{Fano Interference in the B$_{2g}$ symmetry channel \label{sec:Fano}}

In this section we analyze the asymmetric lineshape of the B$_{2g}$-symmetry phonon modes observed for $x<0.7$ in the high temperature phase. 
We describe the asymmetric lineshape and its temperature dependence in SubSec.~\ref{subsec:FanoOV}. 
To fit the spectra, we develop a generalized Fano model, discussed in SubSec.~\ref{subsec:FanoAT}, in which the asymmetric broadening results from the coupling of the phonon modes to an excitonic continuum.
Besides the optical phonon modes, the excitonic continuum also couples to the acoustic phonon mode. 
We discuss how such coupling enhances the transition temperature \Tcx, and leads to softening of the \B2g-symmetry acoustic mode in SubSec.~\ref{subsec:FanoAcoustic}.
The appearance of a quasi-periodic domain wall structure below \Tc introduces additional complexity, and in SubSec.~\ref{subsec:FanoBT}, we discuss how the fitting model must be modified to account for low-energy spectral features which are absent above \Tc.
The fitting results for Ta$_2$NiSe$_5$, relevant to the excitonic continuum, are presented and interpreted in SubSec.~\ref{subsec:FanoSe} (those related to the intrinsic phonon properties are shown later in Sec.~\ref{sec:Para}). 
It is demonstrated that the strengths of the interactions between individual phonons and excitonic continuum are consistent with the displacement patterns deduced from {\it ab initio} calculations. 
For comparison, the fitting results for the alloy compositions are given in SubSec.~\ref{subsec:FanoDoped}.

\subsection{Temperature dependence of asymmetric Fano lineshapes \label{subsec:FanoOV}}

\begin{figure}
\includegraphics[width=0.48\textwidth]{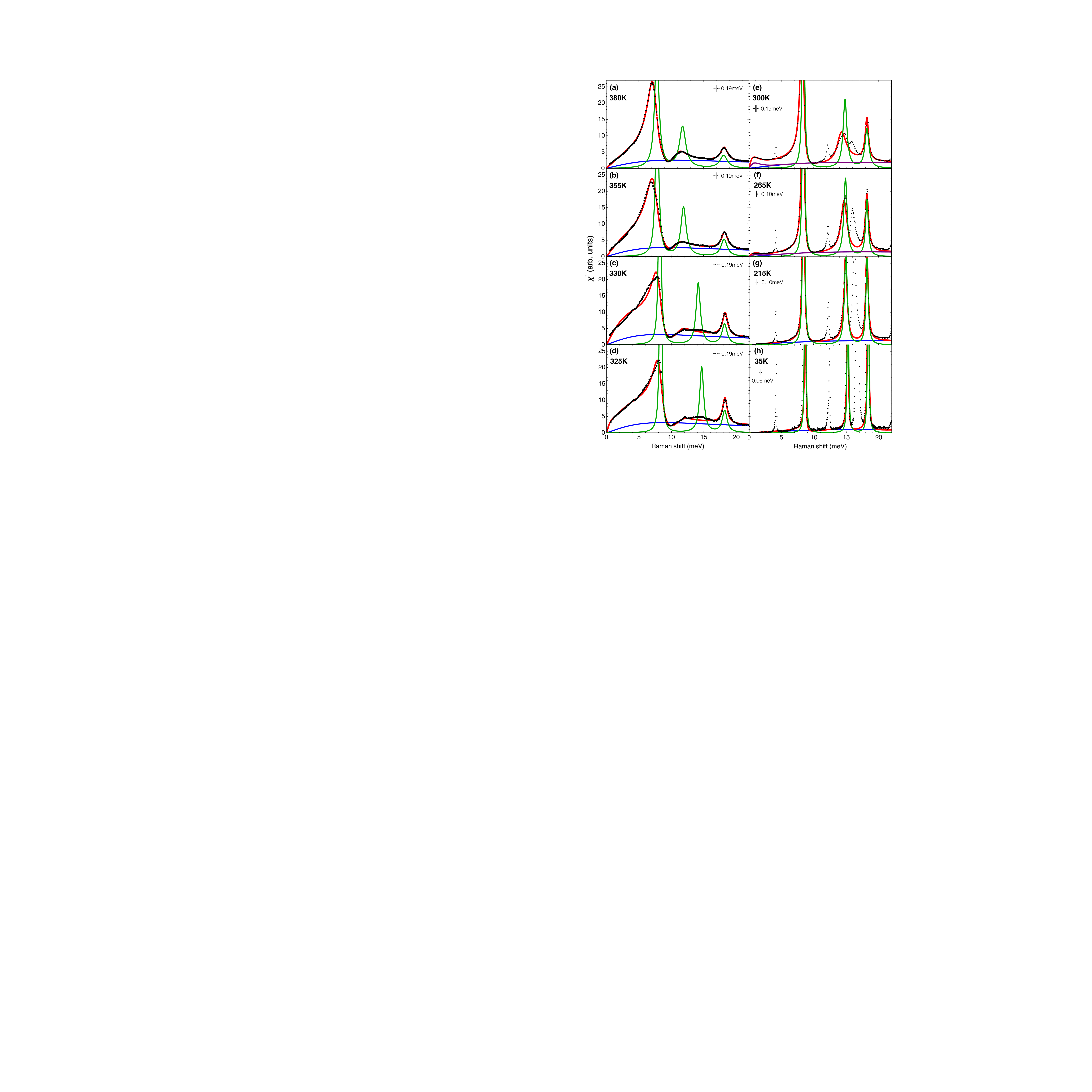}
\caption{\label{fig:SeCompo} 
Temperature dependence of Raman response $\chi^{\prime\prime}$ in the $ac$ scattering geometry for Ta$_2$NiSe$_5$. 
The data is shown by black dots, along with the Fano fits [Eq.~(\ref{eq:fanoChi})] shown as red curves. The spectral resolution is 0.19\,meV for panels (a-e), 0.10\,meV for panels (f-g), and 0.06\,meV for panel (h).
The Fano model describes the coupled dynamics of the B$_{2g}^{(1)}$-B$_{2g}^{(3)}$ phonon modes and the overdamped excitonic continuum. The bare phonon modes [Eq.~(\ref{eq:ChiP})] are shown by green curves, and the bare excitonic continuum [Eq.~(\ref{eq:ChiE})] is shown by blue curves. 
Below \Tc, especially at 300\,K, additional spectral weight is observed at low energies, resulting from the emergence of finite-energy acoustic modes due to scattering off structural domain walls (see Fig.\ref{fig:Domain}). The purple curves show the coupled response of the acoustic modes and the excitonic continuum.
As the excitonic continuum is suppressed on cooling below \Tc, the lineshapes of the phonons become essentially symmetric.
}
\end{figure}

The temperature dependence of the spectra measured in $ac$ scattering geometry for Ta$_2$NiSe$_5$ is shown in Fig.\,\ref{fig:SeCompo}. 
Above \Tc, the B$_{2g}^{(1)}$ and B$_{2g}^{(2)}$ modes have noticeably asymmetric lineshapes, 
while the lineshape of the B$_{2g}^{(3)}$ mode remains quite symmetric. 
Upon approaching \Tc on cooling, the spectral weight of low-frequency continuum
displays an enhancement, and the apparent linewidths of the B$_{2g}^{(1)}$ and B$_{2g}^{(2)}$ modes increase.  
Below \Tc, the modes recover the conventional Lorentzian lineshapes.

An asymmetric lineshape arises due to Fano interference when a sharp phonon interacts with a broad continuum. 
In this case, the continuum is due to the overdamped exciton fluctuations \cite{Pavel2020,Pavel2021}, and the exciton-phonon interaction is caused by modulation of electronic bands near the Fermi surface by the B$_{2g}$-symmetry lattice vibrations.

\subsection{The generalized Fano model for data above \Tc \label{subsec:FanoAT}}

We describe the Fano model that includes the coupling between the B$_{2g}^{(1-3)}$ phonon modes and the continuum; more details can be found in Appendix~\ref{sec:AFano}.

We first discuss the Raman response of the phonons. 
The Raman scattering intensity is related to the fluctuation spectrum of the phonon coordinate $\langle Q_{\omega} Q_{-\omega}\rangle$~\cite{barker1972},
which is given by the imaginary part of the phonon coordinate susceptibility $\chi^{\prime\prime}_{p}(\omega)$ times the Bose factor.
The phononic Raman response function consequently has the form: 
\begin{equation}
\chi^{\prime\prime(0)}_{p}(\omega)=\frac{4t_{p}^2\omega \omega_{p} \gamma_{p}}{(\omega^2-\omega_{p}^2)^2 + 2\gamma_{p}^2(\omega^2 + \omega_{p}^2)+\gamma_{p}^4}\,, 
\label{eq:ChiP}
\end{equation}
where $t_{p}$ is the light-scattering vertex for the phonon, $\omega_{p}$ is the mode's bare frequency, and $\gamma_{p}$ is the half width at half maximum (HWHM) which is related to the oscillator lifetime.

An exciton in a gapped system (a semiconductor or insulator) is expected to have a sharply peaked response at the exciton energy described by a bosonic response function similar to the  Eq.\,\eqref{eq:ChiP}. 
The most important difference for a semimetal is the presence of a gapless continuum of interband particle-hole excitations enabling the decay of the ${\bf q}=0$ exciton into unbound particle-hole pairs (Landau damping). 
To describe the continuum in data, we find it sufficient to assume a purely relaxational dynamics (corresponding to strong overdamping) for the excitonic response. 
This leads to the form
\begin{equation}
\chi^{\prime\prime(0)}_{e}(\omega)=\frac{t_{e}^2\omega}{(\omega_{e}^2/\gamma_{e})^2+\omega^2}~, 
\label{eq:ChiE}
\end{equation}
where $t_{e}$ controls the overall intensity determined by the light-scattering vertex and the excitonic density of states, $\omega_{e}$ is the frequency of the over-damped excitation, and $\gamma_{e}$ is the relaxation rate, typically much larger than $\omega_{e}$.  
Note that for the purpose of the fitting description below $\Omega_e(T) = \omega_{e}^2/\gamma_{e}$ is a single parameter denoting the temperature-dependent peak frequency in the broad response function $\chi^{\prime\prime}_{e}(\omega)$. 
For a purely excitonic transition this peak represents the soft mode:  
$\Omega_e(T_c^{el})=0$ is the condition for bare excitonic transition temperature at $T_c^{el}$, where the static excitonic susceptibility $\chi_{e}(0)$ diverges (see Eq. \eqref{eq:susE}).
	
In the absence of the exciton-phonon interaction between the modes the respective phononic and excitonic responses [Eqs.\eqref{eq:ChiP} and \eqref{eq:ChiE}] simply sum up. 
The exciton-phonon interaction, however, couples these dynamical responses: in particular, a bilinear coupling of the exciton and the \B2g phonon coordinates is expected \cite{Theory2013} (note that the bare phonons are diagonal normal modes and thus no bilinear interaction between them is present).
The resulting response can be obtained by solving the coupled equations of motion [Appendix~\ref{sec:AFano}]. 
Each of the responses $\chi^{\prime\prime}_{p}(\omega)$ and $\chi^{\prime\prime}_{e}(\omega)$ get renormalized by the interaction, and in addition, the response of the phonon coordinate to the exciton one, $\chi^{\prime\prime}_{int}(\omega)$, becomes finite. 
The latter leads to an essential interference term in the total Raman response function~\cite{fano1961,klein1983,Blumberg1994}.

To appreciate the effect of renormalization, we first consider a simplified case in which the excitonic continuum couples to a single sharp phonon mode ($\gamma_{p} \ll \omega_{p}$) [Appendix~\ref{sec:AFanoMore}]. 
Then, solving the coupled dynamical equations, one obtains the full Raman response function
\begin{multline}
\chi^{\prime\prime}(\omega)=\\
\frac{\omega [2 t_p v \omega_p - t_e (\omega^2 - \omega_p^2)]^2}
{(\Omega_e^2 + \omega^2) (\omega^2 - \omega_p^2)^2  
+ 4 \Omega_e v^2 \omega_p (\omega^2 - \omega_p^2) + 4 v^4 \omega_p^2 }
\label{eq:ChiFull}
\end{multline}
could be broken down into a sum of three contributions: 
\begin{equation}
\chi^{\prime\prime}(\omega)=\chi^{\prime\prime}_{p}(\omega)+\chi^{\prime\prime}_{e}(\omega)+\chi^{\prime\prime}_{int}(\omega),
\label{eq:Chi}
\end{equation}
where the first two terms correspond to the phonon response proportional to square of light coupling vertex $t_p^2$ and the excitonic continuum proportional to $t_e^2$, respectively, while the third one, that is proportional to the $t_p t_e$ combination, is the interference term appearing due to the exciton-phonon coupling with strength $v$~\cite{Blumberg1994}. 

We first discuss the phonon response in the presence of the coupling with the excitonic continuum. 
As we will show now, the coupling to the continuum of overdamped excitonic excitations renormalizes the bare linewidth to 
\begin{equation}
\gamma_p^v=\frac{ v^2 \omega_p}{\omega_p^2+\Omega_e^2}\,, 
\label{eq:gamma}
\end{equation}
thus, the bare $\gamma_{p}$ can be neglected altogether~\footnote{Note that the interaction-induced linewidth $\gamma_p^v$ is often  significantly larger than the bare width $\gamma_p$, which is the case for presented data above \Tc. Consequently, an analysis of the phonon lineshape that does not explicitly include the interaction with the continuum (e.g. by simply adding the intensities of Fano-shaped phonons and the continuum) would lead to an erroneous large intrinsic phonon linewidth.} The renormalized phonon response is given by
\begin{multline}
\chi^{\prime\prime}_{p}(\omega)= \\
\frac{4t_{p}^2 \omega \omega_p \gamma_p^v}
{\left[\omega^2-(\omega_{p}^2 - 2 \Omega_e \gamma_p^v)\right]^2
	+4 \omega_p^2 \gamma_p^{v 2}
+\frac{(\omega^2-\omega_p^2)^3}{\omega_p^2+\Omega_e^2}}.  
\label{eq:ChiPhrenorm}
\end{multline}
Comparing Eq.\,\eqref{eq:ChiPhrenorm} to Eq.\,\eqref{eq:ChiP} for the bare phonon, the phonon frequency renormalization is 
$\omega_p^2 \to \omega_p^2 - 2 \Omega_e \gamma_p^v$~\footnote{Note that the cubic term in the denominator can be neglected as long as the frequency is not far from $\omega_p$.}.
On increasing coupling strength $v$, the phononic response maximum shifts to 
$\omega_{p,max}=\omega_{p}-\frac{v^2(2\omega_{p}\Omega_{e}-v^2)}{2(v^2\Omega_{e}+\omega_{p}\Omega_{e}^2+\omega_{p}^3)}$. 

For the excitonic response $\chi^{\prime\prime}_e(\omega)$ the renormalization effects are most significant. 
Those are best illustrated by the low-frequency slope of the excitonic response function $\left.\frac{\partial \chi^{\prime\prime}_e(\omega)}{\partial \omega}\right|_{\omega \to 0}$. 
In the absence of exciton-phonon interaction the low frequency slope for the bare response is equal to $t_e^2/\Omega_e^2$ (see Eq.\,\eqref{eq:ChiE}), while the interaction $v$ is rapidly enhancing the slope as 
\begin{equation}
\left.\frac{\partial \chi^{\prime\prime}_e(\omega)}{\partial \omega}\right|_{\omega \to 0} = 
\frac{t_e^2}{{\left[\Omega_{e}(T) - \frac{2v^2}{\omega_{p}}\right]^{2}}}, 
\label{eq:ChiSlope}
\end{equation}
leading to a critical value at $v_{cr} = \sqrt{\Omega_{e}\omega_{p}/2}$, when the excitonic spectral weight is pushed to the lowest frequencies and thus causing the divergence in the static susceptibility 
\begin{equation} 
\chi(\omega \to 0, T) = 
\frac{t_e^2  - 4 t_e t_p \frac{v}{\omega_{p}} + 2 t_p^2 \frac{\Omega_{e}(T)}{\omega_{p}}}{\Omega_{e}(T) - \frac{2v^2}{\omega_{p}}}, 
\label{eq:Chi0}
\end{equation}
signifying the enhancement of the excitonic phase transition temperature, caused by coupling to the optical phonon.

Finally, the mutual response of the exciton and phonon coordinates appearing due to the coupling $v$ between them, leads to the sign-changing interference term 
\begin{equation}
\chi^{\prime\prime}_{int}(\omega)
=
\frac{4t_{e}t_{p}v\omega\omega_p \frac{\omega_p^2-\omega^2}{\Omega_e^2+\omega_p^2}}
{\left[\omega^2-(\omega_{p}^2 - 2 \Omega_e \gamma_p^v)\right]^2
	+4 \omega_p^2 \gamma_p^{v 2}
+\frac{(\omega^2-\omega_p^2)^3}{\omega_p^2+\Omega_e^2}}.
\label{eq:ChiV}
\end{equation}
The sign of this term depends on phase difference between the exciton and phonon oscillators. 
Because the phase of driven by light phononic oscillator is flipping to the opposite one at the resonant frequency, the sign of this term changes close to the bare phonon frequency $\omega_p$. 
Thus, this term is chiefly responsible for a skewed, asymmetric shape of the resulting Fano feature. 
Depending on the sign of $v$ the peak is skewed to the left or to the right of the original phonon frequency. 
Note that even for weak coupling $v\ll\omega_p$ this term can have an appreciable magnitude close to the phonon energy. 

\begin{figure}
\includegraphics[width=0.44\textwidth]{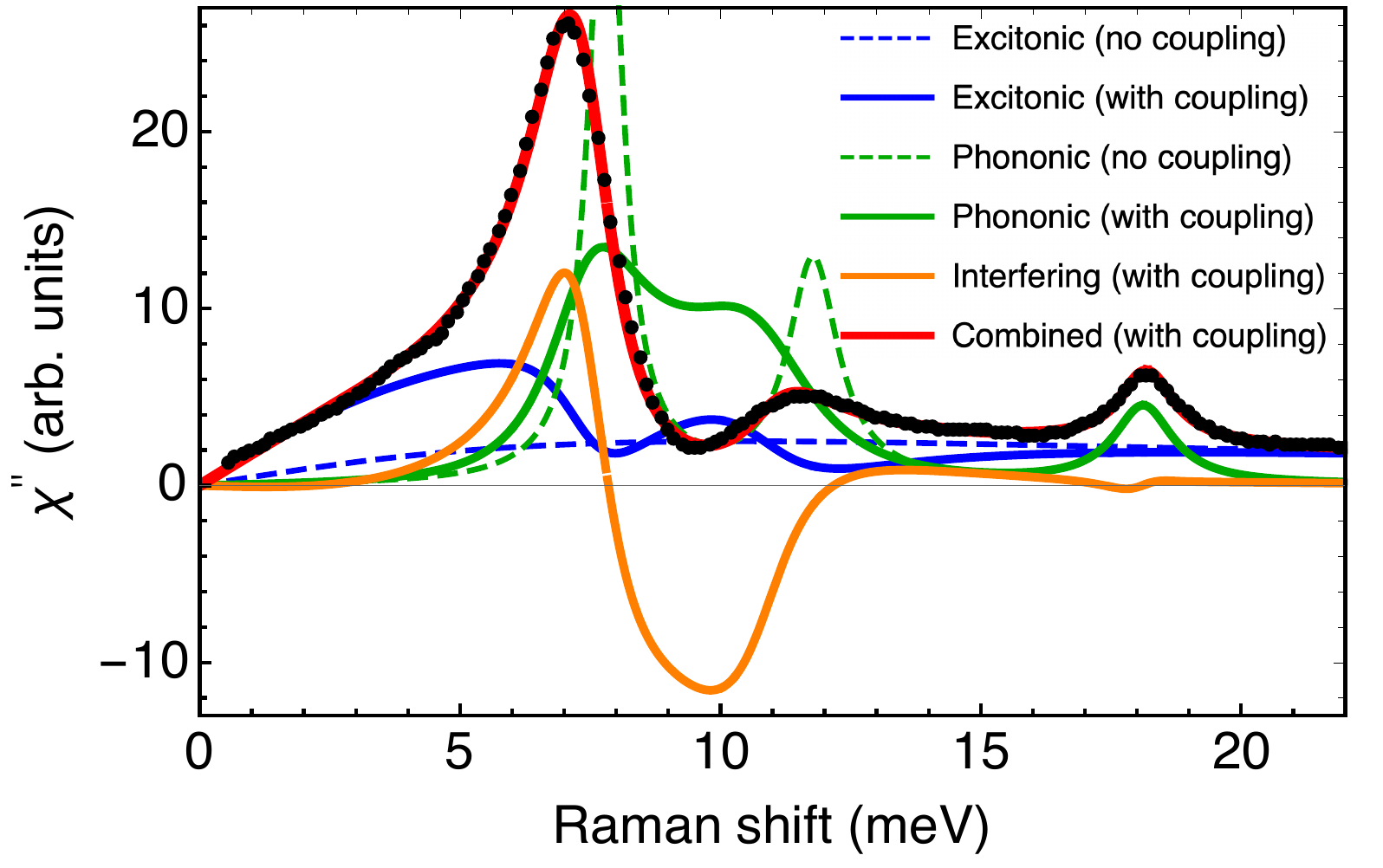}
\caption{\label{fig:SeFit} 
The effect of coupling on Raman spectrum, relevant to the data measured for Ta$_2$NiSe$_5$ at 380\,K. 
The Raman response functions are shown for bare (dashed lines) and renormalized by interaction (solid lines), $\chi^{\prime\prime(0)}_{e,p}(\omega)$ and $\chi^{{}\prime\prime}_{e,p}(\omega)$ correspondingly. 
The interference term $\chi^{\prime\prime}_{int}(\omega)$ is shown by solid orange line. 
The solid red line corresponds to the total sum of the solid blue, solid green, and solid orange lines, Eq.\,\eqref{eq:Chi}. 
The black dots represent the experimentally measured spectrum.}
\end{figure}

To analyze the actual measured low-frequency data shown in Fig.\,\ref{fig:SeCompo}, we use the model with three \B2g optical phonon modes, all interacting with the excitonic continuum, see for details Appendix~\ref{sec:AFano}.  
As above, the total Raman response $\chi^{\prime\prime}(\omega)$ can be similarly decomposed into three contributions, Eq.\,\eqref{eq:Chi}:
the first two describing the renormalized phononic and excitonic responses, respectively, with the third one arising due to the interference effects. 
In Fig.\,\ref{fig:SeFit}, an example of such decomposition is shown for the data taken at 380\,K, along with the deduced bare responses (without the effect of coupling), $\chi^{\prime\prime(0)}_{p}(\omega)$ and $\chi^{\prime\prime(0)}_{e}(\omega)$.
The renormalized phonon features become broader due to the interaction with the continuum and their frequencies shift. 
Most importantly, the coupling noticeably increases the excitonic response at low frequencies. 
The sign-changing interference term enhances the asymmetric lineshape of the combined response.

To fit the $ac$ spectra, first, we determine the $t_e$ parameter of the excitonic continuum (Eq.\,\eqref{eq:ChiE}) 
by fitting the phonon free region of spectra between 25 and 40\,meV. 
As expected, this quantity remains temperature-independent [Fig.\,\ref{fig:SeFano}(b)].
Second, we perform a global fitting of the  spectral region below 22\,meV for all measured temperatures.
We assume (consistently with the expectation due to anharmonic effects, see Eq. (\ref{eq:energyTwo}-\ref{eq:gammaTwo}) below) the linewidth of the three B$_{2g}$-symmetry phonon modes to be a linear function of temperature above \Tc. For spectra below \Tc, we fit the "leaked" phonon modes with the Lorentzian lineshapes, and account for their contribution in the global fitting.

\subsection{The coupling between excitonic continuum and acoustic modes \label{subsec:FanoAcoustic}}

Besides the optical phonon modes, the excitonic continuum also couples to the acoustic phonon modes. 
The acoustic branch has a linear dispersion $\omega_s(q) = c_s q$ ($c_s$ is sound velocity) at small wavectors $q$; its coupling to light and to the excitonic continuum both vanish in the long-wavelength limit~\cite{gallais2016}: $t_s = \tau_s \sqrt{q}$ and $v_s = \beta_s \sqrt{q}$ ($\tau_s$ and $\beta_s$ are constants). 
However, the acoustic mode has important contribution to the enhancement of the phase-transition temperature~\cite{Bohmer2014}. 
The the diverging condition for static Raman susceptibility $\chi(\omega \to 0, T)$ is defined by coupling to all modes, optical and acoustic:
	\begin{equation} 
	\tilde{\Omega}_{e}(T_c) - \frac{2\beta_s^2}{c_s} = 0,
	\label{eq:Chi0All}
	\end{equation}
	in which
	\begin{equation} 
	\tilde{\Omega}_{e}(T) = \Omega_{e}(T)-\sum_{i=1}^3\frac{2 v_i^2}{\omega_{pi}}.  
	\label{eq:Chi0AllD}
	\end{equation}
This equation determines the symmetry-breaking phase transition temperature \Tc.

Because the spectroscopic Fano feature associated with the coupling between the long-wavelength longitudinal acoustic mode and the excitonic continuum lies at frequencies far below accessibility of Raman experiment, the constant $\beta_s$ cannot be derived directly from an interference feature of a typical Raman spectra. 
Nevertheless, the coupling constant $\beta_s$ has observable consequences already above $T_c$. 
In particular, the acoustic mode has been shown to soft on cooling towards \Tc~\cite{XRay2018}. 
This softening does not result from intrinsic instability of the acoustic modes; rather, it is caused by the coupling of the acoustic mode to the softening excitonic excitations [Eq.\,\eqref{eq:ChiE}].

To demonstrate this point, we solve for the lowest-energy pole of the Green's function for the interacting phononic and excitonic excitations assuming an infinitesimal $q$. 
The energy of the pole gives the renormalized (experimentally-measured) sound velocity $\tilde{c}_{s}$ above \Tc [Eq.\,\eqref{eq:A3} of Appendix~\ref{sec:AFano}]:
\begin{equation}
\tilde{c}_{s}^2 = c_{s}^2-\frac{2\beta_s^2c_{s}}{\tilde{\Omega}_{e}(T)},
\label{eq:Acoustic1}
\end{equation}
in which $c_{s}$ is the bare sound velocity in the absence of coupling to the excitons.
Eq.\,\eqref{eq:Chi0All} enables us to rewrite Eq.\,\eqref{eq:Acoustic1} as
\begin{equation}
\tilde{c}_{s}(T) = c_{s}\sqrt{1-\frac{\tilde{\Omega}_{e}(T_{c})}{\tilde{\Omega}_{e}(T)}}.
\label{eq:Acoustic5}
\end{equation}
From Eq.\,\eqref{eq:Acoustic5}, it is clear that the sound velocity softens to zero at \Tc in the absence of any intrinsic ferroelastic instability. The renormalization of the sound velocity above \Tc can be estimated by using Eq.\,\eqref{eq:Acoustic5}. 
Using the low-temperature value of the sound velocity $c_s$~\cite{XRay2018}, we can then calculate $\tilde{c}_{s}(T)$ with the parameters obtained from the Fano fits, i.e. without free parameters. 
The result yields $\tilde{c}_{s}(T=400K)/c_{s}\approx 0.65$, which is in remarkable agreement with the experimental ratio $\tilde{c}_{s,exp}(T=400K)/c_{s,exp}\approx 0.65$. 
The agreement between the estimated and experimentally-determined values suggests that in Ta$_2$NiSe$_5$, the softening of the acoustic mode results solely from the coupling to the softening excitonic excitations.

\subsection{The generalized Fano model for data below \Tc \label{subsec:FanoBT}}

\begin{figure}
\includegraphics[width=0.40\textwidth]{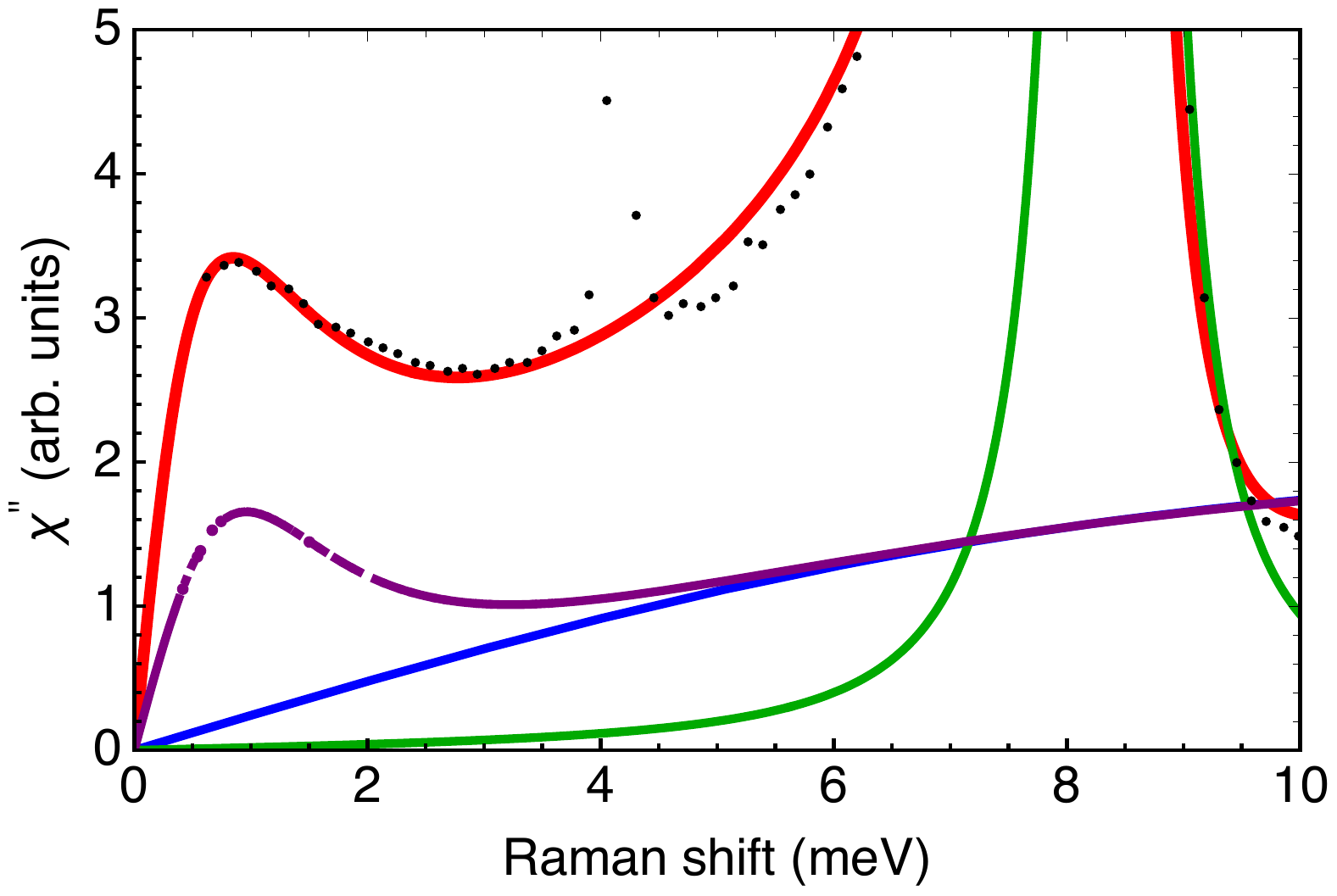}
\caption{\label{fig:Acoustic} 
The contribution from the acoustic modes to low-frequency Raman spectrum in $ac$ polarization, relevant to the data measured for Ta$_2$NiSe$_5$ at 300\,K. 
The data is shown by black dots; 
the total Fano fit is shown as red curve; 
the bare excitonic and phononic components are presented in blue and green, respectively; 
the coupled response of the acoustic and excitonic components is shown in purple.}
\end{figure}

Below \Tc, especially around 300\,K, this model appears to be insufficient to account for an additional low-frequency spectral weight, see Fig.\,\ref{fig:SeCompo}(e) and Fig.\,\ref{fig:Acoustic}.
To understand the origin of this spectral feature below a few meV, we recall that below $T_c$ formation of quasi-periodic structural domains have been observed by transmission electron microscopy (TEM)~\cite{Structure1986,Fast2021}.
Such quasi-periodic structure can take a recoil of quasi-momenta, enabling Raman coupling to acoustic mode at finite momentum $q_d= \frac{2\pi}{d}$, where $d$ is the periodicity of the domain pattern, causing the appearance of the additional spectral feature at frequency $\omega_d \approx c_s q_d$.
Therefore, fitting this ultra-low-frequency spectral feature provides an independent approach to determine the coupling constant $v_d = \beta_s \sqrt{q_d}$ between the acoustic lattice excitations and the excitonic continuum, and hence define the constant $\beta_s$ that controls the enhancement of the transition temperature to \Tc due to coupling between excitons and longitudinal strain fields of the same symmetry. 
This low-energy feature due to recoil on the quasi-periodic structure of domain walls is best seen about 30\,K below \Tc, when the excitonic continuum is sufficiently suppressed but still adequate to provide a Fano-interference feature.

\begin{figure}
\includegraphics[width=0.40\textwidth]{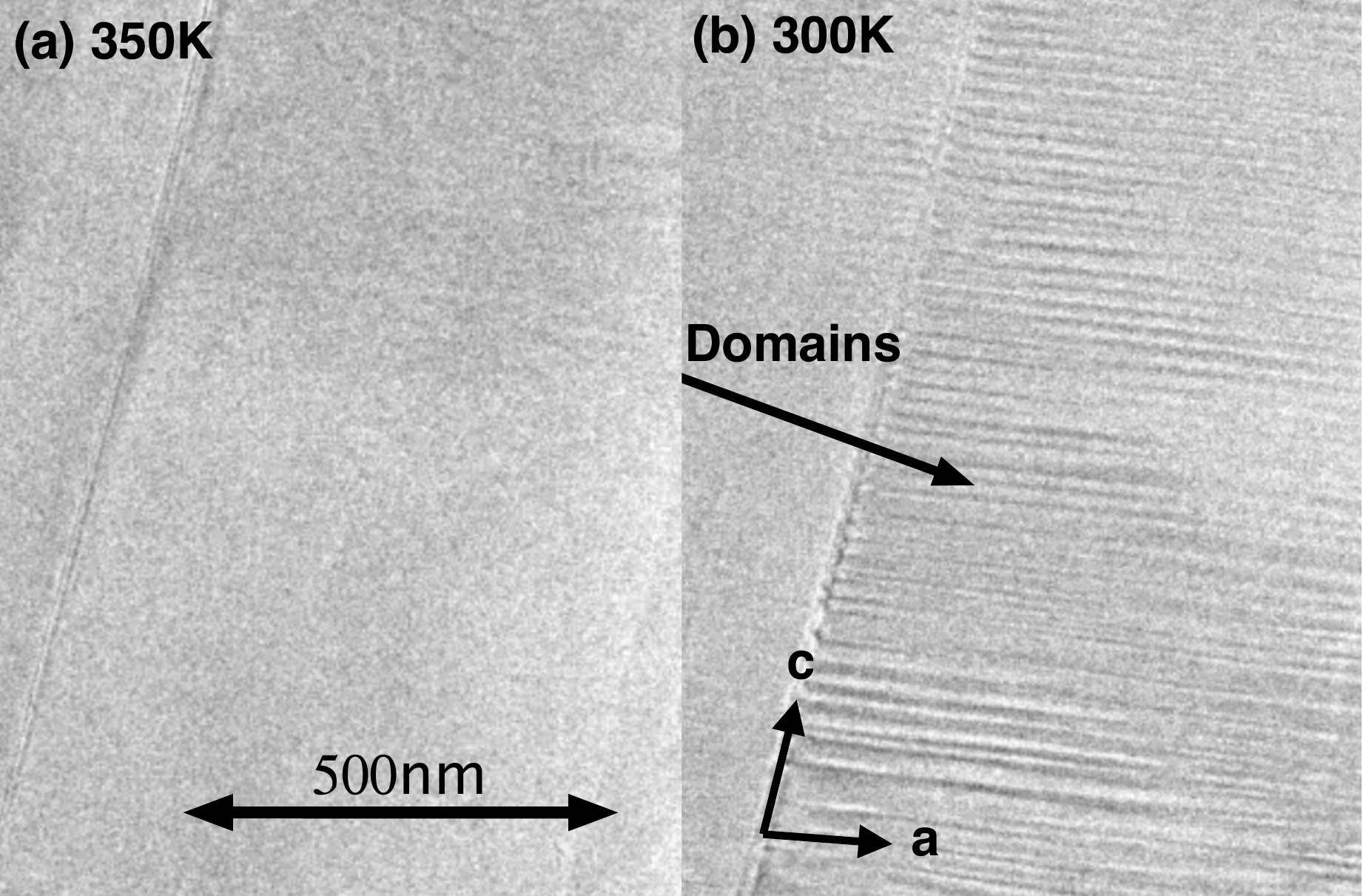}
\caption{\label{fig:Domain} 
Two images of the Ta$_2$NiSe$_5$ recorded from the $ac$ plane by transmission electron microscopy at (a) 350\,K and (b) 300\,K.}
\end{figure}

In Fig.\,\ref{fig:Domain} we compare two TEM images from $ac$ plane of Ta$_2$NiSe$_5$ measured above and below \Tc.
The domains have stripe shape, aligned parallel to the $a$-axis. 
The average spacing $\bar{d}$ between these stripes along $c$-axis direction is on the order of 200\,$\AA$. 
For the purpose of spectral analysis, we will assume that the spacing between stripes $i$ in units of the crystal unit cell constant $c$ follows the Poisson distribution
\begin{equation} 
P(i; \mu = \frac{\bar{d}}{c}) = \frac{\mu^i e^{-\mu}}{i!}, 
\label{eq:Poisson}
\end{equation}
where $\mu$ is average inter-domain distance in the number of $c$-direction unit cells. 

We deduce the unrenormalized speed of sound from the dispersion of the acoustic mode measured by inelastic x-ray scattering at temperatures far away from \Tc: $c_s \approx 30$\,meV\,\AA, see Ref.~\cite{XRay2018}. 
We also note that at high temperatures the acoustic excitations away from long-wavelength limit are significantly broadened~\cite{XRay2018}.

Recognizing that the low-frequency feature is inhomogeneously broadened by random distribution of the stripe distances, for the fitting procedure to the measured response function we perform summation over the distribution function:
\begin{equation} 
\chi^{\prime\prime}(\omega)=\sum_{i=1}^{\infty}P(i; \mu)\chi^{\prime\prime}_i(\omega),
\label{eq:ChiAcous}
\end{equation}
in which each $\chi^{\prime\prime}_i(\omega)$ is the full response function containing coupling to acoustic mode for domain walls $i$ lattice constants apart, see Eq.~\eqref{eq:fanoChii} in Appendix.
As a reminder, for each individual component, the frequency and HWHM are proportional to $q_i$; the light-scattering vertex and coupling are proportional to $\sqrt{q_i}$.

\subsection{The fitting results for Ta$_2$NiSe$_5$ \label{subsec:FanoSe}}

The fits to the data, as well as the phononic and excitonic components, are shown in Fig.~\ref{fig:SeCompo}. 
For the 300\,K and 265\,K spectra below \Tc we also include the acoustic contribution due to coupling via quasi-periodic structure of domain walls.
In Fig.~\ref{fig:Acoustic} we show the acoustic contribution at 300\,K. 
The coupling of the excitonic continuum to the acoustic components enhances the low-energy response.
The parameter $\beta = 7.7$\,meV\,\AA$^{\frac{1}{2}}$ is consistent with Eq.~(\ref{eq:Chi0All}) for \Tc.

\begin{figure}
\includegraphics[width=0.99\linewidth]{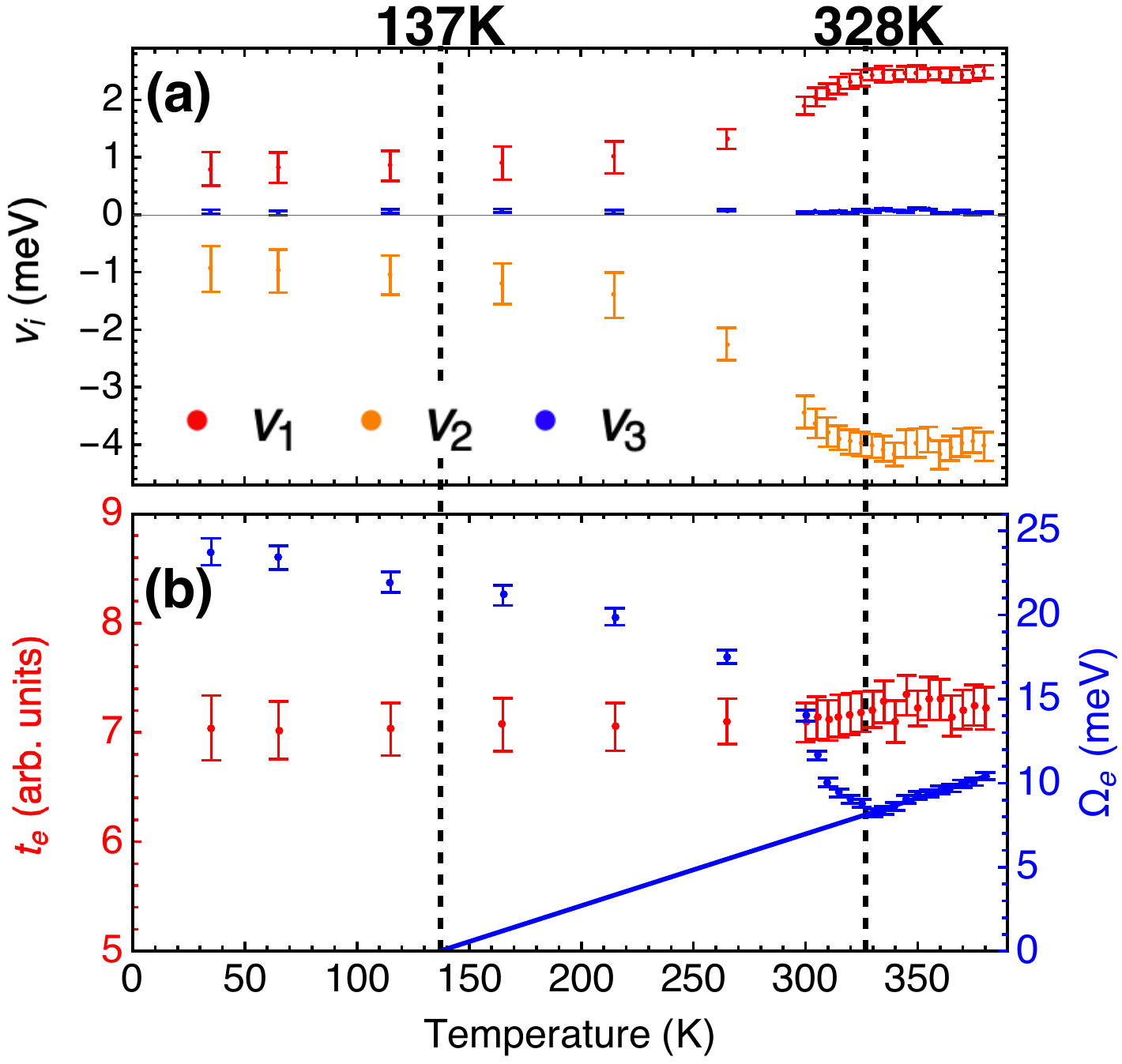}
\caption{\label{fig:SeFano} 
Temperature dependence of the Fano fitting parameters for the $ac$ Raman spectra of Ta$_2$NiSe$_5$. 
(a) The coupling strength $v_i$ (i=1,2,3) in Eq.\,\eqref{eq:fanoV}. 
(b) The light-scattering vertex $t_{e}$, and the frequency at which the excitonic continuum has maximum intensity $\Omega_e(T) = \omega_{e}^2/\gamma_{e}$, Eq.\,\eqref{eq:ChiE}. 
The blue line is a linear fit to the $\Omega_{e}(T)$ data above \Tc.
The dashed lines indicate transition temperature \Tc and bare excitonic transition temperature \Tcex.
}
\end{figure}

The spectral parameters: frequency, FWHM and integrated intensity, -- of the three B$_{2g}$ phonon modes are shown in SubSection.\,\ref{subsec:ParaA}. 
The rest of the fitting parameters are given in Fig.\,\ref{fig:SeFano}.

The temperature dependence of the coupling strength between the individual phonon modes and the excitonic continuum are detailed in Fig.\,\ref{fig:SeFano}(a).
The ratios of the coupling strength to the phonon frequency for the B$_{2g}^{(1)}$ and B$_{2g}^{(2)}$ modes above \Tc are 0.32 and -0.34, respectively. 
These ratios are an order of magnitude larger than the typical values for stable  systems with similar phonon frequency~\cite{Kung2017,Chauviere2011,Blumberg1994}.
On the contrary, the exciton-phonon interaction with the B$_{2g}^{(3)}$ mode is weak.
The signs of the coupling for the B$_{2g}^{(1)}$ and B$_{2g}^{(2)}$ modes are opposite. For these two modes, the magnitude of coupling is temperature independent above \Tc, but decreases and saturates on cooling below \Tc. The temperature dependence of the coupling strength is influenced mainly by two factors: the screening effect of free carriers, reduced below $T_c$ due to the emergence of a pronounced spectral gap \cite{IR2017,Pavel2020}, and the change of the crystal structure below $T_c$.

\begin{figure}[b]
\includegraphics[width=0.8\linewidth]{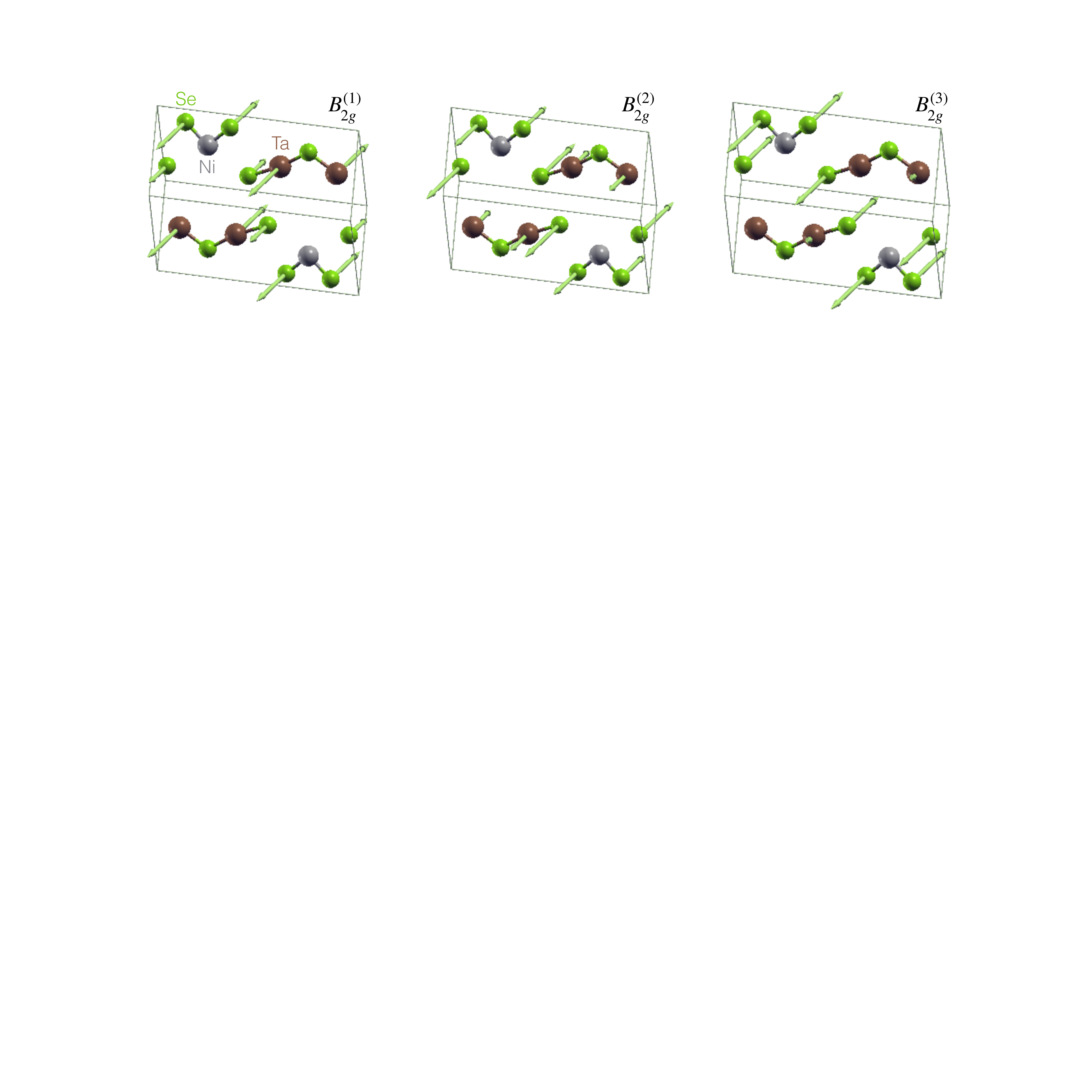}
\caption{\label{fig:Pattern}The calculated vibrational patterns corresponding to the three B$_{2g}$-symmetry phonon modes of Ta$_2$NiSe$_5$.}
\end{figure}
The difference in magnitude of the exciton-phonon coupling for B$_{2g}^{(1,2)}$ and B$_{2g}^{(3)}$ modes can be readily understood from the corresponding displacement patterns obtained from the DFT calculations for \TNS and shown in Fig.\,\ref{fig:Pattern}. 
The calculated patterns are consistent with those reported in other studies~\cite{XRay2018,Baldini2020}. 
While both the B$_{2g}^{(1,2)}$ modes involve displacements of Ta and Se, the B$_{2g}^{(3)}$ mode involve a displacement of Se almost exclusively (the Ni atoms are at the inversion centers and thus do not contribute to the Raman-active phonon modes). 
As the electronic bands in \TNS close to the Fermi level are believed to be predominantly formed by Ta and Ni electrons~\cite{Theory2013,millis2019,ARPES2020}, with much smaller contribution of Se, it appears natural that the B$_{2g}^{(3)}$ mode does not couple to the low-energy electronic degrees of freedom. 
Furthermore, the vibration of the Ta atoms for B$_{2g}^{(1)}$ and B$_{2g}^{(2)}$ displacements are in anti-phase, which can explain the opposite signs of the exciton-phonon coupling for these modes.

The temperature dependence of the light-scattering vertex $t_{e}$ and the frequency $\Omega_{e}$ at which the continuum has maximum intensity are shown in Fig.\,\ref{fig:SeFano}(b).
Above \Tc, the quantity $\Omega_{e}$ linearly decreases on cooling; below \Tc, it rapidly increases and saturates at low temperature.

\subsection{The fitting results for Ta$_2$Ni(Se$_{1-x}$S$_x$)$_5$ with $x$ = 0.25 and 0.67 \label{subsec:FanoDoped}}

\begin{figure}[b]
\includegraphics[width=0.32\textwidth]{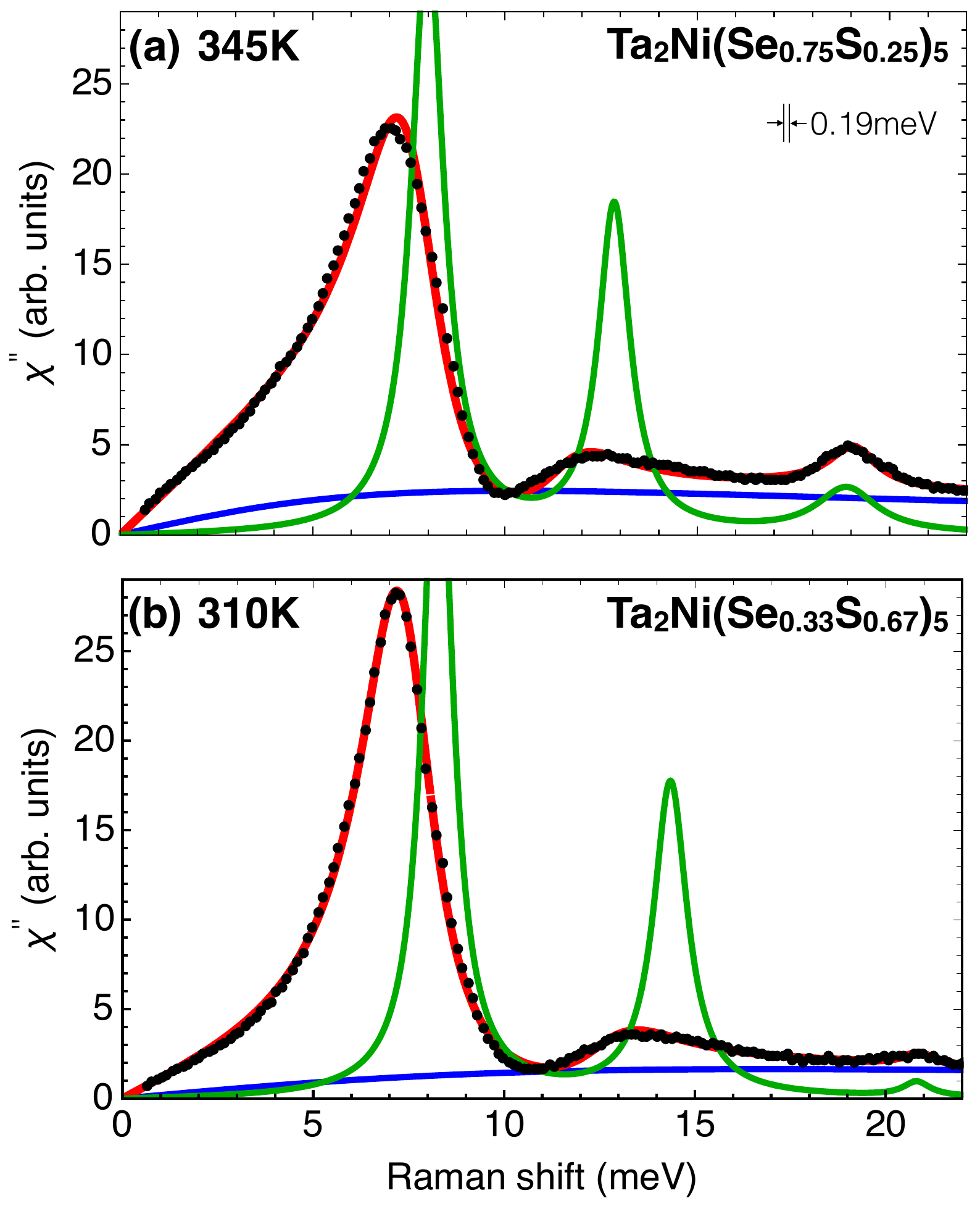}
\caption{\label{fig:DopedCompo} Raman response $\chi^{\prime\prime}$, represented by black dots, in the ac scattering geometry for Ta$_2$Ni(Se$_{0.75}$S$_{0.25}$)$_5$ and Ta$_2$Ni(Se$_{0.33}$S$_{0.67}$)$_5$ with the Fano fits [Eq.~(\ref{eq:fanoG})] shown as red curves. The spectral resolution is 0.19\,meV. The phonon modes [Eq.~(\ref{eq:ChiP})] and excitonic continuum [Eq.~(\ref{eq:ChiE})] are represented by green and blue curves, respectively.}
\end{figure}

\begin{figure}
\includegraphics[width=0.46\textwidth]{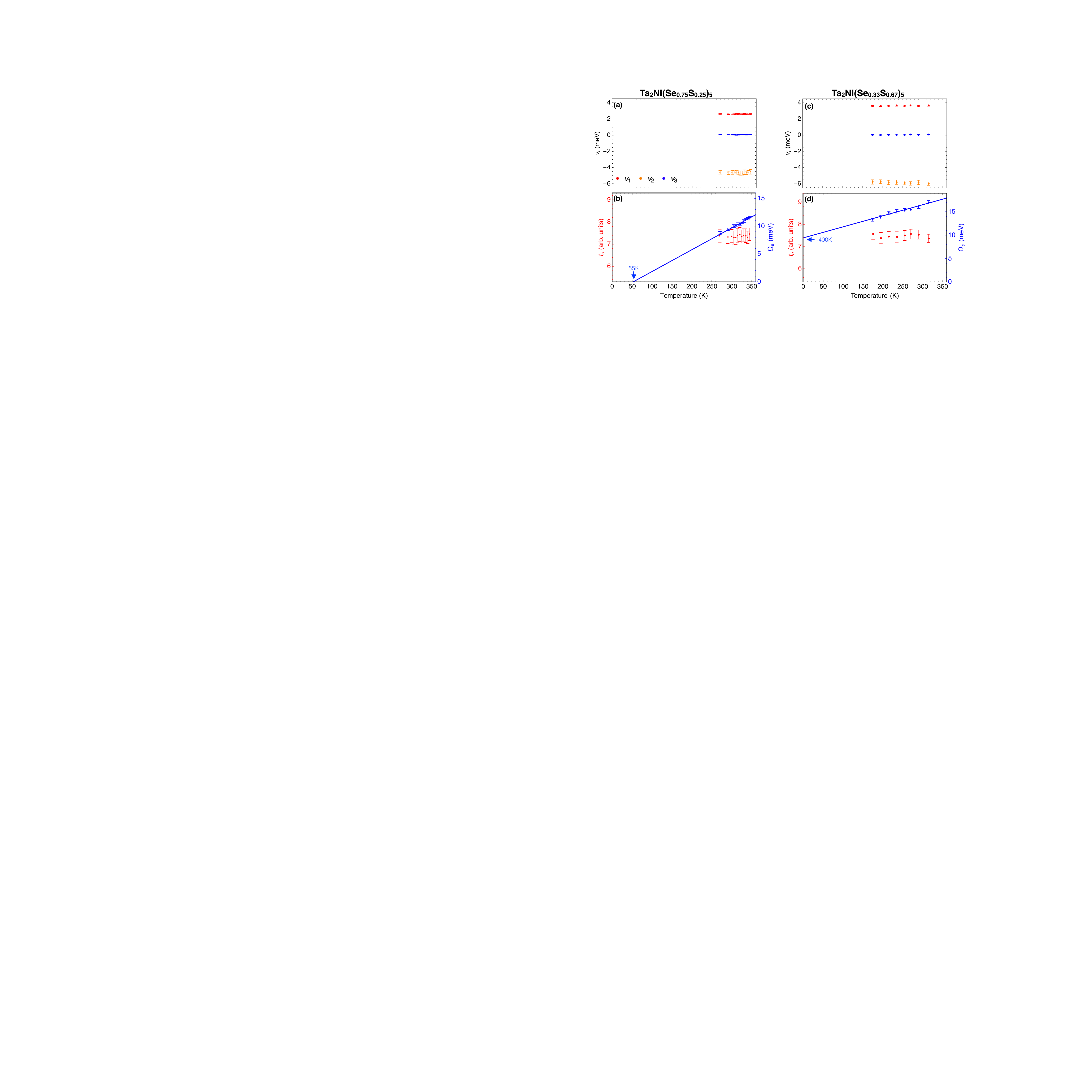}
\caption{\label{fig:DopedFano} Temperature dependence of the Fano fitting parameters for the ac Raman spectra of Ta$_2$Ni(Se$_{0.75}$S$_{0.25}$)$_5$ (a-b) and Ta$_2$Ni(Se$_{0.33}$S$_{0.67}$)$_5$ (c-d). (a,c) The coupling strength $v_i$ (i=1,2,3) in Eq.~(\ref{eq:fanoV}). (b,d) the light-scattering vertex $t_{e}$, and the energy at which the excitonic continuum has maximum intensity $\Omega_{e} = \omega_{e}^2/\gamma_{e}$ of the overdamped electronic mode, Eq.~(\ref{eq:fanoE}). The blue line is a linear fit to the $\Omega_{e}$ data.}
\end{figure}

To compare with the results for Ta$_2$NiSe$_5$, we also fit the $ac$ polarization spectra for the alloy compositions ($x$ = 0.25 and 0.67) above their transition temperature. 
We are not able to perform a Fano analysis below \Tc for two reasons: (a) we have no knowledge of the domain structure and the sound velocity; and (b) we cannot properly subtract the leakage of the A$_{g}$ modes.
In Fig.\,\ref{fig:DopedCompo} we show one characteristic Fano fit for Ta$_2$Ni(Se$_{0.75}$S$_{0.25}$)$_5$ and Ta$_2$Ni(Se$_{0.33}$S$_{0.67}$)$_5$, respectively. 
The fitting parameters are given in Fig.\,\ref{fig:DopedFano}.

The results are similar to the case of Ta$_2$NiSe$_5$: for the coupling strength, we find positive $v_1$, negative $v_2$,
and negligible $v_3$; the light-scattering vertex $t_{e}$ has no temperature dependence above the transition temperature; the energy at which the excitonic continuum has maximum intensity, $\Omega_{e} = \omega_{e}^2/\gamma_{e}$, decreases linearly on cooling above the transition temperature. Extrapolating to zero value of $\Omega_{e}$, we obtain $55\pm18$\,K for Ta$_2$Ni(Se$_{0.75}$S$_{0.25}$)$_5$ and $-400\pm70$\,K for Ta$_2$Ni(Se$_{0.33}$S$_{0.67}$)$_5$.

In Table.\,\ref{tab:v} we present the doping dependence of the relevant physical quantities above the transition temperature for Ta$_2$Ni(Se$_{1-x}$S$_x$)$_5$ family. Because the coupling strength increases with sulfur doping, determining the intrinsic phonon parameters, especially the linewidth, becomes more difficult for Ta$_2$Ni(Se$_{0.75}$S$_{0.25}$)$_5$ and Ta$_2$Ni(Se$_{0.33}$S$_{0.67}$)$_5$. 

\begin{table}[b]
\caption{\label{tab:v}The doping dependence of the scattering vertex and the coupling above the transition temperature for the interaction between the excitonic continuum and phonon modes in the B$_{2g}$ scattering geometry of Ta$_2$Ni(Se$_{1-x}$S$_x$)$_5$ family. The quantities $t_e$ is the vertex of light scattering process for the excitonic continuum. The quantities $v_i$ (i=1,2) are the coupling between the excitonic continuum and the B$_{2g}^{(i)}$ phonon mode.}
\begin{ruledtabular}
\begin{tabular}{lccc}
              &x=0&x=0.25&x=0.67\\
\hline
$t_e$ (arb. units)        &7.22&7.36&7.48\\
\hline
$v_1$ (meV)         &2.45&2.61&3.61\\
$v_2$ (meV)         &-4.03&-4.61&-5.87\\
\end{tabular}
\end{ruledtabular}
\end{table}

\section{Temperature dependence of the intrinsic phonon parameters\label{sec:Para}}

In this section we present the temperature dependence of intrinsic phonon-mode parameters: energies, intensities and linewidths. 
The results for B$_{2g}$-symmetry modes are given in SubSec.~\ref{subsec:ParaB}, and those for A$_{g}$-symmetry modes are given in SubSec.~\ref{subsec:ParaA}.

\subsection{B$_{2g}$-symmetry Phonon Modes\label{subsec:ParaB}}

\subsubsection{Ta$_2$NiSe$_5$\label{subsubsec:ParaBSe}}

\begin{figure}[b]
\includegraphics[width=0.9\linewidth]{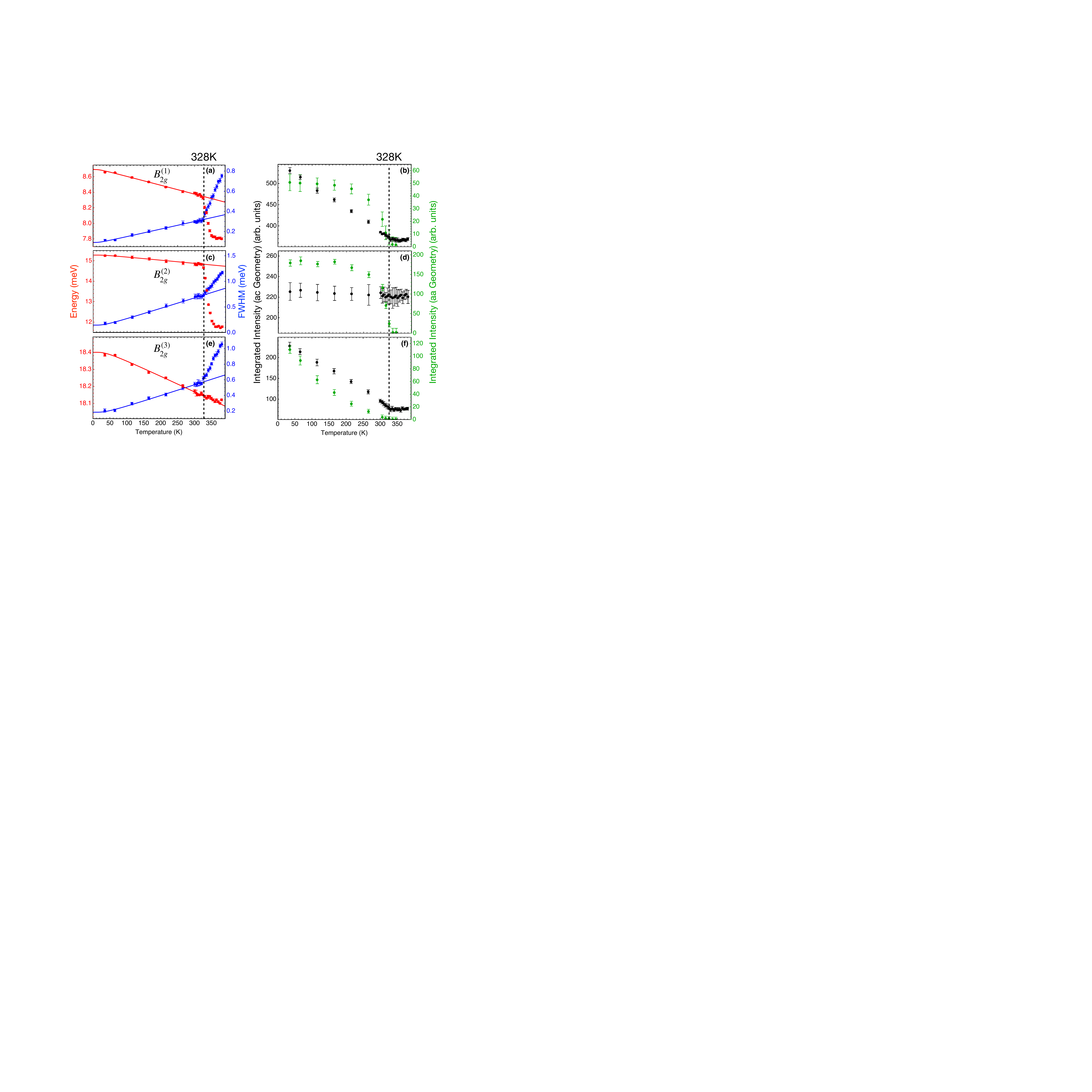}
\caption{\label{fig:ParaSe} 
Temperature dependence of the spectral parameters: the bare frequency, FWHM, and the integrated intensity, -- for the B$_{2g}$-symmetry phonon modes of Ta$_2$NiSe$_5$. 
For each mode, the left panel presents the frequency and FWHM; and the right panel presents the integrated intensity in the allowed and 'forbidden' scattering geometries. 
The solid lines represent the fits to the anharmonic decay model [Eqs.\,(\ref{eq:energyTwo}-\ref{eq:gammaTwo})].}
\end{figure}
In Fig.\,\ref{fig:ParaSe} we show the temperature dependence of the spectral parameters
for B$_{2g}$-symmetry phonon modes in Ta$_2$NiSe$_5$: the bare phonon frequency, FWHM, and integrated intensity.

Below \Tc, the temperature dependence of both frequency $\omega_p(T)$ and FWHM $\Gamma_p(T)=2\gamma_p(T)$ of the phonon modes can be accounted for by standard model assuming anharmonic decay into two phonons with identical frequencies and opposite momenta~\cite{Klemens1966}: \begin{equation}
\omega_p(T)=\omega_0-\omega_2[1+\frac{2}{e^{\hbar\omega_0/2k_B T}-1}],
\label{eq:energyTwo}
\end{equation}
and
\begin{equation}
\Gamma_p(T)=\Gamma_0+\Gamma_2[1+\frac{2}{e^{\hbar\omega_0/2k_B T}-1}].
\label{eq:gammaTwo}
\end{equation}
We note that a low value of $\Gamma_0$ indicated high quality of the crystals.

Above \Tc, however, the slope of the FWHM increases for the three phonon modes, and the FWHM becomes larger than the value predicted by the anharmonic decay model. The appearance of additional decay channels above \Tc is related to the presence of gapless particle-hole pairs. Indeed, above $T_c$ \TNS is gapless spectroscopically\cite{Pavel2020}, while below $T_c$ a gap rapidly develops at low energies, suppressing the damping due to particle-hole pairs. Because the Fano model used for fitting takes into account the damping due to exciton-phonon interaction, existence of additional damping above \Tc suggests the importance of the interactions beyond that model, i.e. nonlinear ones.

More noticeable is that the frequencies of the B$_{2g}^{(1)}$ and B$_{2g}^{(2)}$ modes exhibit a large increase on cooling from around 350\,K to \Tc. 
The frequency of the B$_{2g}^{(3)}$ mode, however, does not show such anomaly. 
Only B$_{2g}^{(1)}$ and B$_{2g}^{(2)}$ modes exhibit this anomaly; interestingly (1) they are allowed to couple to interband scattering by symmetry selection rules and (2) their corresponding vibration involve motion of Ni atoms and hence they couple to the electronic bands near Fermi level. 
Although B$_{2g}^{(3)}$ mode can also couple to the interband scattering, its vibrational pattern almost does not contain motion of Ni atoms.

As for the integrated intensity, B$_{2g}^{(1)}$ and B$_{2g}^{(3)}$ modes have around 2-fold increase of intensity on cooling below \Tc, while B$_{2g}^{(2)}$ mode shows temperature-independent intensity. 
Two factors could influence the intensity. First, the interaction-induced gap opening up below \Tc should play a major role. 
The system gradually changes from a semimetal at high temperature to an insulator at low temperature. The screening effect, which reduces the light-scattering vertex, and in turn, the intensity for phonon modes, is suppressed on cooling. 
This factor therefore favors increase of intensity on cooling. 
Second, the structural change below \Tc could play also a role. 
The ion positions are shifted within the unit cell below \Tc, and the vibrational patterns of the three B$_{2g}$ modes are modified. Therefore, the polarizability induced by lattice vibrations, which is proportional to the phonon intensity, also change with temperature. The fact that B$_{2g}^{(2)}$ mode shows temperature-independent intensity might be related to its unique vibrational pattern.

\subsubsection{The alloy compositions ($x$=0.25, 0.67)}\label{subsubsec:ParaBDoped}

\begin{figure}
\includegraphics[width=0.9\linewidth]{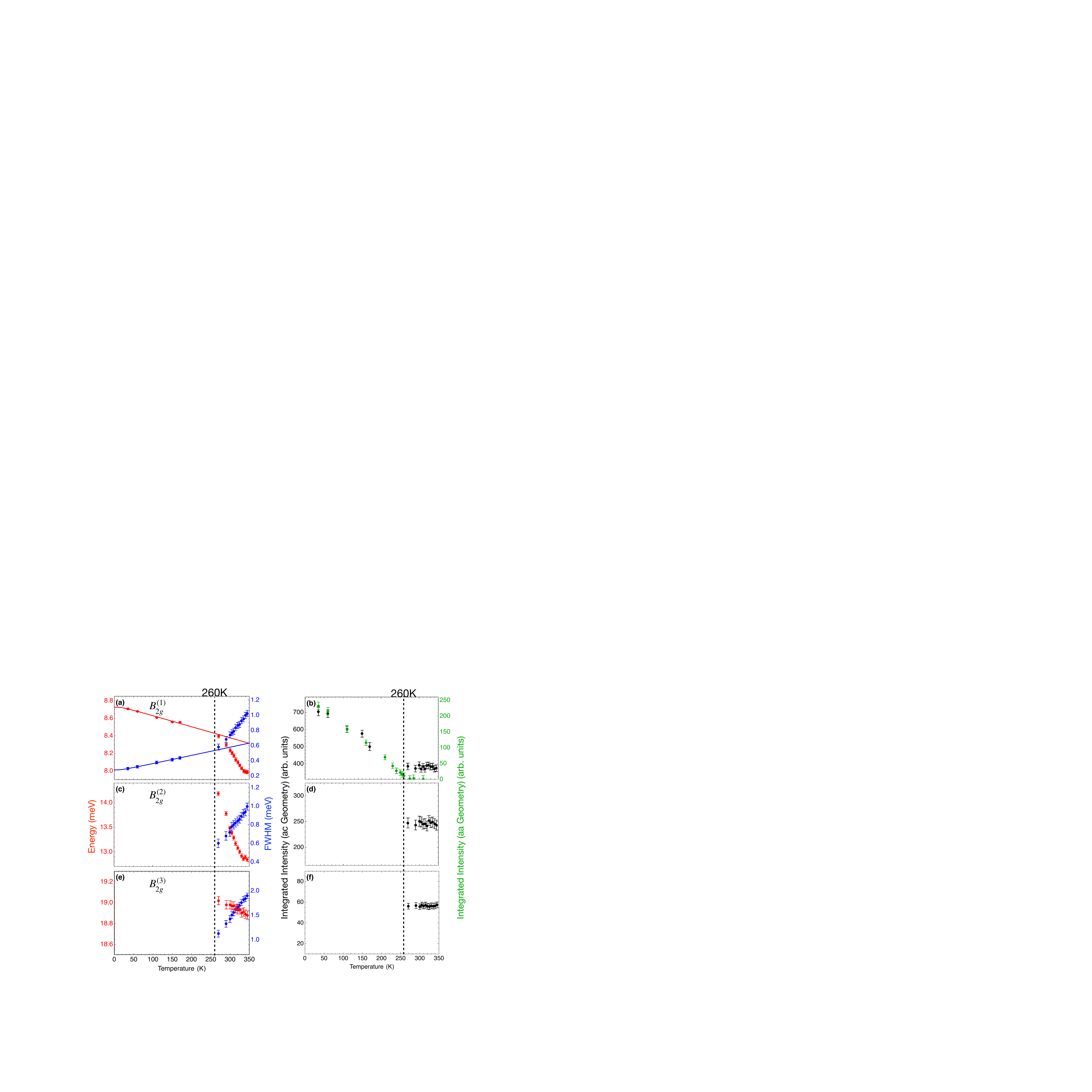}
\caption{\label{fig:Light} 
Temperature dependence of the spectral parameters: frequency, FWHM, and integrated intensity, -- for the B$_{2g}$-symmetry phonon modes of Ta$_2$Ni(Se$_{0.75}$S$_{0.25}$)$_5$. 
For each mode, the left panel presents the energy and FWHM; the right panel presents the integrated intensity in the allowed and 'forbidden' scattering geometries.}
\end{figure}

\begin{figure}
\includegraphics[width=0.9\linewidth]{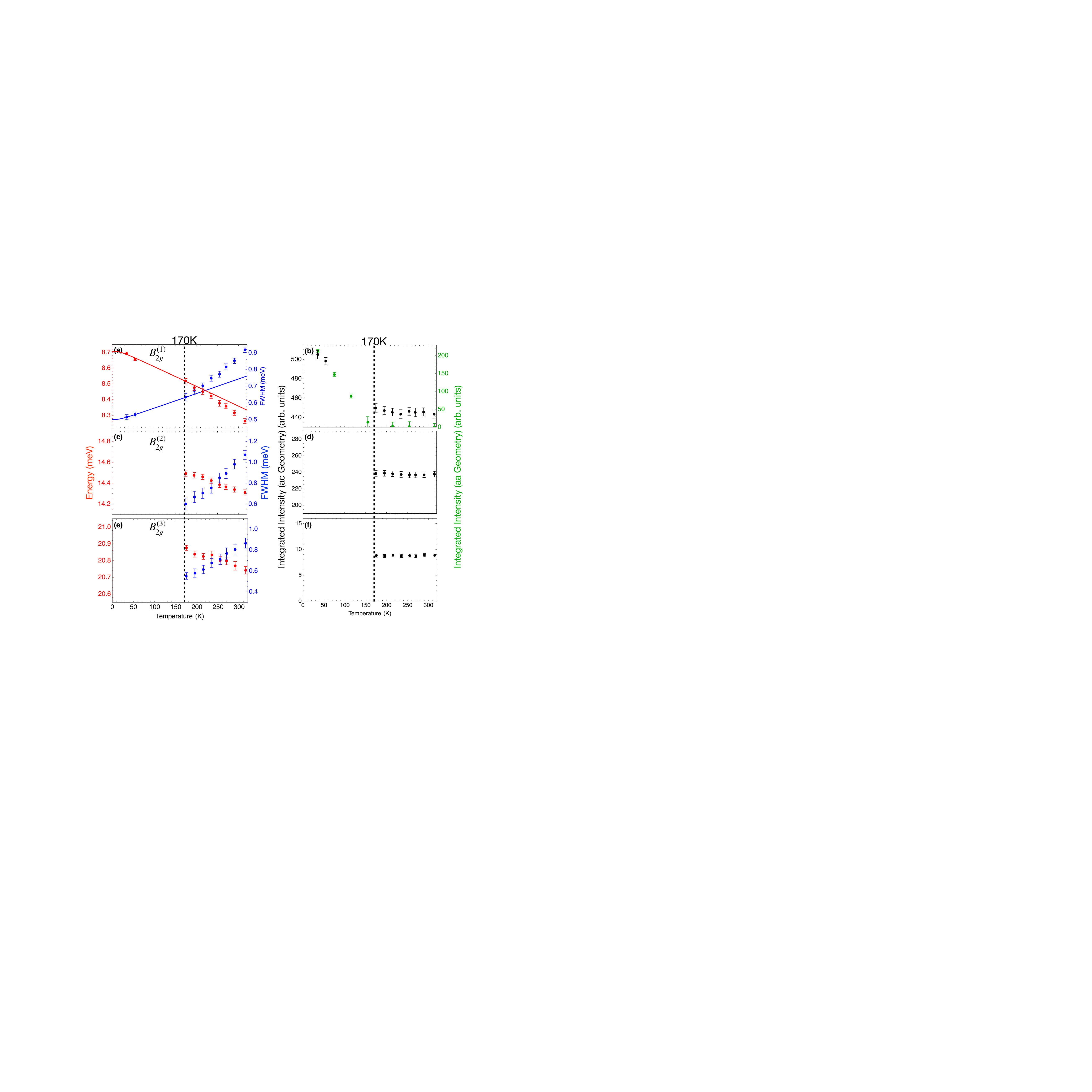}
\caption{\label{fig:Heavy} 
Temperature dependence of the spectral parameters: frequency, FWHM, and integrated intensity, -- for the B$_{2g}$-symmetry phonon modes of Ta$_2$Ni(Se$_{0.33}$S$_{0.67}$)$_5$. 
For each mode, the left panel presents the energy and FWHM; the right panel presents the integrated intensity in the allowed and 'forbidden' scattering geometries.}
\end{figure}

In Fig.\,\ref{fig:Light} and  \ref{fig:Heavy} we show the temperature dependence of the B$_{2g}$-mode spectral parameters for Ta$_2$Ni(Se$_{0.75}$S$_{0.25}$)$_5$ and Ta$_2$Ni(Se$_{0.33}$S$_{0.67}$)$_5$, respectively. Although we do not perform Fano analysis for the data below \Tc, we use Lorentzian lineshape to fit the B$_{2g}^{(1)}$ mode at low-enough temperatures, at which the lineshape is essentially symmetric. For the B$_{2g}^{(2)}$ and B$_{2g}^{(3)}$ modes, because they are not well separated from the leakage of A$_{g}$ modes, it is difficult to reliably obtain their spectral parameters.

For the alloy compositions, the increase of the FWHM slope above \Tc is not as strong as in the case of Ta$_2$NiSe$_5$. Moreover, although for Ta$_2$NiSe$_5$ the large frequency increase of the B$_{2g}^{(1)}$ mode on cooling happens in a 20\,K energy range, the frequency increase for the alloy compositions happens in a much larger energy range.

\subsubsection{Ta$_2$NiS$_5$\label{subsubsec:ParaBS}}

\begin{figure}
\includegraphics[width=0.7\linewidth]{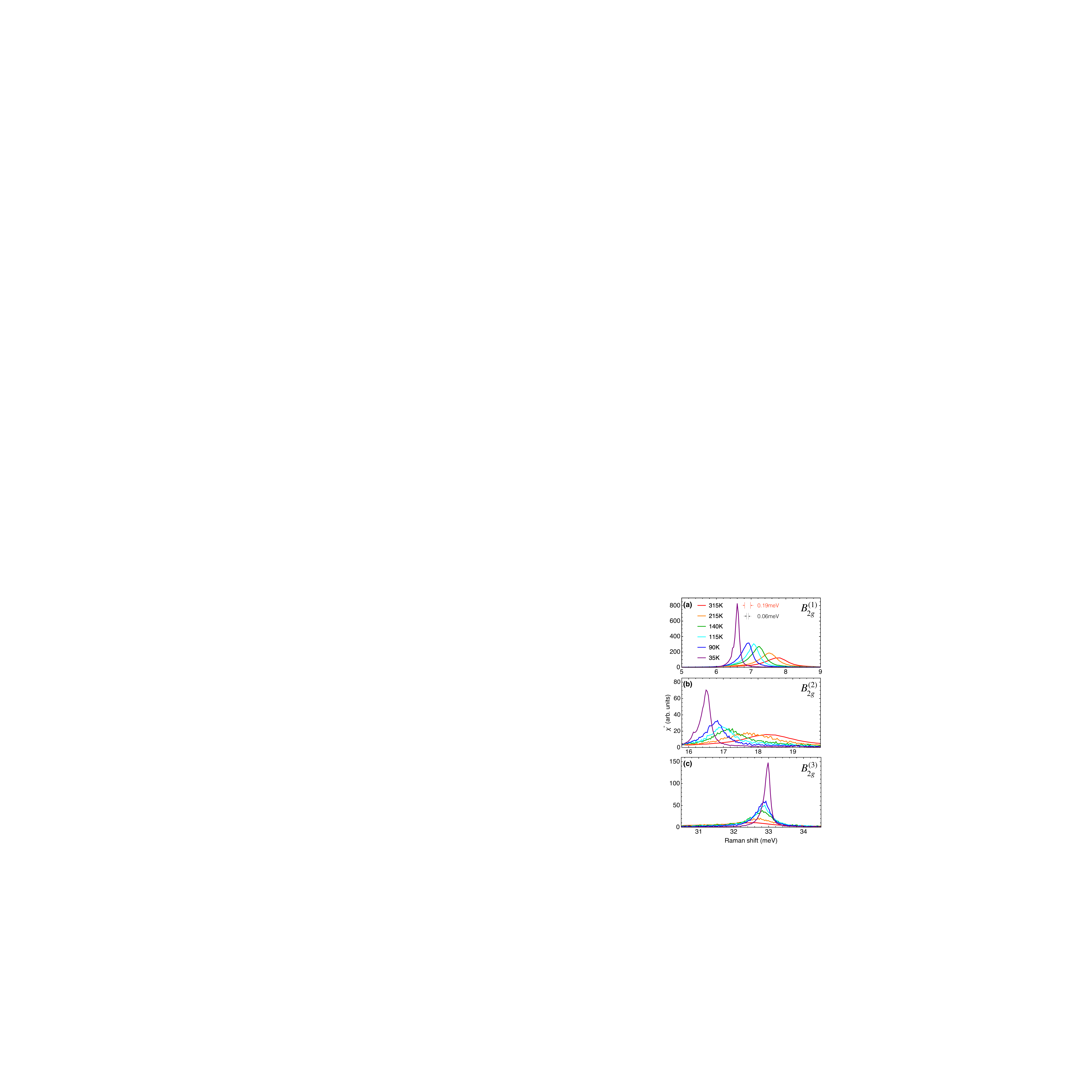}
\caption{\label{fig:SCompo} 
Temperature dependence of B$_{2g}$ phonons 
%Raman response $\chi^{\prime\prime}\(\omega)$ 
in the $ac$ scattering geometry for Ta$_2$NiS$_5$.
The spectral resolution is 0.19\,meV for the 315\,K data, and 0.06\,meV for the other data.
}
\end{figure}

Ta$_2$NiS$_5$ is a semiconductor that, in contrast to \TNS, does not show signatures of an excitonic insulator like the flattening of valence band dispersion~\cite{ARPES2018S,ARPES2019S}.
Consistent with its semiconductor nature, we observe no excitonic continuum and in turn no asymmetric lineshape for the phonon modes, Fig.\,\ref{fig:OV}(c-d).  
In Fig.\,\ref{fig:SCompo} we zoom in on the temperature dependence of the phonons in $ac$ scattering geometry. % for Ta$_2$NiS$_5$. 
The three B$_{2g}$-symmetry phonon modes exhibit conventional Lorentzian lineshapes, and the linewidth has almost 4-fold decrease on cooling from 315\,K to 35\,K. 
Moreover, the B$_{2g}^{(1)}$ and B$_{2g}^{(2)}$ modes show softening behavior on cooling.

\begin{figure}
\includegraphics[width=0.95\linewidth]{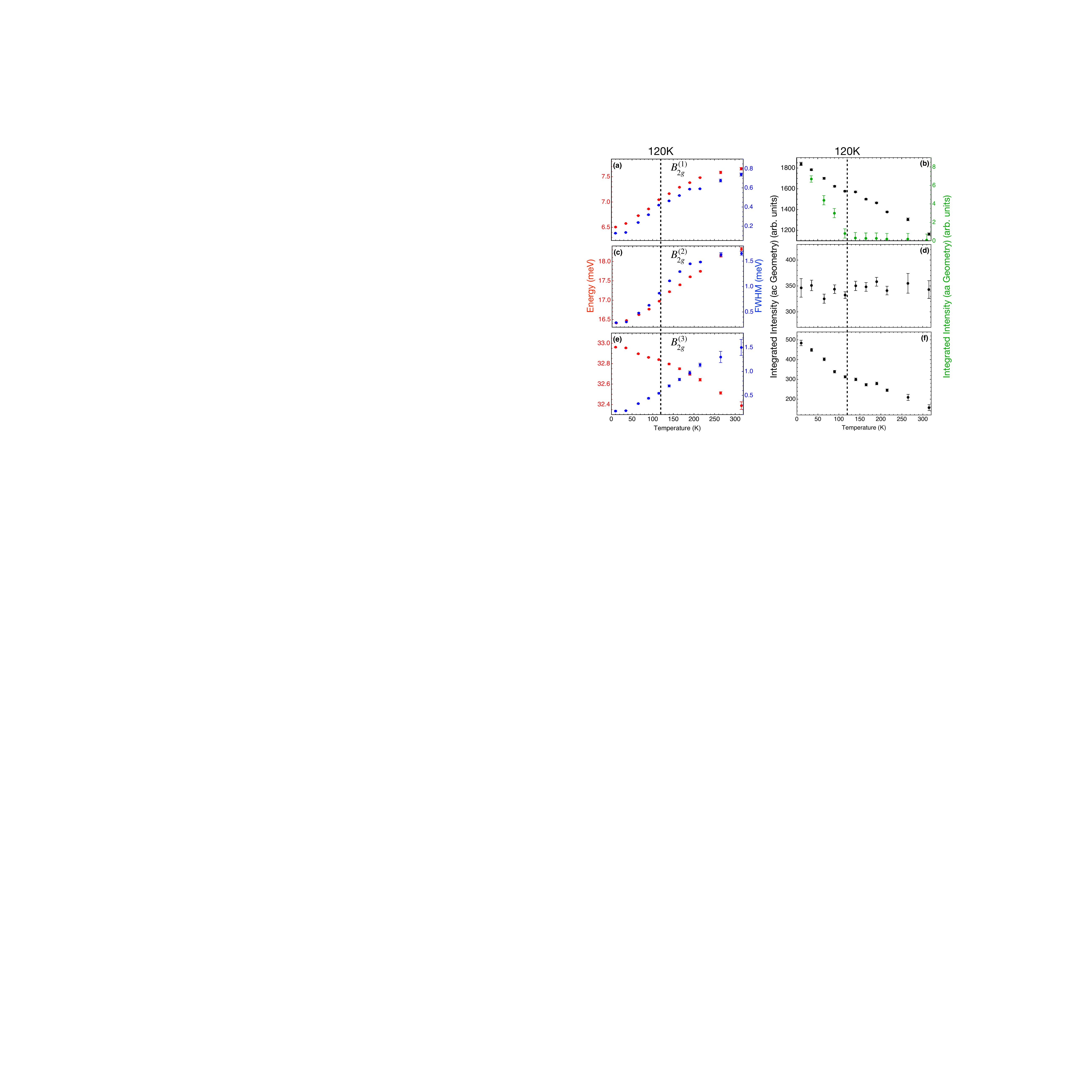}
\caption{\label{fig:ParaS} 
Temperature dependence of the spectral parameters: frequency, FWHM, and integrated intensity, -- for the B$_{2g}$-symmetry phonon modes of Ta$_2$NiS$_5$. 
For each mode, the left panel presents the energy and FWHM; the right panel presents the integrated intensity in the allowed and 'forbidden' scattering geometries.}
\end{figure}

In Fig.\,\ref{fig:ParaS} we show the temperature dependence of the spectral parameters for the B$_{2g}$-symmetry phonon modes for Ta$_2$NiS$_5$: the frequency, FWHM, and the integrated intensity. 
These spectral parameters are obtained by fitting the measured spectral features with Lorentzian lineshapes.

The behavior of the B$_{2g}$-symmetry modes is not consistent with the anharmonic decay model: for B$_{2g}^{(1)}$ and B$_{2g}^{(2)}$ modes, the frequency anomalously decreases on cooling; for B$_{2g}^{(3)}$ mode, the decrease of FWHM on cooling is too steep to be accounted by the anharmonic decay model.
Hence, we suggest that there is a change in the phonon self energy, which must affect the the apparent mode frequency. 
For the B$_{2g}^{(1)}$ and B$_{2g}^{(2)}$ modes, whose frequency is below 30\,meV, the energy decreases on cooling; for the B$_{2g}^{(3)}$ mode, whose frequency is above 30\,meV, the energy increases on cooling.

The integrated intensity of B$_{2g}^{(2)}$ mode is temperature-independent, while that of the other modes increases on cooling. However, different from the case of Ta$_2$NiSe$_5$, in which the increase of the intensity happens below \Tc, the intensity increase of the Ta$_2$NiS$_5$ modes is through the whole measured temperature range.

\subsection{A$_{g}$-symmetry phonon modes\label{subsec:ParaA}}

\begin{figure}
	\includegraphics[width=0.34\textwidth]{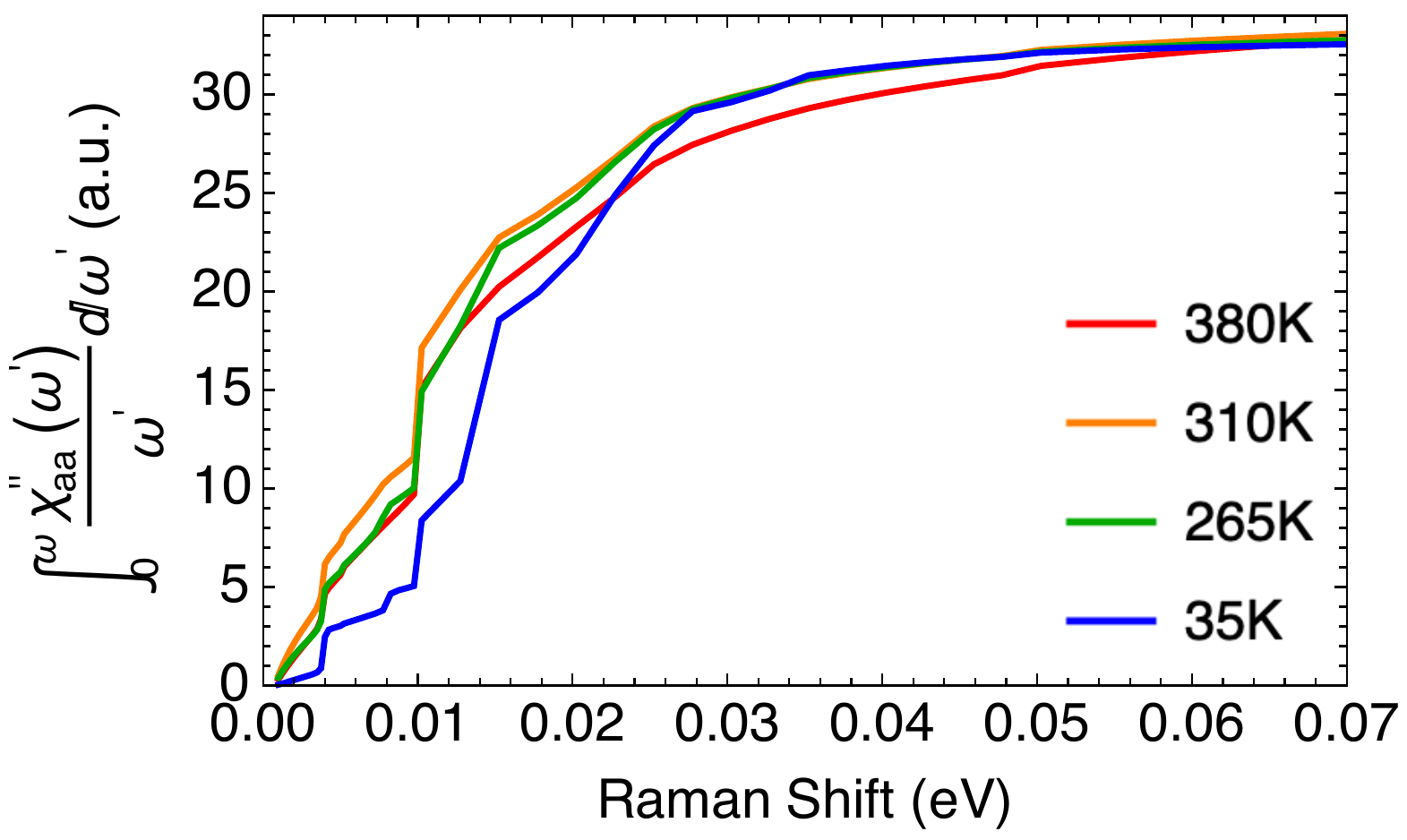}
	\caption{\label{fig:Conserve}The integral $I_{aa}(\omega)=\int_0^{\omega} \frac{\chi_{aa}^{\prime\prime}(\omega^{\prime})}{\omega^{\prime}}\,d\omega^{\prime}$ as a function of the Raman shift $\omega$ for Ta$_2$NiSe$_5$ in $aa$ scattering geometry.}
\end{figure}

\begin{figure}[b]
\includegraphics[width=0.4\textwidth]{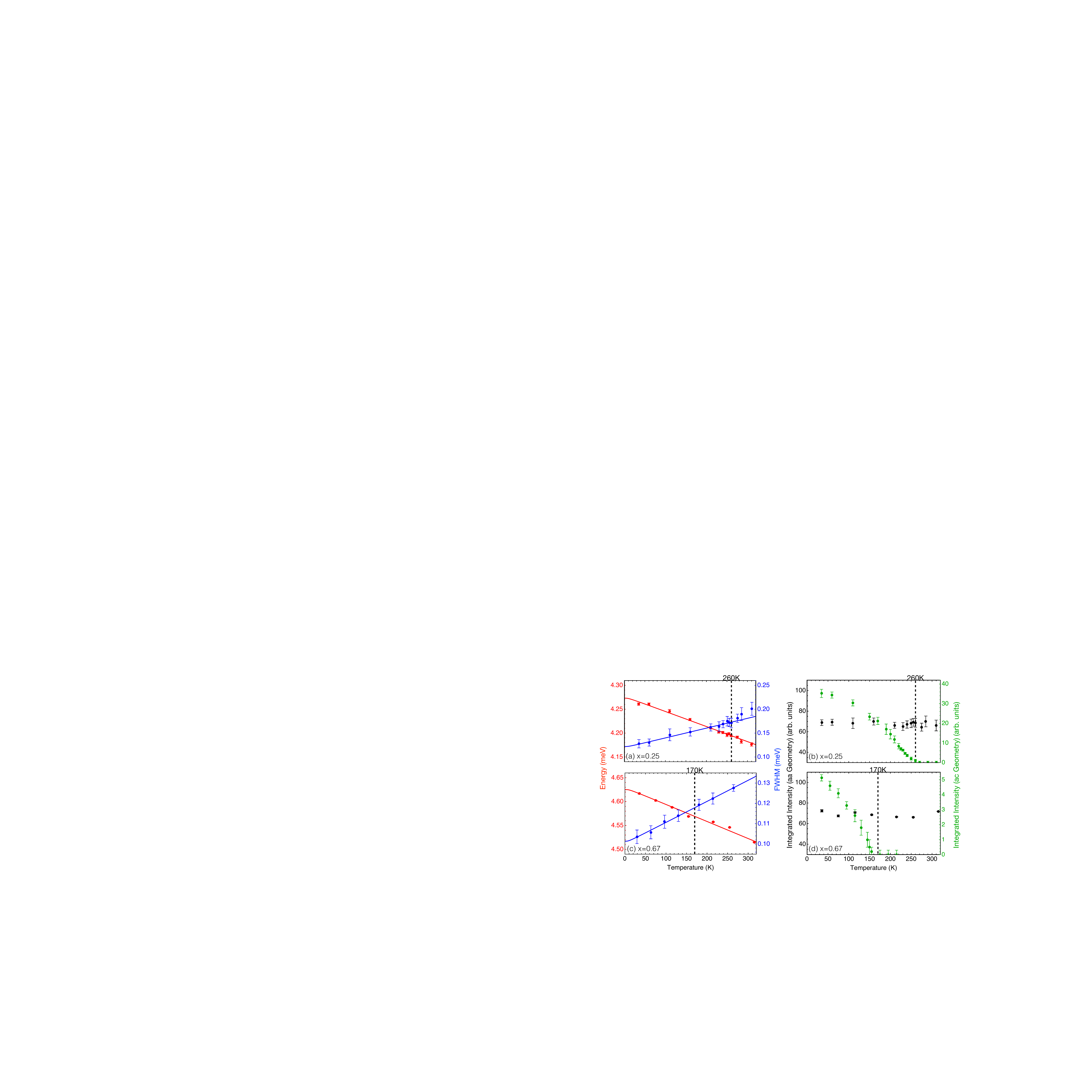}
\caption{\label{fig:DopedAg} Temperature dependence of the spectral parameters: frequency, FWHM, and integrated intensity, -- for the A$_{g}^{(1)}$ phonon mode of Ta$_2$Ni(Se$_{1-x}$S$_x$)$_5$ with $x = 0.25$ (a-b) and 0.67 (c-d). For each sample, the left panel presents the energy and FWHM; the right panel presents the integrated intensity. The solid lines represent the fits to the anharmonic decay model [Eqs.~(\ref{eq:energyTwo}-\ref{eq:gammaTwo})].}
\end{figure}

\begin{figure*}
\includegraphics[width=0.98\textwidth]{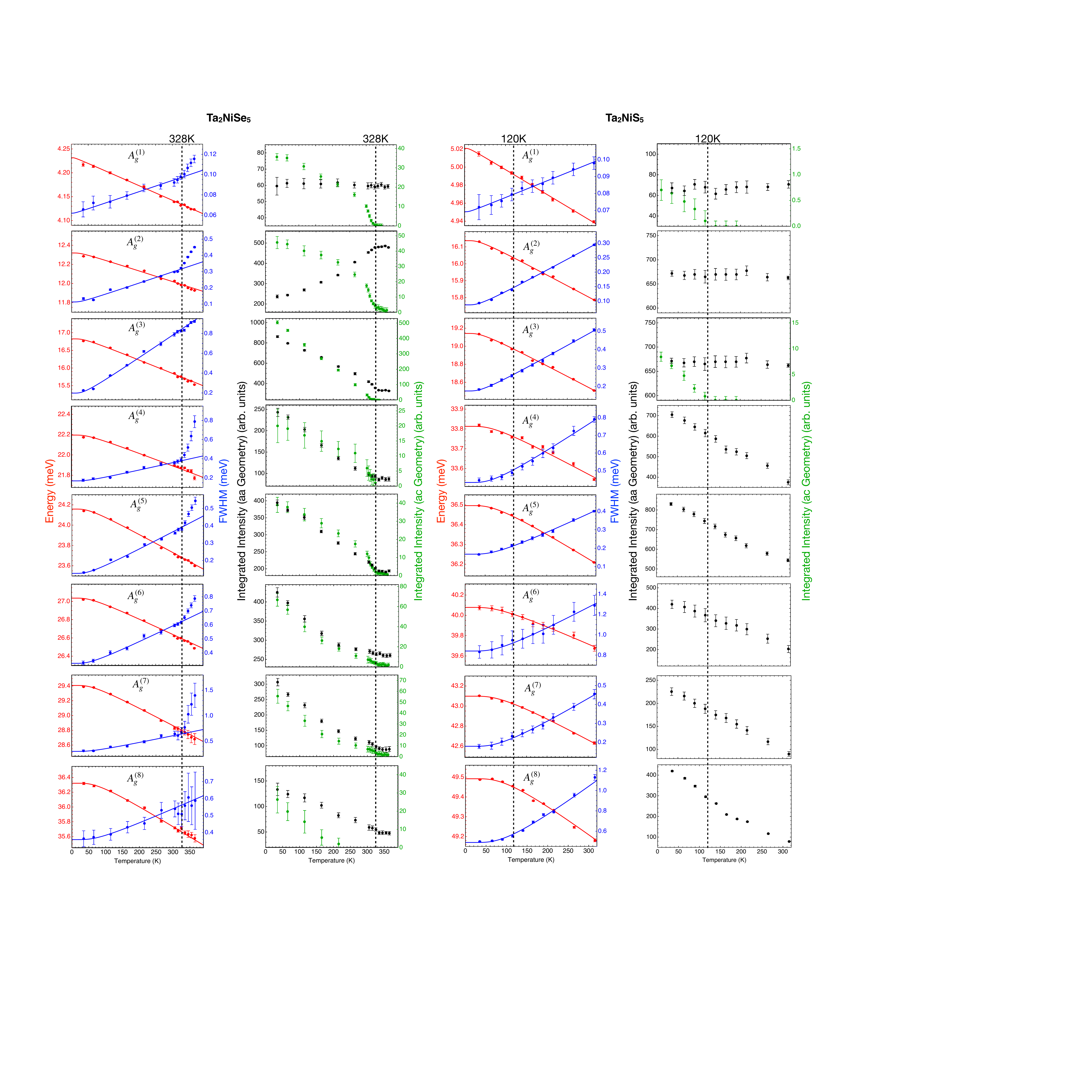}
\caption{\label{fig:ParaLarge} 
Temperature dependence of the spectral parameters: frequency, FWHM, and integrated intensity, -- for the A$_{g}$-symmetry phonon modes for Ta$_2$NiSe$_5$ and Ta$_2$NiS$_5$. For each mode, the left panel presents the energy and FWHM; the right panel presents the integrated intensity. The solid lines represent the fits to the anharmonic decay model [Eqs.~(\ref{eq:energyTwo}-\ref{eq:gammaTwo})].}
\end{figure*}

In this subsection we discuss the properties of the full-symmetric phonon modes; the relevant results are shown in Fig.\,\ref{fig:Conserve} with respect to the overall spectral weight distribution in the A$_g$ channel, and in Figs.\,\ref{fig:DopedAg} and \ref{fig:ParaLarge} regarding the spectral parameters of the individual A$_g$ modes. In Fig.\,\ref{fig:ParaLarge} we compare the temperature dependence of the spectral parameters (frequency, FWHM, and integrated intensity) of the A$_{g}$-symmetry phonon modes for Ta$_2$NiSe$_5$ and Ta$_2$NiS$_5$ crystals. 
The temperature dependence of both frequency and FWHM above \Tc for these modes can be accounted by the anharmonic decay model [Eq.\,(\ref{eq:energyTwo}-\ref{eq:gammaTwo})].

The FWHM of the A$_{g}$-symmetry modes (except for A$_{g}^{(3)}$ and A$_{g}^{(8)}$) for Ta$_2$NiSe$_5$ exhibit an anomalous increase above the transition temperature, which we attribute to enhanced electron-phonon scattering rate in its semimetal phase. However, the modes of Ta$_2$NiS$_5$ show no anomaly of FWHM, because it is a semiconductor with a direct gap observed throughout the measured temperature range.

The intensity of most modes has more than 2-fold increase on cooling, with a few exceptions: for Ta$_2$NiSe$_5$, the intensity of the A$_{g}^{(1)}$ mode is independent of temperature, while the intensity of the A$_{g}^{(2)}$ phonon mode decreases on cooling; for Ta$_2$NiS$_5$, the intensities of the A$_{g}^{(1)}$ and A$_{g}^{(2)}$ modes are temperature independent. 

The enhancement of intensity on cooling in the semimetallic samples could in principle be related to the change of electronic structure (less screening effect) below $T_c$. 
However, an interesting insight can be further obtained by computing the integral 
$I_{aa}(\omega)=\int_0^{\omega} \frac{\chi_{aa}^{\prime\prime}(\omega^{\prime})}{\omega^{\prime}}\,d\omega^{\prime}$ (Fig.\,\ref{fig:Conserve}). 
In the limit $\omega \to \infty$ this integral is proportional to the static susceptibility in the $aa$ scattering geometry, which is expected to be approximately constant as a function of temperature since there is no instability in A$_{g}$ channel. 
On cooling, the phonon intensity is enhanced [Fig.\,\ref{fig:ParaLarge}], while the electronic continuum is suppressed [Fig.\,\ref{fig:OV}(a)]. 
However, the integral $I_{aa}$ is essentially temperature independent at 70\,meV. 
This conservation implies that the enhancement of the phonon intensity is balanced by the reduction of the electronic intensity. 
Note that  these effects occur at temperatures below $T_c$, where the continuum in \B2g symmetry is strongly suppressed, indicating the opening of a gap. 
On the other hand, strong correlation effects have been shown to lead to a non-zero intensity within the gap~\cite{Pavel2020}. 
Thus, the peculiar balance between the electronic and phononic contributions at low energies implies an unconventional coupling of the electronic modes to the \Ag phonons in the correlated excitonic insulator state.

For comparison, in Fig.\,\ref{fig:DopedAg} we show the temperature dependence of the spectral parameters for the A$_{g}^{(1)}$ mode for Ta$_2$Ni(Se$_{1-x}$S$_x$)$_5$ with $x = 0.25$ and 0.67. 
Because of the difficulty caused by broad lineshape and two-frequency behavior of the phonon modes for alloy compositions, only the lowest-frequency A$_{g}^{(1)}$ mode, which are well separated from other modes in frequency, renders reliable fitting results.  
For Ta$_2$Ni(Se$_{0.75}$S$_{0.25}$)$_5$, some linewidth broadening is observed above \Tc, which is weaker than for Ta$_2$NiSe$_5$. 
However, for Ta$_2$Ni(Se$_{0.33}$S$_{0.67}$)$_5$ almost no broadening is observed.

\section{Relation to time-resolved experiments\label{sec:Time}}

In this section we convert Raman response from frequency domain to time domain, and discuss relevant time-resolved studies on Ta$_2$NiSe$_5$.

\begin{figure}
	\includegraphics[width=0.48\textwidth]{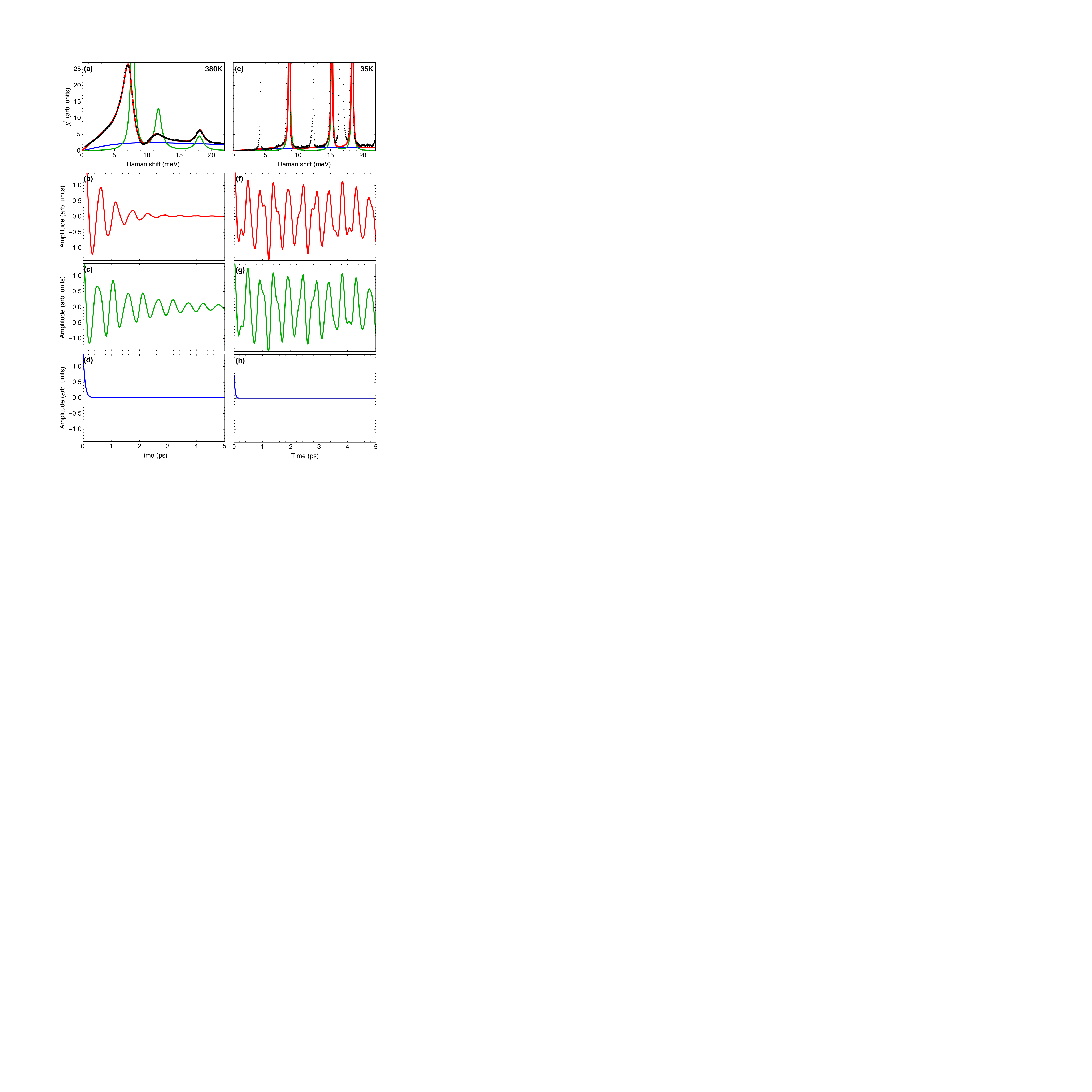}
	\caption{\label{fig:TimeAC} 
		Time-domain signal for the B$_{2g}$-symmetry excitations of Ta$_2$NiSe$_5$. (a) The Raman response for Ta$_2$NiSe$_5$ measured in ac geometry at 380\,K. The Fano fit [Eq.~(\ref{eq:fanoChi})] is shown as the red curve; the bare phonon modes [Eq.~(\ref{eq:ChiP})] are shown by the green curve; the bare excitonic continuum [Eq.~(\ref{eq:ChiE})] is shown by the blue curve. The time-domain signal corresponding to the Fano fit, bare phonon modes, and bare excitonic continuum, obtained by inverse Fourier transform, are shown in (b), (c), and (d), respectively. (e) The Raman response for Ta$_2$NiSe$_5$ measured in ac geometry at 35\,K. The corresponding time-domain signals are shown in (f), (g), and (h), respectively.}
\end{figure}

We first discuss the appearance of B$_{2g}$-symmetry excitations in the time domain, obtained by inverse Fourier transform of the Raman response, see Fig.\,\ref{fig:TimeAC}. 
Above the transition temperature, the phonon-exciton interaction significantly broadens the spectral features [Fig.\,\ref{fig:TimeAC}(a)]; as a result the deduced time-resolved response decays fast and essentially dies out before 5\,ps [Fig.\,\ref{fig:TimeAC}(b)]. 
For the bare phononic response, the time-domain signal beyond 4\,ps is dominated by the oscillation of 2\,THz mode [Fig.\,\ref{fig:TimeAC}(c)], corresponding to the B$_{2g}^{(1)}$ phonon. 
The signal of the bare excitonic response shows a pure relaxational behavior [Fig.\,\ref{fig:TimeAC}(d)]. Comparing Fig.\,\ref{fig:TimeAC}(b) and (c), we note that it is difficult to identify the Fano interference feature from time-domain signal, because such interference exhibits no distinct characters in the time domain, except for that the oscillations are strongly damped. 
Indeed, in ultrafast studies, the interference nature is not revealed even after Fourier transform of the time-domain data to the frequency domain\,\cite{Fast2018a,Ning2020,Baldini2020}. The likely reason for that is the high pump fluence used, causing the exciton and phonon responses decouple already on the very early time scales. 

Below the transition temperature, the energy at which the continuum has maximum response moves to higher energy, and the phonon-exciton coupling strength is reduced. 
Consequently, the interference effect is suppressed and the combined response is essentially the same as the bare phononic response [Fig.\,\ref{fig:TimeAC}(f-g)]. 
The bare excitonic response still has a pure relaxational time dependence, with very short lifetime [Fig.\,\ref{fig:TimeAC}(d) and (h)].

\begin{figure}
	\includegraphics[width=0.48\textwidth]{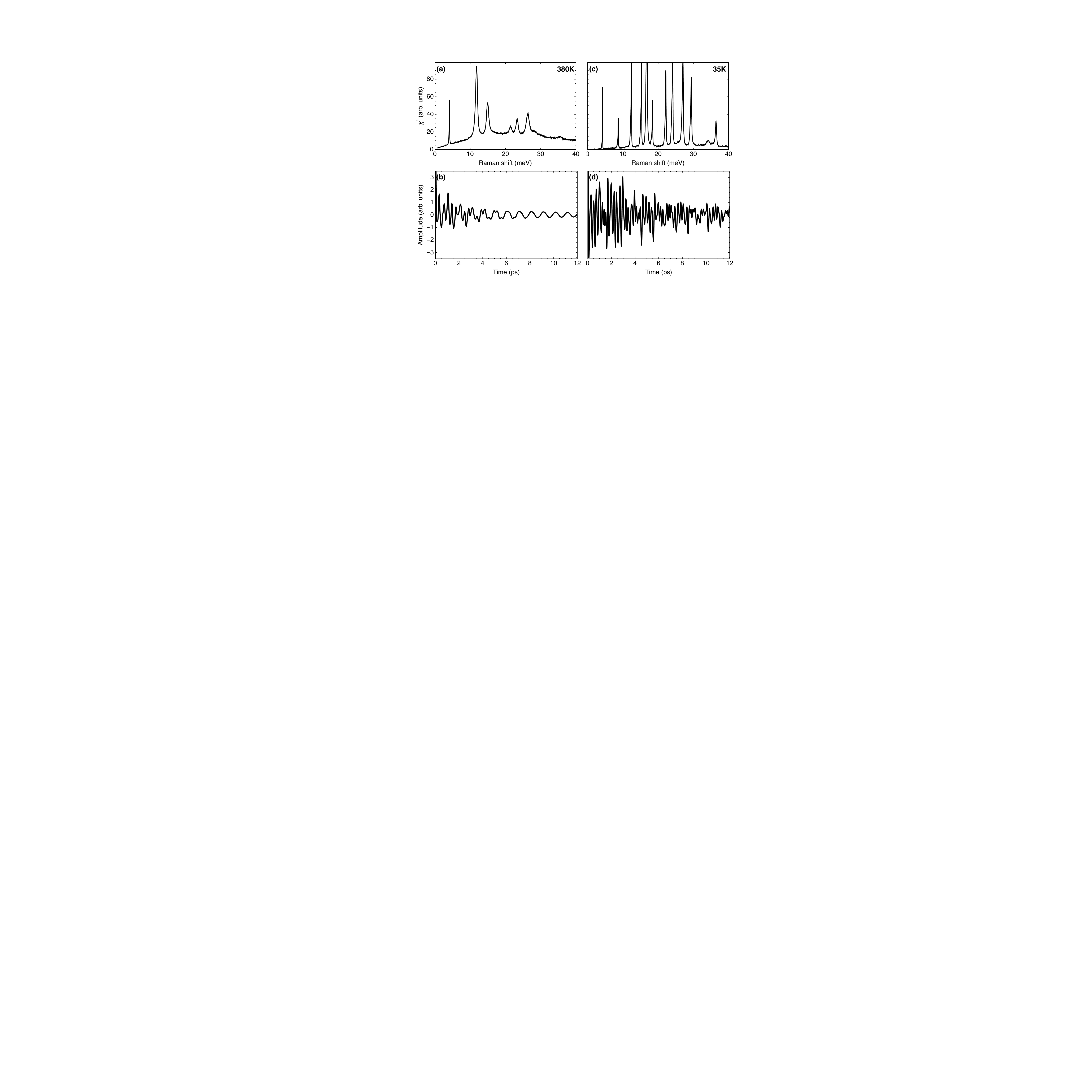}
	\caption{\label{fig:TimeAA} 
		Time-domain signal for the A$_{g}$-symmetry excitations of Ta$_2$NiSe$_5$. (a) The Raman response for Ta$_2$NiSe$_5$ measured in aa geometry at 380\,K. The time-domain signal, obtained by inverse Fourier transform of the data, is shown in (b). (c) The Raman response for Ta$_2$NiSe$_5$ measured in aa geometry at 35\,K. The corresponding time-domain signal is shown in (d).}
\end{figure}

For comparison, in Fig.\,\ref{fig:TimeAA} we present the time-domain signal for the A$_{g}$-symmetry excitations. 
Above the transition temperature, although multiple oscillations are simultaneously present at short time scale, beyond 10\,ps the time-domain signal is dominated by the oscillation of 1\,THz [Fig.\,\ref{fig:TimeAA}(b)], corresponding to the A$_{g}^{(1)}$ phonon mode whose spectral width is much smaller than others [Fig.\,\ref{fig:TimeAA}(a)]. 
At low temperature, the difference between the spectral widths of various modes is reduced [Fig.\,\ref{fig:TimeAA}(c)]; hence, these modes have comparable lifetime and multiple oscillations survive beyond 10\,ps [Fig.\,\ref{fig:TimeAA}(d)].

\begin{figure}
	\includegraphics[width=0.40\textwidth]{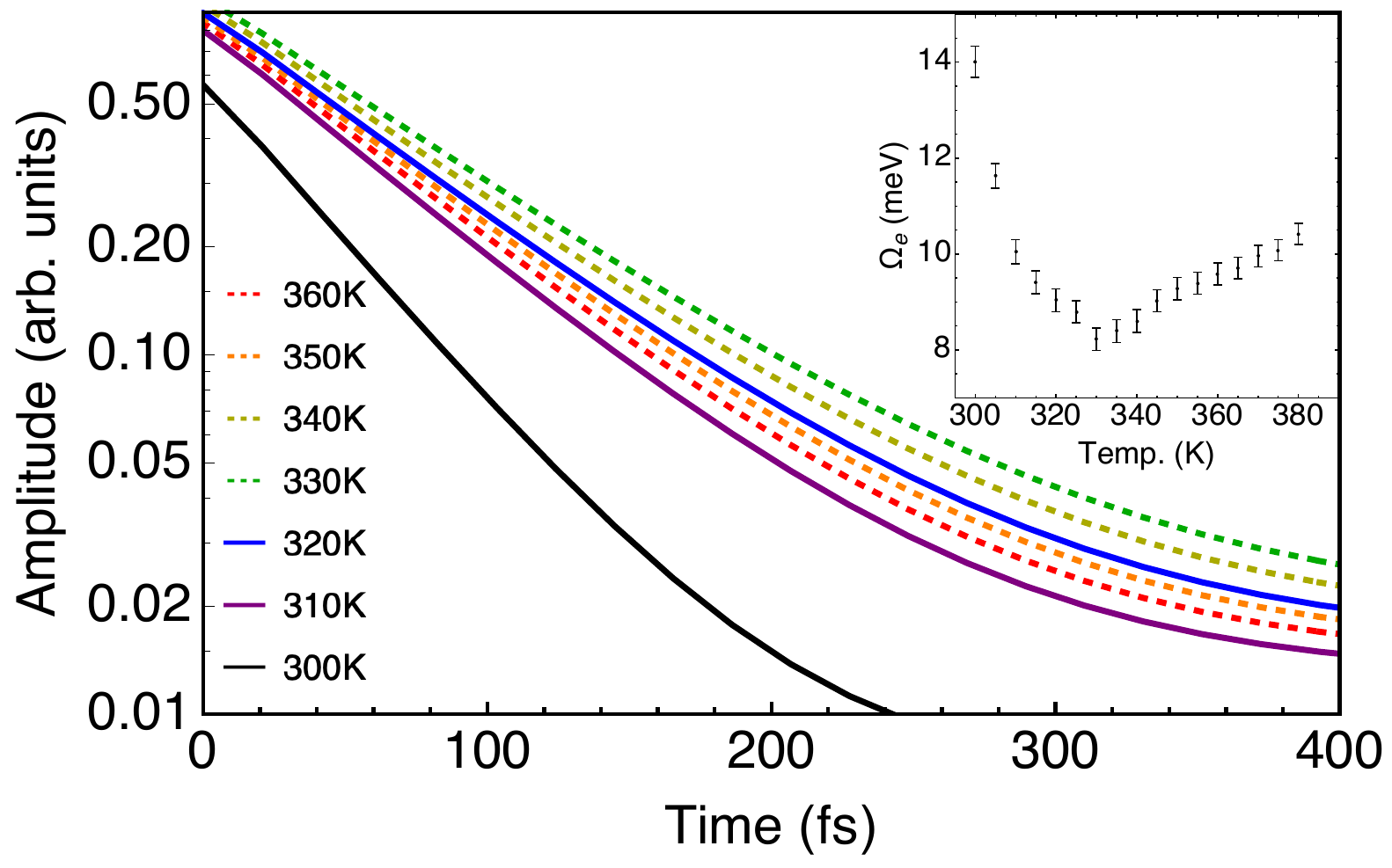}
	\caption{\label{fig:TimeDrude} 
		Time-domain signal for the excitonic continuum of Ta$_2$NiSe$_5$ above and below the transition temperature, plotted in semi-log scale. The data above \Tc are dashed for clarity. The relaxation time starts to decrease just below \Tc. Inset: temperature dependence of the energy at which the excitonic continuum has maximum intensity, $\Omega_{e}(T)$.}
\end{figure}

In relation to the ultra-fast studies of Ta$_2$NiSe$_5$~\cite{Fast2018a,andrich2020,Bretscher2021,Fast2021}, the Raman results are consistent with the ultra-fast data under low fluence. 
One ultra-fast work~\cite{Fast2018a} provides evidence for coupling between the A$_{g}^{(1)}$ mode and the amplitude mode of the condensate under high-fluence pumping. 
We note that in our studies we can unambiguously identify the amplitude mode with the excitonic continuum we observe. 
Above the transition temperature, the overdamped excitonic mode softens on cooling towards \Tc; just below the transition temperature, its energy increases on further cooling. 
Such temperature dependence of mode energy is characteristic of an amplitude mode while the optical \B2g phonons only harden on cooling. 
In Fig.\,\ref{fig:TimeDrude} we present the time-domain signal corresponding to the excitonic mode around the transition temperature.

Additionally, a mode associated with the phase of the excitonic order parameter may exist. 
The electron-phonon coupling and exchange interactions modify the otherwise sombrero-shape free-energy landscape, reducing the continuous U(1) symmetry to a discrete Z$_2$ symmetry, leading to a finite energy of the phase mode. 
A recent pump-probe microscopy study suggests that the phase-mode energy is smaller than the A$_{g}^{(1)}$-mode energy (around 4\,meV)~\cite{andrich2020}. 
However, we do not observe any mode down to 0.4\,meV at low temperature [Fig.\,\ref{fig:TimeAC}(e)]. Furthermore, the phase oscillations of the order parameter should harden above $T_c$, which would have 
manifested itself in the low-energy range of the spectra. 
Such signatures are not observed, either [Fig.~\ref{fig:SeCompo}]. 
Therefore, existence of the phase mode below 4\,meV is not supported by the Raman data.

\section{Conclusions\label{sec:Con}}

We have employed polarization resolved Raman spectroscopy to conduct a systematic spectroscopic study of the lattice dynamics  in Ta$_2$Ni(Se$_{1-x}$S$_x$)$_5$ ($x=0$, ..., 1)  family of the excitonic insulators.

We identify and classify by symmetry all the Raman active phonon modes of \Ag and \B2g symmetries. 
A change in selection rules is detected at temperature $T_c(x)$, indicating the orthorhombic-to-monoclinic structural phase transition related to the excitonic insulator state~\cite{Pavel2020,Pavel2021}. 
We find that the symmetry breaking transition persists for entire Ta$_2$Ni(Se$_{1-x}$S$_x$)$_5$ family with $T_c(x)$ monotonically decreasing from 328\,K for $x=0$ to 120\,K for $x=1$. 
Its signatures in the resistivity data, however, become weak or undetectable for large sulfur concentration $x$ that we attribute to the weakness of the symmetry breaking at large $x$, demonstrated by the decrease of phonon intensity in the 'forbidden' scattering geometry. 

For $x < 0.7$, the two lowest-frequency $ac$-quadrupole-symmetry B$_{2g}^{(1)}$ and B$_{2g}^{(2)}$ phonon modes show strongly asymmetric lineshapes at high temperatures, which is indicative of coupling between the phonons and an excitonic continuum of the same symmetry~\cite{Pavel2020,Pavel2021}. 
Within the framework of extended Fano model, we develop a quantitative description of the observed lineshapes, enabling us to disentangle the excitonic and phononic contributions to the spectra, to derive the intrinsic phonon parameters and determine the exciton-phonon interaction strength, that affects the transition temperature \Tcx \cite{Pavel2021}. 
The displacement patterns obtained from {\it ab-initio} calculations explain the trends in the deduced exciton-phonon coupling values. 
At $T<T_c$ the remnant excitonic continuum demonstrates signatures of coupling to the finite-momentum \B2g-symmetry acoustic phonons, allowed due to scattering on a quasi-periodic structural of domain walls. 
We have also shown that the coupling to the excitonic continuum explains the acoustic mode softening observed in \cite{XRay2018,Pavel2021}.

The intrinsic parameters of the optical phonons for Ta$_2$NiSe$_5$ and Ta$_2$NiS$_5$ at low temperatures are mostly in good agreement with anharmonic decay model and DFT calculations. 
For alloy compositions a two-mode behavior is found for most modes, where signatures appear at two frequencies, corresponding to the ones in Ta$_2$NiSe$_5$ and Ta$_2$NiS$_5$. 
However, several types of anomalous behavior have been observed in the temperature dependencies. 
For $x<0.7$, the B$_{2g}^{(1)}$ and B$_{2g}^{(2)}$ modes anomalously harden close to $T_c$ (e.g., Fig.\,\ref{fig:ParaSe}), indicating an interaction with continuum beyond Fano model description~\cite{Pavel2020}. 
For Ta$_2$NiS$_5$ (Fig.\,\ref{fig:ParaS}) the B$_{2g}^{(1)}$ and B$_{2g}^{(2)}$ modes do not show Fano shapes, but the frequencies of these modes anomalously soften on cooling, although never going critical. 
Finally, the intensity of the most modes is strongly temperature dependent. 
Interestingly, for Ta$_2$NiSe$_5$ we observe that the sum of electronic and phononic contributions to the $aa$ static susceptibility is conserved (Fig.\,\ref{fig:Conserve}), pointing to an unexpected electron-phonon coupling mechanism within the excitonic insulator state.

In addition, based on our results we have provided an interpretation of recent time-resolved pump-probe experiments and discussed the signatures expected from the modes we have observed, see Figs.\,\ref{fig:TimeAC},\,\ref{fig:TimeAA} and \ref{fig:TimeDrude}. 
Overall, this work provides a comprehensive study lattice dynamics in Ta$_2$Ni(Se$_{1-x}$S$_x$)$_5$; more generally, the unconventional behaviors we observed point to the importance of the effects of electron-phonon coupling in correlated semimetals.

\begin{acknowledgments}
M.Y. and P.A.V. contributed equally to this work. 
We acknowledge discussions with K.\,Haule. 
We are grateful to Y.\,Kauffmann for help with TEM measurements.
The spectroscopic work conducted at Rutgers was supported by NSF Grant No. DMR-1709161 (M.Y. and G.B). 
P.A.V. acknowledges the Postdoctoral Fellowship support from the Rutgers University Center for Materials Theory. 
The sample growth and characterization work conducted at the Technion was supported by the Israel Science Foundation Grant No. 320/17 (H.L., I.F. and A.K.). 
H.L. was supported in part by a PBC fellowship of the Israel Council for Higher Education. 
The work at NICPB was supported by the Estonian Research Council Grant No. PRG736 an by the European Research Council (ERC) under Grant Agreement No. 885413. 
M.K. was supported by NSF DMREF grant DMR-1629059.
\end{acknowledgments}

\appendix

\section{Estimation of heating in excitation laser spot \label{sec:AHeat}}

We use three methods to estimate the heating rate, a measure of the temperature increase per unit laser power in the focused laser spot (K$\slash$mW):
(i) Stokes/anti-Stokes Raman scattering intensity ratio analysis;
(ii) monitoring laser power that is inducing the phase transition; and 
(iii) a thermoconductivity model calculation. 
We use data for Ta$_2$NiSe$_5$ as an example to illustrate these methods.

For the first method, we note that the Stokes scattering cross section $I_S$ and Anti-Stokes cross section $I_{AS}$ are related by the detailed balance principle~\cite{Hayes2004}
\begin{equation}
n I_S(\omega)=(n+1)I_{AS}(\omega)~,
\label{eq:SAS}
\end{equation}
in which $n$ stands for the Bose factor
\begin{equation}
n(\omega,T)=\frac{1}{exp(\hbar\omega/k_BT)-1}~,
\label{eq:n}
\end{equation}
where $\hbar$ is the reduced Planck's constant, $\omega$ is frequency, $k_B$ is the Boltzmann's constant, and $T$ is the temperature in the laser spot.

Using Eq.~(\ref{eq:SAS}-\ref{eq:n}), the temperature can be derived as 
\begin{equation}
T=\frac{\hbar\omega}{k_B \ln(I_S/I_{AS})}~,
\label{eq:T}
\end{equation}
or, numerically,
\begin{equation}
T[K]=\frac{11.605\;\omega}{\ln(I_S/I_{AS})}[meV]~.
\label{eq:Tn}
\end{equation}

\begin{figure}
\includegraphics[width=0.40\textwidth]{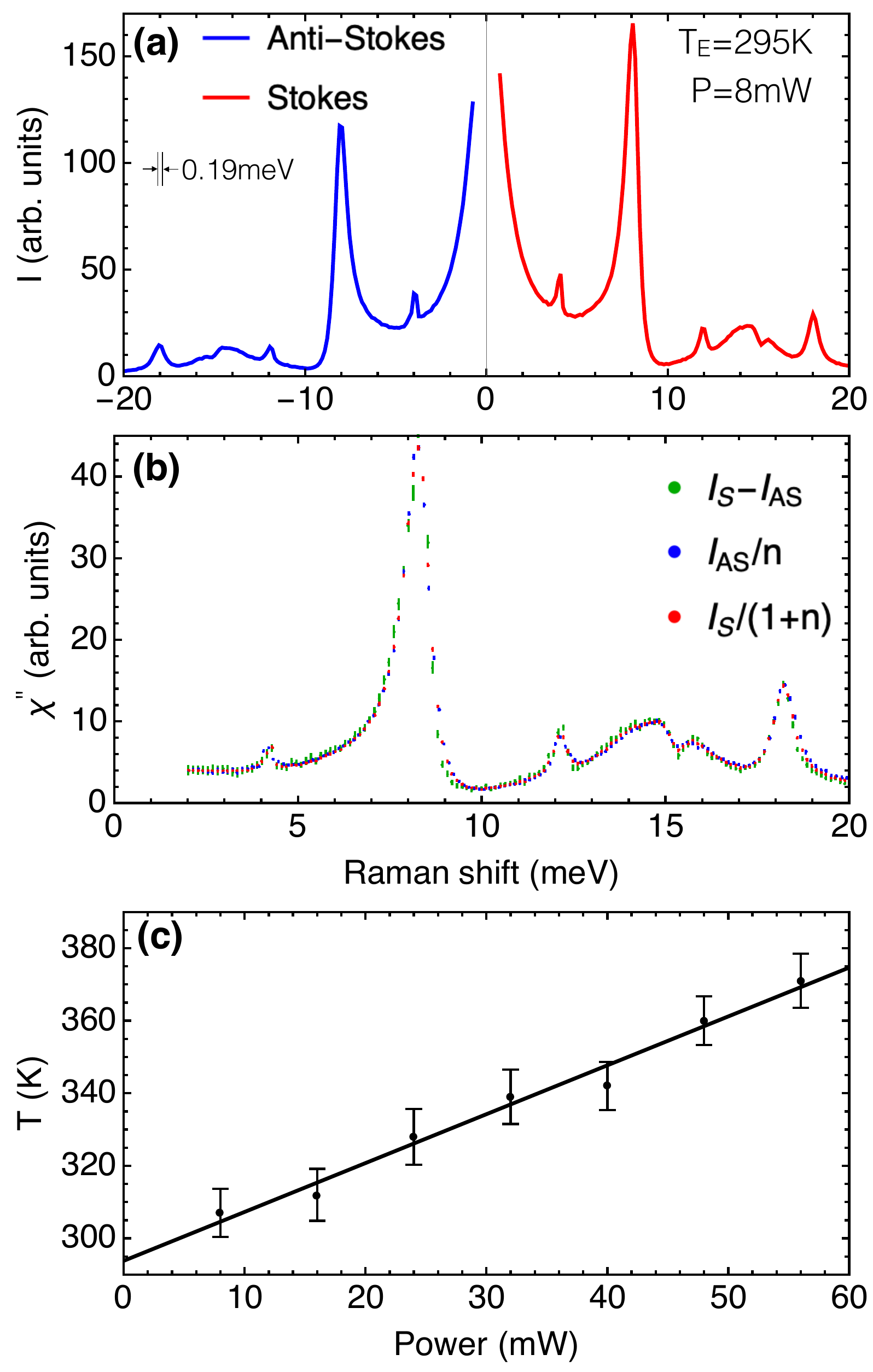}
\caption{\label{fig:SAS} 
Stokes-Anti-Stokes analysis of the $ac$ spectra for Ta$_2$NiSe$_5$. 
(a) The Stokes and Anti-Stokes cross sections measured at environmental temperature $T_E$\,=\,295\,K with laser power P\,=\,8\,mW. (b) The Raman response calculated from the measured spectra in (a). The laser-spot temperature, which appears in the Bose factor $n$, is 307\,K. The spectral resolution is 0.19\,meV for panels (a-b). (c) The laser-power dependence of the laser-spot temperature. The straight line represents a linear fit. }
\end{figure}

We studied the Stokes and Anti-Stokes cross section relation for the $ac$ scattering geometry at 295\,K environmental temperature with various laser power. 
In Fig.~\ref{fig:SAS} we show the results measured with 8\,mW laser power as an example. 
The Raman responses calculated from Stokes and Anti-Stokes cross sections match well with the temperature at the laser spot being 307\,K. 
We use Eq.~(\ref{eq:Tn}) with the phonon intensity integrated from 7.5 to 8.5\,meV to calculate the temperature at the laser spot. 
By a linear fit to the power dependence of the laser-spot temperature, we find the heating rate to be 1.29$\pm$0.17\,K$\slash$\,mW.

For the second method, we gradually increase laser power at 295\,K environmental temperature. We find that when the laser power is in the range of 26$\pm$4\,mW, the domain stripes, which are visible across the sample surface under laser illumination, disappear inside the laser spot. Moreover, the temperature dependence of phonon width and intensity has a sudden change. We assume that at this laser power, the temperature at the laser spot reaches the transition temperature 328\,K \cite{Structure1986,Transport2017}, yielding the heating rate of 1.25$\pm$0.20\,K$\slash$\,mW. 
We note that for Ta$_2$NiS$_5$, we do not observe domain stripes below its transition temperature 120\,K. Its transition temperature is identified by the sudden change of phonon intensity, see Fig.\,\ref{fig:ParaS}(b) for example.

\begin{figure}
\includegraphics[width=0.40\textwidth]{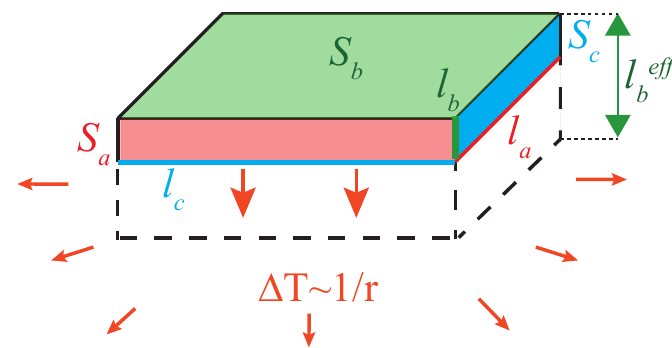}
\caption{\label{fig:Volume} 
The geometry of the heated volume. 
The lengths $l_a$ and $l_c$ are determined by the spot size, while $l_b$ is approximated by the skin depth; the $S_b$ is in the cleave plane facing the vacuum, while the other faces are in the bulk. 
Within the region marked by dashed lines, the heat flow is mostly along $b$ axis, while outside the region heat flows in all directions comparably, such that the temperature increase above the base temperature $\Delta T$ decays as $1/r$, $r$ being the distance from the heated region.}
\end{figure}

For the third method, we consider a heated region shown in Fig.\,\ref{fig:Volume} of a rectangular shape, with the dimensions given by the beam spot size and the penetration depth. 
The static heat equation takes the form:
	\begin{equation}
	\begin{gathered}
	-\kappa_x \frac{\partial^2 T}{\partial x^2}
	-\kappa_y \frac{\partial^2 T}{\partial y^2}
	-\kappa_z \frac{\partial^2 T}{\partial z^2}
	=\mathcal{P}({\bf r}),
	\\
	\left.\frac{\partial T({\bf r})}{\partial z}\right|_{y=0} = 0;
	\\
	\partial T(x\to\pm\infty,y\to-\infty,z\to\pm\infty) = T_0.
	\end{gathered}
	\label{eq:heateq}
	\end{equation}
where $\mathcal{P}({\bf r})$ is the power density. 
We ignore the cooling provided by the helium gas flow above the sample surface, so that there is no heat flow at the $y=0$ boundary, while deep inside the bulk the temperature should reach the base temperature $T_0$. 
Equation \eqref{eq:heateq} can be solved using Green's function of the Laplace equation and image method to satisfy the first boundary condition. In the latter, we extend the equation formally to $y>0$ half-space and add an "image" power density
\begin{equation}
\mathcal{P}(x,y,z)\theta(-y)\to\mathcal{P}(x,y,z)\theta(-y)+\mathcal{P}(x,-y,z)\theta(y).
\end{equation}
The solution is then given by:
\begin{equation}
T({\bf r}) - T_0 = \int \frac{d \tilde{x}' d\tilde{y}' d\tilde{z}'}{4\pi} 
\frac{
	\mathcal{P}(\sqrt{\kappa_x}\tilde{x}',...)\theta(-\tilde{y}')+ \tilde{y}'\to-\tilde{y}'
}
{|\tilde{\bf{r}}-{\bf \tilde{r}}'|},
\end{equation}
where $\tilde{\bf{r}} = (x/\sqrt{\kappa_x},y/\sqrt{\kappa_y},z/\sqrt{\kappa_z})$. The solution decays as $1/r$ at large distances and thus satisfies the second boundary condition of Eq.\,\eqref{eq:heateq}. 
An analytical result can be obtained for
\begin{equation}
\mathcal{P}(x,y,z) = \frac{t P}{l_al_bl_c}\theta(l_a/2-|x|)\theta(l_b-|y|)\theta(l_c/2-|z|)
\end{equation}
at ${\bf r} = 0$ in the limit $l_b^2/\kappa_b\ll l_{a,c}^2/\kappa_{a,c}$,  where $P$ is the laser power and $t$ is the transmission coefficient of the sample. Performing the integral one obtains
\begin{equation}
T(0) - T_0 \approx \frac{
	t P
	\left(\frac{l_a}{\sqrt{\kappa_a}} 
	\ln
	\frac{
		\frac{l_c}{\sqrt{\kappa_c}}+
		\sqrt{
			\frac{l_c^2}{\kappa_c}+\frac{l_a^2}{\kappa_a}
		}}
	{l_a/\sqrt{\kappa_a}}
	+a\leftrightarrow c
	\right)
}
{\sqrt{\kappa_b} \pi l_al_c}.
\label{eq:heatGF}
\end{equation}
Qualitatively, the result can be understood as follows. Due to the anisotropic shape of the heated region, right below it most heat is transferred along the $b$ axis. However, the one-dimensional heat equation would result in a linear solution $T(y)$, depending on the cutoff scale $l_c^{eff}$. It can be estimated from the condition of the heat flow to the lateral direction being equal to the heat flow along $b$. Approximating the temperature gradient along $a,b,c$ as $\Delta T/l_{a,b,c}$ one gets the condition
\begin{equation}
2 l_b^{eff} l_a \kappa_c \frac{\Delta T}{l_c}+2 l_b^{eff} l_c \kappa_a \frac{\Delta T}{l_a} = l_a l_c \kappa_b \frac{\Delta T}{l_b^{eff}},
\end{equation}
that results in the estimate
\begin{equation}
l_b^{eff}=\sqrt{\frac{l_a^2 l_c^2 \kappa_b}{2l_a^2 \kappa_c + 2l_c^2 \kappa_a}}.
\end{equation}
Equating the total heat flow outside this region to the input power one obtains the estimate
\begin{equation}
\Delta T \approx \frac{tP}{\sqrt{8 \kappa_b(l_a^2 \kappa_c+l_c^2 \kappa_a)}}.
\label{eq:heatest}
\end{equation}
Overall, one notices that the temperature increase scales with the linear size of the spot. The laser heating power $R$ is then determined as $\Delta T/P$; for the actual estimate we thus need to find the transmission coefficient $t$ first.

The transmission coefficient can be calculated from the complex index of refraction $n$ by the relationship $t=1-|(n-1)/(n+1)|^2$:
\begin{equation}
t=\frac{4n_1}{(n_1+1)^2+n_2^2}~,
\label{eq:t}
\end{equation}
in which $n_1$ and $n_2$ are the real and imaginary part of $n$. The complex index of refraction can be obtained from the complex dielectric constant $\epsilon$ by the relationship $\epsilon=n^2$:
\begin{equation}
n_1=\sqrt{\frac{\sqrt{\epsilon_1^2+\epsilon_2^2}+\epsilon_1}{2}}~,
n_2=\sqrt{\frac{\sqrt{\epsilon_1^2+\epsilon_2^2}-\epsilon_1}{2}}~,
\label{eq:ior}
\end{equation}
where $\epsilon_1$ and $\epsilon_2$ are the real and imaginary part of $\epsilon$. The imaginary part $\epsilon_2$ can be calculated from the real part of the optical conductivity $\sigma_1$:
\begin{equation}
\epsilon_2=\frac{60}{\omega}\sigma_1~,
\label{eq:e}
\end{equation}
in which the unit of $\omega$ is cm$^{-1}$ and that of $\sigma_1$ is $\Omega^{-1}$cm$^{-1}$. The polarization of the incoming light is along a-axis of the sample. Because the quantities $\epsilon_1$ and $\sigma_1$ are 8.13 and 2.47\,$\Omega^{-1}$cm$^{-1}$ respectively at the laser wavelength~\cite{IR2017}, we find t=0.64.

The experimentally measured thermal conductivity~\cite{Thermal2021} at 295\,K is $\kappa_a$ = 208\,mWK$^{-1}$cm$^{-1}$ and $\kappa_c$ = 57.4\,mWK$^{-1}$cm$^{-1}$.

The length scales $l_a=5\times10^{-3}$\,cm and $l_c=1\times10^{-2}$\,cm is determined by the $50\times100$\,$\mu$m$^{2}$ laser spot. The length $l_b$ is the skin depth, which is calculated from the imaginary part of the index of refraction $n_2$:
\begin{equation}
l_b=\frac{1}{2\pi\omega n_2}~,
\label{eq:lb}
\end{equation}
in which the unit of $\omega$ is cm$^{-1}$ and that of $l_b$ is cm. At the laser wavelength, $n_2$=1.5, and we find $l_b=7\times10^{-6}$\,cm (70\,nm). 

The known values for the variables in Eq.\,\eqref{eq:heatGF} are summarized in Table\,\ref{tab:Heat}. 
Because $\kappa_b$ is unknown, we cannot use Eq.\,\eqref{eq:heatGF} to calculate the laser heating rate. However, if the heating rate is 1.29\,K$\slash$\,mW at 295\,K, as determined from the Stokes/anti-Stokes analysis, the value for $\kappa_b$ should be 1.38\,mWK$^{-1}$cm$^{-1}$.

\begin{table}
\caption{\label{tab:Heat}The various parameters used in calculating the laser heating rate at 295\,K. The transmission coefficient is 0.64.}
\begin{ruledtabular}
\begin{tabular}{cccc}
Quantity (unit)               & a-axis& c-axis\\
$\kappa$ (mWK$^{-1}$cm$^{-1}$)& 208  & 57.4  \\
%$S$ (10$^{-8}$cm$^{2}$)       & 7     & 3.5\\
$l$ (10$^{-3}$cm)             & 5    & 10    \\
\end{tabular}
\end{ruledtabular}
\end{table}

An estimate for the electronic contribution to thermal conductivity $\kappa^{(e)}$ could be  calculated from the electric resistivity $\rho$ by virtue of Wiedemann-Franz law:
\begin{equation}
\kappa^{(e)}=\frac{LT}{\rho}~,
\label{eq:kappa}
\end{equation}
in which $L=2.44\times10^{-5}$\,mW$\Omega$K$^{-2}$ is the Lorenz number. The resistivity along b axis at 295\,K is $\rho_b=2.34\times10^{-1}$\,$\Omega$cm~\cite{Arima2020}, and we find $\kappa^{(e)}_b$ = 0.0308\,mWK$^{-1}$cm$^{-1}$. The ratio of $\kappa^{(e)}_b$ to $\kappa_b$ is consistent with the ratios obtained along a and c axes~\cite{Thermal2021}. Moreover, that $\kappa_b \ll \kappa_{a,c}$ is consistent with the quasi-2D structure of the system~\cite{Graphite2009}.

\begin{figure}[b]
\includegraphics[width=0.4\textwidth]{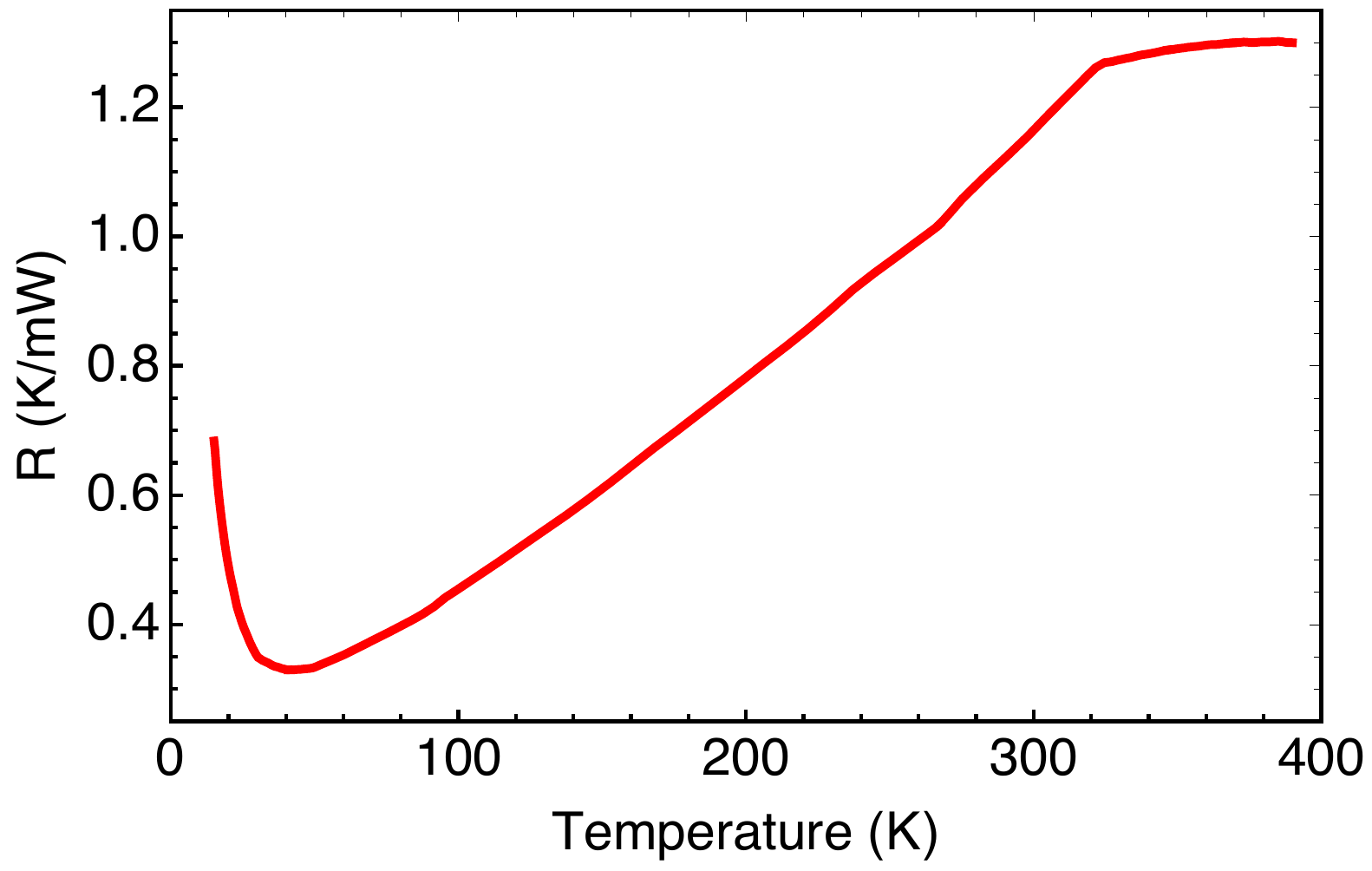}
\caption{\label{fig:Heating}Temperature dependence of the laser heating rate R, calculated by Eq.\,\eqref{eq:heatGF} with the thermal conductivity data taken from Ref.~\cite{Thermal2021}.}
\end{figure}

To illustrate that the laser heating rate is strongly temperature dependent, we use Eq.\,\eqref{eq:heatGF} to calculate the heating rate as a function of temperature. The temperature dependence of $\kappa_a$ and $\kappa_c$ are taken from Ref.~\cite{Thermal2021}. We assume that the ratio of $\kappa_b$ to $\kappa_c$ is temperature independent, and fix this ratio to be $1.38 / 57.4 = 2.40\%$. In Fig.\,\ref{fig:Heating} we show the calculated heating rate. Because of such assumption, the laser heating rate at 295\,K is the same as that determined from the Stokes/anti-Stokes analysis.

\section{The fitting model for the Fano interference\label{sec:AFano}}

Above \Tc, the B$_{2g}^{(1)}$ and B$_{2g}^{(2)}$ phonon modes of Ta$_2$NiSe$_5$ exhibit  strongly asymmetric Fano lineshapes. 
To analyze the physics of this Fano interference features, we propose a model describing three phonon modes individually coupled to broad excitonic continuum. 
The total Raman response is described in the following way:
\begin{equation}
\chi^{\prime\prime}\sim\Im T^{T}GT~,
\label{eq:fanoChi}
\end{equation}
where $T^{T}=(\begin{array}{cc}T_{ph}^{T}&t_{e}\end{array})$ denotes the vertices for light scattering process (the subscript "T" denotes "Transpose"), in which $T_{ph}^{T}=(\begin{array}{ccc}t_{p1}&t_{p2}&t_{p3}\end{array})$. $G$ comprises the Green's functions for the interacting phononic and excitonic excitations that can be obtained by solving the Dyson equation:
\begin{equation}
G=(G_0^{-1}-V)^{-1}.
\label{eq:fanoG}
\end{equation}
In Eq.~(\ref{eq:fanoG})
\begin{equation}
G_0=\begin{pmatrix}G_{ph}^{0} & 0\\
0 & G_{e}^{0}\end{pmatrix}~
\label{eq:fanoG0}
\end{equation}
is the bare Green's function and 
\begin{equation}
V=\begin{pmatrix} 0 & V_{e-ph}\\
V_{e-ph}^T & 0\end{pmatrix}~
\label{eq:fanoV}
\end{equation}
is the exciton-phonon interaction. 
Here the bare phononic Green's function
\begin{multline}
G_{ph}^{0}= \operatorname{diag}[G_{pi}^{0}]\\
= \operatorname{diag}[-(\frac{1}{\omega-\omega_{pi}+i\gamma_{pi}}-\frac{1}{\omega+\omega_{pi}+i\gamma_{pi}})]~,
\label{eq:fanoP}
\end{multline}
correspond to the B$_{2g}^{(i)}$-symmetry phonon modes (i=1,2,3), 
\begin{equation}
G_e=\frac{1}{\omega_{e}^2/\gamma_{e}-i\omega}~
\label{eq:fanoE}
\end{equation}
represents the excitonic continuum, and 
$V_{e-ph}^{T}=(\begin{array}{ccc}v_{1}&v_{2}&v_{3}\end{array})$  
denotes exciton interaction strength with the corresponding phonons. For phononic equation, the parameters $\omega_{pi}$ and $\gamma_{pi}$ have the meaning of the mode frequency and the half width at half maximum (HWHM), respectively. For the excitonic equation, $\omega_{e}$ is the energy of the overdamped excitations and $\gamma_{e}$ represents the relaxation rate. The quantity $\Omega_{e} = \omega_{e}^2/\gamma_{e}$ denotes the energy at which the excitonic continuum has maximum intensity.

The bare phononic and excitonic responses are
\begin{equation}
\chi^{\prime\prime(0)}_{p}=\Im T_{ph}^{T}G_{ph}^0T_{ph}=\sum_{i=1}^3t_{pi}^2\Im G_{pi}^0\, 
\label{eq:ChiP1}
\end{equation}
and 
\begin{equation}
\chi^{\prime\prime(0)}_{e}=t_{e}^2\Im G_e^0\,. 
\label{eq:ChiE1}
\end{equation}
These two expressions are related to Eqsw.~(\ref{eq:ChiP}-\ref{eq:ChiE}) in the main text.

The renormalized phononic and excitonic responses, $\chi^{\prime\prime}_{p}$ and $\chi^{\prime\prime}_{e}$, as well as the interference term $\chi^{\prime\prime}_{int}$ are calculated from the renormalized Green's function $G$:
\begin{equation}
G=\begin{pmatrix}G_{ph} & G_{e-ph}\\
G_{e-ph}^T & G_{e}\end{pmatrix}~,
\label{eq:G}
\end{equation}
where $G_{e-ph}^{T}=(\begin{array}{ccc}G_{1e}&G_{2e}&G_{3e}\end{array})$ represents the interference between the phononic modes and the continuum.

The phononic term is calculated from the 3x3 phononic block:
\begin{equation}
\chi^{\prime\prime}_{p}=\Im T_{ph}^{T}G_{ph}T_{ph}\,;
\label{eq:ChiP2}
\end{equation}
the excitonic term results from the 1x1 excitonic block:
\begin{equation}
\chi^{\prime\prime}_{e}=t_{e}^2\Im G_{e}\,;
\label{eq:ChiE2}
\end{equation}
and the interference term is given by the off-diagonal block
\begin{equation}
\chi^{\prime\prime}_{int}=t_{e} \sum_{i=1}^3t_{pi}\Im G_{ie}\,.
\label{eq:ChiINT2}
\end{equation}
These three terms, as defined, satisfy Eq.\,\eqref{eq:Chi}:
\begin{equation}
\chi^{\prime\prime}=\chi^{\prime\prime}_{p}+\chi^{\prime\prime}_{e}+\chi^{\prime\prime}_{int}.
\label{eq:ChiAgain}
\end{equation}

For the $ac$ spectra below \Tc, especially at about 300\,K, this model cannot properly account for the low frequency Raman response. 
This additional spectral feature is related to Raman coupling to the longitudinal acoustic excitations at finite momenta in the presence of quasi-periodic domain walls, detailed in the main text. 
The Green's function of each individual acoustic mode has form similar to the phononic Green's function:
\begin{equation}
G_{s}(\omega, q)=-(\frac{1}{\omega-c_s q+ir_s q}-\frac{1}{\omega+c_s q+ir_s q})~.
\label{eq:fanoS}
\end{equation}
Its light-scattering vertex $t_s(q)$ and coupling to the excitonic continuum $v_s(q)$ are both proportional to the square root of wavevector: $t_s = \tau_s \sqrt{q}$ and $v_s = \beta_s \sqrt{q}$. The frequency $\omega_s$ and the HWHM $\gamma_s$ are both proportional to the wavevector: $\omega_s = c_s q$ and $\gamma_s = r_s q$.

After introducing this mode, Eqs.~(\ref{eq:fanoG0}-\ref{eq:fanoV}) extend to
\begin{equation}
G_0'=\begin{pmatrix}G_{ph}^0&0&0\\0&G_{s}&0\\0&0&G_{e}\end{pmatrix}\,,
\label{eq:fanoG0LT}
\end{equation}
\begin{equation}
V'=\begin{pmatrix}
0&0&V_{e-ph}\\
0&0& \beta_s \sqrt{q}\\
V_{e-ph}^T & {\beta_s} \sqrt{q} &0\end{pmatrix}\,,
\label{eq:fanoVLT}
\end{equation}
and the vertex for light-scattering process becomes $T^{'T}=\left(\begin{array}{ccc}T_{ph}^{T}&\tau_s \sqrt{q}&t_{e}\end{array}\right)$.

Then the total response function containing coupling to acoustic mode for domain walls $i$ lattice constants apart is given by
\begin{equation}
\chi^{\prime\prime}_i\sim\Im T^{'T}G'T'~.
\label{eq:fanoChii}
\end{equation}

Due to the requirement of causality, the real part of susceptibility should be an even function while the imaginary part should be an odd function. Therefore, the static Raman susceptibility is purely real. The total static Raman susceptibility is given by
\begin{equation}
\chi(\omega=0)\sim\Re T^{T}G(\omega=0)T~.
\label{eq:susT}
\end{equation}
The bare phononic static Raman susceptibility can be obtained from
\begin{equation}
\chi_p^{(0)}(\omega=0)\sim\Re T_{ph}^{T}G_{ph}^0(\omega=0)T_{ph}=\sum_{i=1}^3\frac{2t_{pi}^2\omega_{pi}}{\omega_{pi}^2+\gamma_{pi}^2}~,
\label{eq:susP}
\end{equation}
Similarly, the bare excitonic static Raman susceptibility is given by
\begin{equation}
\chi_e^{(0)}(\omega=0)\sim t_e^2\Re G_e^0(\omega=0)=\frac{t_e^2}{\Omega_{e}}~,
\label{eq:susE}
\end{equation}
For temperature-independent $t_e$, the temperature dependence of $1/\chi_e$ directly follows that of $\Omega_{e}(T) = \omega_{e}^2/\gamma_e$.

Using the Green's function $G'$, which describes the interacting phononic (both optical and acoustic) and excitonic excitations, we can evaluate the renormalization of sound velocity above \Tc. Because the relevant frequencies for the acoustic mode are much lower than that of the optical or excitonic modes, we can keep $\omega$ only in the bare acoustic mode's Green's function:
\begin{equation}
G^{*}=\begin{pmatrix}\frac{\omega_{p1}}{2}&0&0&0&-v_1\\0&\frac{\omega_{p2}}{2}&0&0&-v_2\\0&0&\frac{\omega_{p3}}{2}&0&-v_3\\0&0&0&-\frac{\omega^2-c_s^2q^2}{2c_sq}&-\beta_s\sqrt{q}\\-v_1&-v_2&-v_3&-\beta_s\sqrt{q}&\Omega_e\end{pmatrix}^{-1}~,
\label{eq:A1}
\end{equation}
in which finite linewidths of phonon modes are neglected. The pole of this Green's function is obtained by requiring the determinant of inverse $G^{*}$ to be zero:
\begin{equation}
\frac{\omega^2 - c_{s}^2q^2}{2c_{s}q} = - \frac{\beta_s^2q}{\Omega_{e}-\sum_{i=1}^3\frac{2 v_i^2}{\omega_{pi}}}.
\label{eq:A2}
\end{equation}
Identifying $\omega^2$ as $\tilde{c}_{s}^2q^2$, in which $\tilde{c}_{s}$ represents the renormalized sound velocity above \Tc, we have
\begin{equation}
\tilde{c}_{s}^2 = c_{s}^2-\frac{2\beta_s^2c_{s}}{\Omega_{e}-\sum_{i=1}^3\frac{2 v_i^2}{\omega_{pi}}}.
\label{eq:A3}
\end{equation}

\section{Illustration of the fitting model for the Fano interference\label{sec:AFanoMore}}

\begin{figure}[b]
\includegraphics[width=0.36\textwidth]{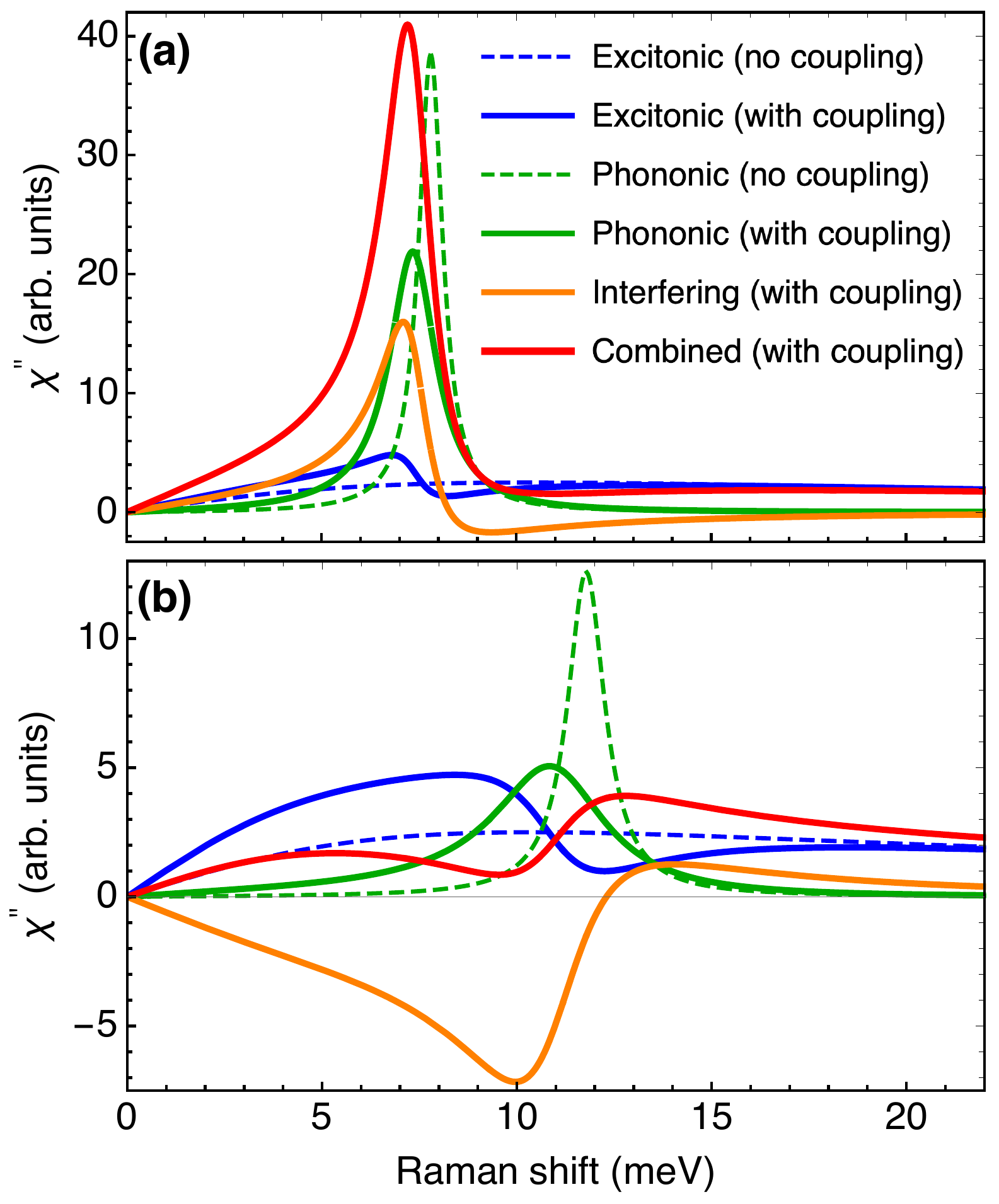}
\caption{\label{fig:FanoD} 
Coupling of the excitonic continuum with (a) the B$_{2g}^{(1)}$ phonon mode and (b) the B$_{2g}^{(2)}$ phonon mode. 
The parameters used for plotting are obtained from fitting the 380\,K spectrum of Ta$_2$NiSe$_5$. 
The Raman response $\chi^{\prime\prime}(\omega)$ of the excitonic and phononic excitations for cases without (dashed lines) and with (solid lines) the effect of coupling are compared. 
The solid red lines correspond to the addition of the solid blue, solid green, and solid orange lines.}
\end{figure}
To provide more insights into the fitting model described in Appendix~\ref{sec:AFano}, we consider a simplified case in which only one phonon mode couples to an excitonic continuum. When exciton-phonon interaction $v$ is zero, the total Raman response is reduced to the sum of the excitonic component and the phononic component:
\begin{equation}
\chi^{\prime\prime}_{0}(\omega)=\frac{t_{e}^2\omega}{\Omega_{e}^2+\omega^2}+\frac{4t_{p}^2\gamma_{p}\omega_{p}\omega}{(\omega^2-\omega_{p}^2)^2 + 2\gamma_{p}^2(\omega^2 + \omega_{p}^2)+\gamma_{p}^4}~,
\label{eq:simpleChi0}
\end{equation}
in which $\Omega_{e}$ stands for $\omega_{e}^2/\gamma_{e}$.

In Fig.~\ref{fig:FanoD} we show the excitonic continuum coupling with the B$_{2g}^{(1)}$ and B$_{2g}^{(2)}$ phonon modes in an individually way to illustrate the effect of coupling. First we discuss the renormalization effect. The renormalized phonon mode shifts in energy and broadens in lineshape. If we neglect the bare phonon width $\gamma_{p}$ for simplicity, the central energy of the renormalized phonon mode, $\omega_{pv}$, can be expressed as
\begin{equation}
\omega_{pv}=\omega_{p}+\frac{v^2(v^2-2\omega_{p}\Omega_{e})}{2(v^2\Omega_{e}+\omega_{p}\Omega_{e}^2+\omega_{p}^3)}~.
\label{eq:wpv}
\end{equation}
For both the B$_{2g}^{(1)}$ and B$_{2g}^{(2)}$ phonon modes, $v^2$ is smaller than $2\omega_{p}\Omega_{e}$; therefore $\omega_{pv}$ is smaller than $\omega_{p}$, meaning the renormalized phonon mode shifts to lower energy. Interestingly, at frequency $\omega_{pv}$ the renormalized excitonic continuum has the same intensity as the unrenormalized continuum; below $\omega_{pv}$ the excitonic response is enhanced while above $\omega_{pv}$ the excitonic response is suppressed.

\begin{figure}[b]
\includegraphics[width=0.36\textwidth]{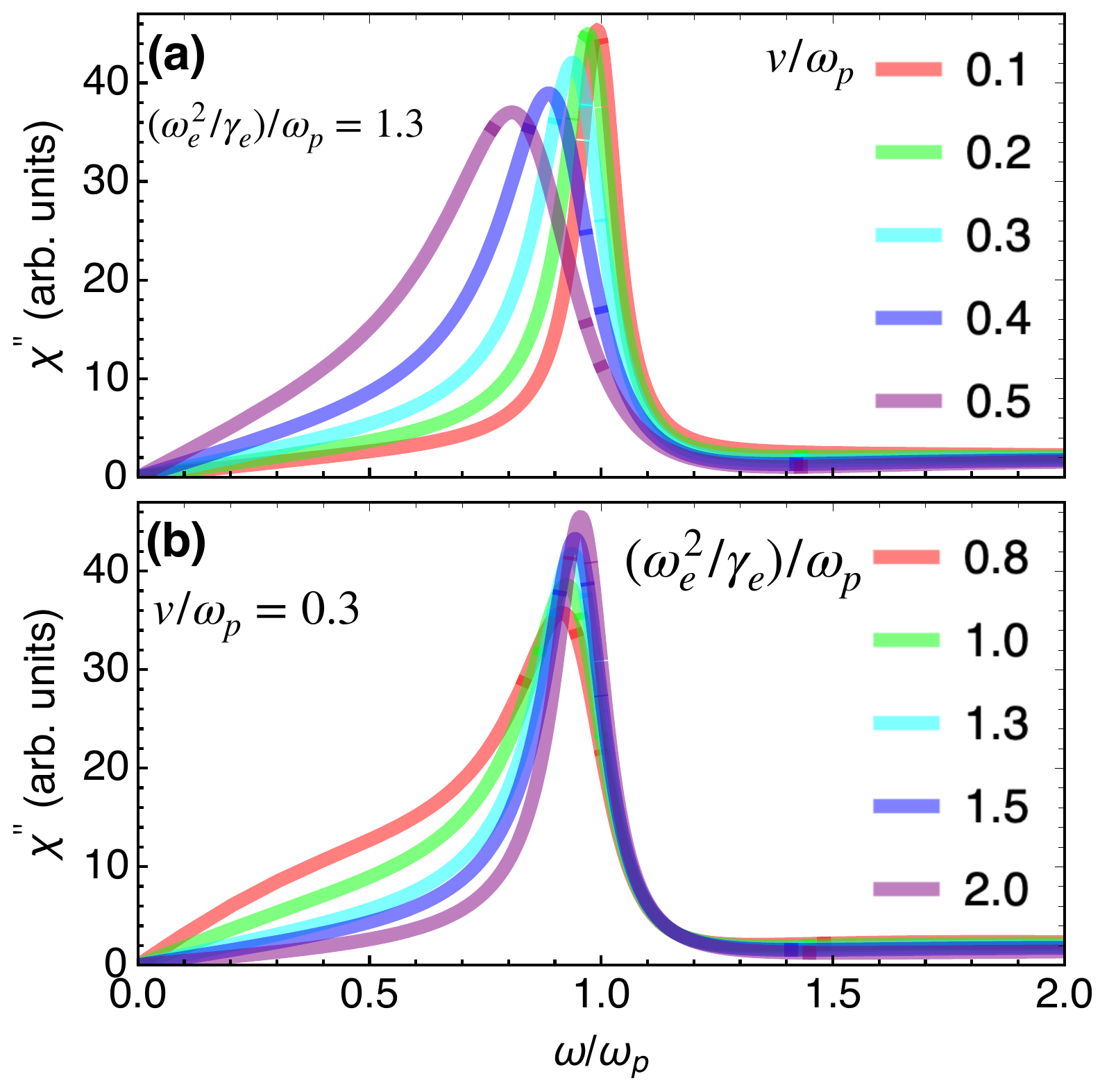}
\caption{\label{fig:FanoP} 
The influence of the coupling strength $v$, and the frequency at which the excitonic continuum has maximum intensity $\Omega_{e} = \omega_{e}^2/\gamma_{e}$ on the Raman response $\chi^{\prime\prime}(\omega)$ of the coupled excitonic and phononic modes.
(a) The combined response for varying $v/\omega_p$ with fixed $\Omega_{e}/\omega_p$=1.3; 
(b) The combined response for varying $\Omega_{e}/\omega_p$ with fixed $v/\omega_p$=0.3. For all panels, $t_p$=3.8\,arb.\,units; 
$t_e$=7.2\,arb.\,units; 
$\omega_p$=7.8\,meV; $\gamma_p$=0.38\,meV. 
The ratios derived from the 380\,K spectrum of Ta$_2$NiSe$_5$ are $\Omega_{e}/\omega_p$=1.3 and $v/\omega_p$=0.3. 
The horizontal axis is normalized to the phonon frequency $\omega_p$.}
\end{figure}
Second we discuss the interference effect. For the interference term, corresponding to the off-diagonal elements of the renormalized Green's function, there is a frequency $\omega_{int}$ at which it changes the sign. 
The frequency $\omega_{int}$ has the following expression
\begin{equation}
\omega_{int}=\sqrt{2\Omega_{e}\gamma_{p}+\gamma_{p}^2+\omega_{p}^2}.
\label{eq:wint}
\end{equation}
We note that $\omega_{int}$ is a bit larger than $\omega_{p}$. 
The shape of the interference term is controlled by sign of $v$: for positive $v$, below the zero-intensity point the interference term has positive intensity while the intensity is negative above the zero-intensity point; for negative $v$, the opposite is true. 
Far away from the resonance, the interference term decays to zero because the phonon mode has negligible intensity at such high energy.

Finally we note that when $\omega$ is much larger than $\omega_p$ and $\gamma_e$, the Raman response is 
In Fig.~\ref{fig:FanoP} we show how the coupling strength $v$, and the energy at which the excitonic continuum has maximum intensity $\Omega_{e} = \omega_{e}^2/\gamma_{e}$ influence the combined Raman response. The lineshape becomes more asymmetric with increasing $v/\omega_p$ [Fig.~\ref{fig:FanoP} (a)]. When $\Omega_{e}$ is varied, we find that the lineshape becomes more asymmetric for smaller $\Omega_{e}$ [Fig.~\ref{fig:FanoP} (b)].

approximately $t_e^2/\omega$. Therefore, $t_e$ can be determined by fitting the data at large $\omega$.

\section{The fitted parameters of the anharmonic decay model for Ta$_2$NiSe$_5$ and Ta$_2$NiS$_5$\label{sec:ATable}}

The temperature dependence of both frequency and FWHM for the most phonon modes of Ta$_2$NiSe$_5$ and Ta$_2$NiS$_5$ modes can be accounted by the anharmonic decay model [Eq.\,(\ref{eq:energyTwo}-\ref{eq:gammaTwo})]. 
The values for the parameters from fitting are summarized in Table\,\ref{tab:Decay}.

\begin{table}
\caption{\label{tab:Decay}
The fitting parameters of the anharmonic decay model [Eq.~(\ref{eq:energyTwo}-\ref{eq:gammaTwo})] for the Raman-active optical phonon modes of Ta$_2$Ni(Se$_{1-x}$S$_x$)$_5$ family. 
The frequencies and FWHM below the phase transition temperature are fitted. 
The anharmonic decay model is not applicable for the B$_{2g}$-symmetry modes of Ta$_2$NiS$_5$. The units are in meV.}
\begin{ruledtabular}
\begin{tabular}{ccccc}
Mode          & $\omega_0$ & $\omega_2$ & $\Gamma_0$ & $\Gamma_2$\\
\hline
&&&&\\
\multicolumn{5}{c}{Ta$_2$NiSe$_5$}\\
&&&&\\
A$_{g}^{(1)}$ & 4.235(16)  & 0.0038(10) & 0.06(4)    & 0.0014(15) \\
B$_{2g}^{(1)}$& 8.721(5)   & 0.0290(11) & 0.073(15)  & 0.019(3)   \\
A$_{g}^{(2)}$ & 12.484(12) & 0.0405(22) & 0.09(3)    & 0.025(6)   \\
B$_{2g}^{(2)}$& 15.35(4)   & 0.070(15)  & 0.05(10)   & 0.09(4)    \\
A$_{g}^{(3)}$ & 17.077(12) & 0.191(5)   & 0.08(5)    & 0.114(23)  \\
B$_{2g}^{(3)}$& 18.45(3)   & 0.050(16)  & 0.11(10)   & 0.08(5)    \\
A$_{g}^{(4)}$ & 22.28(3)   & 0.082(15)  & 0.12(8)    & 0.05(4)    \\
A$_{g}^{(5)}$ & 24.287(11) & 0.129(6)   & 0.05(3)    & 0.072(15)  \\
A$_{g}^{(6)}$ & 27.17(4)   & 0.137(16)  & 0.23(11)   & 0.09(5)    \\
A$_{g}^{(7)}$ & 29.62(6)   & 0.21(3)    & 0.18(16)   & 0.12(9)    \\
A$_{g}^{(8)}$ & 36.75(14)  & 0.30(8)    & 0.3(4)     & 0.09(24)   \\
&&&&\\
\multicolumn{5}{c}{Ta$_2$Ni(Se$_{0.75}$S$_{0.25}$)$_5$}\\
&&&&\\
A$_{g}^{(1)}$ & 4.277(16)  & 0.0039(9)  & 0.12(5)    & 0.0025(34) \\
B$_{2g}^{(1)}$& 8.758(7)   & 0.0322(15) & 0.252(21)  & 0.028(4)   \\
&&&&\\
\multicolumn{5}{c}{Ta$_2$Ni(Se$_{0.33}$S$_{0.67}$)$_5$}\\
&&&&\\
A$_{g}^{(1)}$ & 4.630(12)  & 0.0048(8) & 0.10(2)    & 0.0012(11) \\
B$_{2g}^{(1)}$& 8.739(19)  & 0.032(4)  & 0.479(26)  & 0.022(4)   \\
&&&&\\
\multicolumn{5}{c}{Ta$_2$NiS$_5$}\\
&&&&\\
A$_{g}^{(1)}$ & 5.025(10)  & 0.0040(7)  & 0.068(25) & 0.0014(19)  \\
A$_{g}^{(2)}$ & 16.198(4)  & 0.0610(13) & 0.051(9)  & 0.0359(26)  \\
A$_{g}^{(3)}$ & 19.276(11) & 0.134(4)   & 0.11(3)   & 0.070(11)   \\
A$_{g}^{(4)}$ & 33.92(4)   & 0.111(19)  & 0.28(11)  & 0.15(6)     \\
A$_{g}^{(5)}$ & 36.634(10) & 0.138(5)   & 0.056(28) & 0.112(16)   \\
A$_{g}^{(6)}$ & 40.29(20)  & 0.21(12)   & 0.6(7)    & 0.3(4)      \\
A$_{g}^{(7)}$ & 43.39(6)   & 0.29(4)    & 0.01(14)  & 0.17(9)     \\
A$_{g}^{(8)}$ & 49.72(3)   & 0.23(3)    & 0.04(10)  & 0.45(8)     \\
\end{tabular}
\end{ruledtabular}
\end{table}

\end{document}